\documentclass[%
 reprint,
showpacs,preprintnumbers,
nofootinbib,
 amsmath,amssymb,
 aps,
prd,
]{revtex4-1}

\usepackage{graphicx}
\usepackage{dcolumn}
\usepackage{bm}

\newcommand{\nn}{\nonumber}
\newcommand{\ii}{{\hspace{0.7pt}\mathrm i\hspace{0.7pt}}}

\newcommand{\LPc}{{\mathsf P}}
\newcommand{\LP}{{\mathcal P}}
\newcommand{\LQc}{{\mathsf Q}}
\newcommand{\LQ}{{\mathcal Q}}

\newcommand{\bk}{{\boldsymbol{k}}}
\newcommand{\bn}{{\boldsymbol{n}}}
\newcommand{\bx}{{\boldsymbol{x}}}
\newcommand{\by}{{\boldsymbol{y}}}
\newcommand{\bOmega}{{\boldsymbol{\Omega}}}

\newcommand{\bRR}{{\mathbb{R}}}
\newcommand{\bZZ}{{\mathbb{Z}}}

\newcommand{\rmd}{{\mathrm d}}
\newcommand{\inn}{{\rm in}}
\newcommand{\out}{{\rm out}}
\newcommand{\odd}{{\rm odd}}
\newcommand{\even}{{\rm even}}

\newcommand{\Ree}{{\rm Re\,}}
\newcommand{\Imm}{{\rm Im\,}}

\newcommand{\bra}[1]{\langle #1 \vert}
\newcommand{\ket}[1]{| #1 \rangle}
\newcommand{\bigbra}[1]{\bigl\langle #1 \bigr\vert}
\newcommand{\bigket}[1]{\bigl\vert #1 \bigr\rangle}
\newcommand{\abs}[1]{{\lvert #1 \rvert}}

\allowdisplaybreaks[3]

\begin{document}

\preprint{KUNS-2432 \cr MISC-2013-01}

\title{Propagators in de Sitter space}

\author{Masafumi Fukuma}
\email{fukuma@gauge.scphys.kyoto-u.ac.jp}
\affiliation{
 Department of Physics, Kyoto University  \\
Kyoto 606-8502, Japan
}%
\author{Yuho Sakatani}%
\email{yuho@cc.kyoto-su.ac.jp}
\affiliation{%
Maskawa Institute for Science and Culture,\\ Kyoto Sangyo University, 
Kyoto 603-8555, Japan
}%
\author{Sotaro Sugishita}%
\email{sotaro@gauge.scphys.kyoto-u.ac.jp}
\affiliation{
 Department of Physics, Kyoto University  \\
Kyoto 606-8502, Japan
}%


\date{\today}

\begin{abstract}
In a spacetime with no global timelike Killing vector, 
we do not have a natural choice for the vacuum state of matter fields, 
leading to an ambiguity in defining the Feynman propagators. 
In this paper, taking the vacuum state to be the instantaneous ground state 
of the Hamiltonian at each moment, 
we develop a method for calculating 
wave functions associated with the vacuum 
and the corresponding in-in and in-out propagators. 
We apply this method to free scalar field theory in de Sitter space 
and obtain de Sitter invariant propagators in various coordinate patches. 
We show that the in-out propagator in the Poincar\'{e} patch 
has a finite massless limit 
in a de Sitter invariant form. 
We argue and numerically check that our in-out propagators  
agree with those obtained by a path integral 
with the standard $\ii\varepsilon$ prescription, 
and identify the condition on a foliation of spacetime 
under which such coincidence can happen for the foliation. 
We also show that the in-out propagators satisfy Polyakov's composition law. 
Several applications of our framework are also discussed. 
\end{abstract}

\maketitle


\section{Introduction}

In a spacetime with no global timelike Killing vector, 
we do not have an established prescription 
to define the vacuum state of matter fields. 
The issue exists even at the level of free fields, 
leading to an ambiguity in defining propagators \cite{BD} 
(see also \cite{PT} for recent discussions).

de Sitter space is a typical example of such spacetimes, 
and various vacua have been studied throughout the decades. 
Among them, the Euclidean vacuum 
(or the Bunch-Davies vacuum) \cite{Bunch:1978yq} 
is often used in cosmology 
to describe the physics in the inflationary era.
This is mainly because it is invariant under de Sitter group 
and further satisfies the Hadamard condition, 
the condition essentially stating that a two-point function 
comes to behave in the same way as in flat Minkowski space 
as two points get closer to each other 
(see, e.g., 
\cite{DB, Brunetti:1995rf, Brunetti:1999jn, Hollands:2001nf, Kay:1988mu, Brunetti:2005pr} 
for arguments that physically natural states should satisfy the Hadamard condition). 
Also often studied are a series of vacua 
called the $\alpha$-vacua (or Mottola-Allen vacua) 
\cite{Mottola:1984ar,Allen:1985ux}, 
which are parametrized by a complex number $\alpha$\,. 
They are all de Sitter invariant 
but do not satisfy the Hadamard condition 
except for $\alpha=-\infty$  
which corresponds to the Euclidean vacuum.%
\footnote{
It is pointed out 
on the basis of the in-in formalism (or the Schwinger-Keldysh formalism) 
\cite{Schwinger:1960qe,Keldysh:1964ud}  
that two-point functions for $\alpha$-vacua 
have various pathological behaviors 
(e.g., the breaking of their analyticities) 
when a quantum field has an interaction 
\cite{Goldstein:2003ut, Goldstein:2003qf, Einhorn:2002nu}. 
Note that discussions in favor of the Hadamard condition 
are about the in-in propagators 
and are not applied to the in-out propagators. 
In this paper, we will not further touch on this fundamental issue 
of the Hadamard condition on two-point functions. 
} 

An interesting feature of de Sitter space is 
its thermodynamic property. 
As is pointed out in \cite{Gibbons:1977mu}, 
a particle detector staying in de Sitter space 
and interacting weakly with a scalar field in the Euclidean vacuum 
behaves as if it is in a thermal bath with the temperature 
$T=1/2\pi\ell$\,, where $\ell$ is the de Sitter radius. 
This phenomenon crucially depends on 
the setup where the Euclidean vacuum is taken. 
In fact, other $\alpha$-vacua do not yield such thermal behavior \cite{Bousso:2001mw}.
In this sense, the choice of vacuum is also important 
in understanding the thermodynamic character of curved spacetimes.

In this paper, we take the vacuum of a free scalar field 
to be the instantaneous ground state of the Hamiltonian at each moment. 
We develop a general method to calculate transition amplitudes during finite time intervals
for a quantum mechanical system with time-dependent Hamiltonian, 
and define the propagators 
as the limit of two-point functions 
when the initial and final times are sent to the past and future infinities.%
\footnote{
There had been a study to take the vacuum to be the instantaneous ground state, 
which is sometimes called  the
\emph{instantaneous Hamiltonian diagonalization method} 
(see, e.g., \cite{Grib:1976pw,BD,Castagnino:1985nh} and references therein). 
Our framework has the same principle as that of the method 
in determining the vacuum, 
but has an advantage over the method in that 
it enables us to obtain an explicit form of various propagators for finite time intervals  
and can be applied to a wide class of nonstatic spacetimes.
} 
In our method, wave functions associated with the vacuum 
are automatically determined with no need to consider 
asymptotic boundary conditions such as positive-energy conditions.

We apply the method to construct various Feynman propagators in de Sitter space.  
We treat the principal series (with large mass, $m>(d-1)/2$) 
and the complementary series (with small mass, $m<(d-1)/2$) at the same time, 
and show that the obtained in-in and in-out propagators always take de Sitter invariant forms. 
Furthermore, we show that 
our de Sitter invariant in-out propagator 
has a finite massless limit in the Poincar\'e patch.%
\footnote{
The in-out propagator in the global patch still diverges in the massless limit. 
} 
This is in contrast to the in-in propagators, 
for which the no-go theorem is known 
that there is no de Sitter invariant Fock vacuum 
for massless scalar fields \cite{Allen:1985ux}.

We argue and numerically check that our in-out propagators  
agree with the propagators obtained 
by a path integral with the standard $\ii\varepsilon$ prescription. 
This result is consistent with the de Sitter invariance of the propagators 
since the corresponding path integral is performed 
over a patch which is preserved 
under the infinitesimal action of de Sitter group ${\rm SO}(1,d)$. 
Moreover, our in-out propagators are shown 
to satisfy the composition law \cite{Polyakov:2007mm}, 
which has been claimed by Polyakov recently 
as a principle to be satisfied in order for the propagator 
to be interpreted as representing a sum over paths of a particle moving in a spacetime.

This paper is organized as follows. 
In section \ref{sec:general}, 
we develop a general framework for a given foliation of spacetime 
to calculate wave functions and propagators 
for quantum mechanics with time-dependent Hamiltonian. 
In section \ref{sec:Minkowski} we demonstrate how our prescription works 
in the simplest spacetime, Minkowski space. 
A mathematical detail is given in Appendix \ref{appendix:in-in-matrix}. 
Another well-studied example of asymptotically Minkowski space is investigated 
in Appendix \ref{appendix:asymptotically_Minkowski}. 
We then analyze the de Sitter case in section \ref{sec:deSitter}. 
After giving a brief review on the geometry of de Sitter space 
in subsection \ref{sec:geometry}, 
we discuss the propagators in the Poincar\'e patch in subsection \ref{sec:Poincare} 
and those in the global patch in subsection \ref{sec:global}. 
For the both patches, we first make a mode expansion of a scalar field 
and calculate the propagators for each mode. 
We then make a sum over modes to obtain the propagators in spacetime, 
both of the in-in and in-out types. 
The obtained propagators are found to be 
written with the de Sitter invariant quantity. 
In section \ref{sec:path_integral} 
we introduce the concept of effective noncompactness in the time direction, 
and show that when the foliation meets the condition of effective noncompactness, 
our in-out propagator coincides with that obtained by a path integral 
with the standard $\ii\varepsilon$ prescription. 
We further show that the Poincar\'e and the global patches 
satisfy the condition, 
and confirm the coincidence of the two propagators 
by numerical calculations. 
In section \ref{sec:heat_kernel} 
we prove that our in-out propagators 
have the heat-kernel representation, 
which means that the propagators satisfy Polyakov's composition law 
\cite{Alvarez:2009kq}. 
Section \ref{sec:conclusion} is devoted to discussions and conclusion. 
We leave some of mathematical details 
in Appendices \ref{appendix:propagator_Minkowski}--\ref{appendix:Legendre_integral} 
with useful formulae.  
In Appendix \ref{appendix:alpha} 
we show that each of the in- and out-vacua for the two patches 
can be identified with an $\alpha$-vacuum. 
In Appendix \ref{appendix:iepsilon-different}, 
we consider two possible ways to introduce the $\ii\varepsilon$ prescription, 
and confirm that the two prescriptions give the same analytic expressions 
after taking the limit $\varepsilon\to0$\,.

\section{General framework}
\label{sec:general}

\subsection{Setup}
\label{sec:parameters}

In this paper, we consider 
quantum theory of a free real scalar field $\phi(x)$ 
living in a $d$-dimensional curved spacetime 
with background metric $g_{\mu\nu}$:%
\footnote{
The metric has the signature $(-,+,\dotsc,+)$.
} 
\begin{align}
 S[\phi(x)] = {}- \frac{1}{2}\,\int\rmd^dx\,\sqrt{-g}\,
  \bigl(g^{\mu\nu}\,\partial_\mu\phi\,\partial_\nu\phi +m^2\,\phi^2\bigr)\,.
\label{action}
\end{align}
We assume that the spacetime is globally hyperbolic 
and the foliation of spacetime (i.e., the set of timeslices) 
is already specified. 
We denote the temporal and spatial coordinates 
by $t$ and $\bx$, respectively, 
and the spacetime coordinates by $x=(x^\mu)=(t,\bx)$ $(\mu=0,1,\dotsc,d-1)$. 
We further assume that the metric has the form 
\begin{align}
 \rmd s^2 = {}- N^2(t)\,\rmd t^2 + A^2(t)\, h_{ij}(\bx)\,\rmd x^i\,\rmd x^j 
\nn\\
 (i,j=1,\dotsc ,d-1)\,.
\label{parametrization}
\end{align}
The action is then written as
\begin{align}
 S &= \frac{1}{2}\,\int\rmd t\int\rmd^{d-1}x\,\sqrt{h}\,N \, A^{d-1}\, \nn\\
     &\qquad \times\Bigl(N^{-2}\,\partial_t\phi\,\partial_t\phi 
          +A^{-2}\,\phi\,\Delta_{d-1}\phi -m^2\phi^2\Bigr)\,,
\end{align}
where $\sqrt{h}\equiv \sqrt{\det h_{ij}}$\,, and 
$\Delta_{d-1}\equiv (1/\sqrt{h})\,\partial_i\bigl(\sqrt{h}\,h^{ij}\,\partial_j\bigr)$ 
is the Laplacian for the spatial metric 
$\rmd s_{d-1}^2=h_{ij}(\bx)\,\rmd x^i\,\rmd x^j$\,. 
We have neglected the surface term coming from integration by parts. 
We introduce a complete system $\{Y_n(\bx)\}$ of real-valued 
orthonormal eigenfunctions of $\Delta_{d-1}$ satisfying 
\begin{align}
 \Delta_{d-1}\, Y_n(\bx) = {}- \lambda_n \,Y_n(\bx)\,,\nn\\
 \int\rmd^{d-1}\bx\,\sqrt{h(\bx)}\, Y_n(\bx)\, Y_{n'}(\bx) = \delta_{nn'}\,,
\end{align}
and make a mode expansion of the scalar field as 
\begin{align}
 \phi(x) = \phi(t,\bx) = \sum_n \phi_n(t)\, Y_n(\bx) \,.
\label{mode_expn}
\end{align}
Note that $\phi_n(t)\in\bRR$ 
since $Y_n(\bx)$ are real-valued. 
The action can then be written as a sum of 
the actions for mode functions $\{\phi_n(t)\}$\,: 
\begin{align}
 S_\varepsilon &=\sum_n \int\rmd t \, L_{n,\varepsilon}\bigl(\phi_n(t),\dot\phi_n(t),t\bigr)
\end{align}
with
\begin{align}
 L_{n,\varepsilon}(\phi_n,\dot\phi_n,t)
  &= \frac{\rho(t)}{2}\,\dot\phi_n^2 
  - \frac{\rho(t)\,\omega_n^2(t)}{2}\,\phi_n^2\,,
\\
 \rho(t)&\equiv e^{+\ii\varepsilon}N^{-1}(t)\, A^{d-1}(t) \,,
 \label{rhomega-1} \\
 \omega_n(t) &\equiv e^{-\ii\varepsilon}N(t)\,\sqrt{\lambda_n A^{-2}(t)+m^2} 
\label{rhomega-2}\,.   
\end{align}
This shows that the $n^{\rm th}$ mode function $\phi_n(t)$ behaves as 
a quantum oscillator with time-dependent mass $\rho(t)$ 
and frequency $\omega_n(t)$. 
Here, we have introduced an infinitesimal imaginary part 
$\ii\varepsilon$ $(\varepsilon>0)$ 
in order to discuss the behavior of states near the temporal boundary 
in a well-defined manner.
Note that the combination $\rho(t)\,\omega_n(t)$ is always real.
The quantum oscillator with time-dependent mass $\rho(t)$ 
and frequency $\omega_n(t)$ 
is described by the following time-dependent Hamiltonian
 in the Schr\"{o}dinger picture: 
\begin{align}
 H_{n,\,s}(t) &= H_n\bigl(\phi_{n,\,s},\,\pi_{n,\,s},\,t\bigr)\nn\\
  &=\frac{1}{2\,\rho(t)}\,\pi_{n,\,s}^2
  + \frac{\rho(t)\,\omega_n^2(t)}{2}\,\phi_{n,\,s}^2\,,
\end{align}
where the suffix $s$ indicates that 
the operators are in the Schr\"{o}dinger picture. 
Thus, the theory is reduced to quantum mechanics of 
a set of independent harmonic oscillators 
with time-dependent parameters:
\begin{align}
 H_s(t) &= \sum_n H_{n,\,s}(t)\nn\\
  &=\sum_n \Bigl[\frac{1}{2\,\rho(t)}\,\pi_{n,\,s}^2 
  + \frac{\rho(t)\,\omega_n^2(t)}{2}\,\phi_{n,\,s}^2\Bigr] \,. 
\end{align}
Note that the introduction of $\ii\varepsilon$ in (8) and (9) 
corresponds to the replacement 
$H_{n,\,s}(t)
 =e^{-\ii\varepsilon}\,\bigl[H_{n,\,s}(t)\rvert_{\varepsilon=0}\bigr]$\,,
which makes the Hamiltonian a non-hermitian operator. 
The quantization is accomplished by setting the commutation relations 
\begin{align}
 &[\phi_{n,\,s},\, \pi_{m,\,s}]=\ii\,\delta_{n,m}\,, \\
 &[\phi_{n,\,s},\, \phi_{m,\,s}] = 0 = [\pi_{n,\,s},\, \pi_{m,\,s}]\,.
\end{align}
In the following subsections, we develop a general theory 
to describe the time evolution of states 
for quantum mechanics with such a time-dependent Hamiltonian.

We here make a comment on a subtlety existing in field redefinitions 
(for brevity we set $\varepsilon=0$ below). 
By transforming the mode function $\phi_n(t)$ to 
$\chi_n(t)=\rho^{1/2}(t)\,\phi_n(t)
\equiv e^{\sigma(t)}\,\phi_n(t)$\,, 
one can make the coefficient of the kinetic term to unity: 
\begin{align}
 S[\chi_n(t)] &= \int_{t_i}^{t_f}\!\!\rmd t\,\frac{1}{2}\,
  \bigl[\dot{\chi}_n^2(t)-\Omega_n^2(t)\,\chi_n^2(t)\bigr] \,,
\\
  \Omega_n^2(t) &\equiv \omega_n^2(t)-\bigl(\dot{\sigma}(t)\bigr)^2
  -\ddot{\sigma}(t) \,.
\end{align}
However, it can often happen that $\Omega_n^2(t)$ takes negative values 
for some region of $m^2$ 
even though the original $\omega^2(t)$ is strictly positive.%
\footnote{
A typical example is a scalar field in the Poincar\'{e} patch of de Sitter space. 
One can easily see that $\Omega_n^2(t)$ can be negative 
when the mass is small and $\phi_n(t)$ represents a mode of long wave length. 
} 
Although the physics should be the same 
for the two descriptions using $\phi_n(t)$ and $\chi_n(t)$ 
(as long as $\ii\varepsilon$ is introduced in a consistent way), 
the inverted harmonic potential for $\chi_n(t)$ can easily cause a catastrophe 
when making an analysis based on an approximation such as the WKB approximation. 
In order to avoid this subtlety (and also to keep the original symmetry manifest), 
we will not make such transformations.%
\footnote{
The coefficient of the kinetic term can also be set to unity 
by making a transformation of the time coordinate. 
We will see that physical quantities do not change 
under the transformation (see the last paragraph of subsection \ref{sec:wave}).
} 

\subsection{Quantum mechanics with time-dependent Hamiltonian}
\label{sec:time-dep_Hamiltonian}

To simplify expressions in the following discussions, 
we for a while omit the mode index $n$ and denote the canonical variables 
$\{\phi_{n,\,s},\,\pi_{n,\,s}\}$ in the Schr\"{o}dinger picture 
by $\{q_s,p_s\}$. 
Our Hamiltonian then takes the form
\begin{align}
 H_s(t) = H(q_s, p_s, t) = \frac{1}{2\rho(t)}\,p_s^2 
  + \frac{\rho(t)\,\omega^2(t)}{2}\,q_s^2\,, 
\label{Hamiltonian1}\end{align}
and the system is quantized by setting the commutation relation
\begin{align}
 [q_s,p_s]=\ii\,.
\label{koukan1}
\end{align}
Recall that $\rho(t)=e^{\ii\varepsilon}\,\abs{\rho(t)}$ and 
$\omega(t)=e^{-\ii\varepsilon}\,\abs{\omega(t)}$\,.

We denote by $T_s$ the time at which quantization 
is carried out in the Schr\"odinger picture. 
The Hilbert state ${\cal H}=\{\ket{\psi}\}$ 
with a hermitian inner product $(\psi_1,\psi_2)$ 
is then constructed on the timeslice at $t=T_s$\,,
and the dual space ${\cal H}^\ast=\{\bra\psi\}$ is defined 
with respect to the hermitian inner product 
with the rule $\bra{\psi_1}\equiv \bigl(\ket{\psi_1}\bigr)^\dagger$\,, 
i.e., $\bra{\psi_1}(\ket{\psi_2})=(\psi_1,\psi_2)$. 
The time evolution of a state $\ket{\psi}\in{\cal H}$ 
is governed by the Schr\"odinger equation 
\begin{align}
 \partial_t\, \ket{\psi,t} = -\ii H_s(t)\,\ket{\psi,t}  
\label{Sch_eq}
\end{align}
with the initial condition $\ket{\psi,T_s}=\ket{\psi}$. 
The Schr\"odinger equation can be integrated to the form 
\begin{align}
 \ket{\psi,t} &= U(t,T_s)\,\ket{\psi} \,,
\end{align}
where  $U(t,T_s)$ is the time-evolution operator 
expressed as the time-ordered exponential of $H_s(t)$, 
\begin{align}
 &U(t,T_s) \nn\\
 &\equiv {\rm T} \exp\Bigl(-\ii \int_{T_s}^{t} \rmd t'\, H_s(t')\Bigr)
\nn\\
 &\equiv \lim_{\Delta t_k\to 0}\bigl(1-\ii\Delta t_N H_s(t_N)\bigr)
  \bigl(1-\ii\Delta t_{N-1} H_s(t_{N-1})\bigr)\cdots\nn\\
 &\quad\qquad\times\bigl(1-\ii\Delta t_1 H_s(t_1)\bigr) 
\nn\\
 &\qquad\qquad\Bigl(\begin{array}{c}t=t_N>t_{N-1}>\cdots>t_1>t_0=T_s \cr
  \Delta t_k=t_{k}-t_{k-1}\end{array} \Bigr)\,.
\end{align}
The hermitian conjugate of $\ket{\psi,t}$ is given by
\begin{align}
 \bra{\psi,t} = \bra{\psi}\,U^\dagger(t,T_s)\,.
\end{align}
In addition, we introduce a one-parameter family of states 
for a given state $\overline{\bra{\psi}}\in{\cal H}^\ast$ as 
\begin{align}
 \overline{\bra{\psi,t}} \equiv \overline{\bra{\psi}}\,U^{-1}(t,T_s)\,,
\end{align}
which satisfy
\begin{align}
 \partial_t\,\overline{\bra{\psi,t}} = +\ii\,\overline{\bra{\psi,t}}\,H_s(t)\,,
  \quad \overline{\bra{\psi,T_s}}=\overline{\bra{\psi}}\,.
\end{align}
Note that the pairing of $\overline{\bra{\psi_1}}$ and $\ket{\psi_2}$ 
does not change under the time evolution, 
\begin{align}
 \overline{\langle \psi_1,t \vert} \psi_2, t \rangle
  = \overline{\langle \psi_1 \vert} \psi_2 \rangle\,,
\end{align}
although this is not the case for $\bra{\psi_1,t} \psi_2,t \rangle$ 
when $\varepsilon\neq 0$ 
because the time evolution operator is then not unitary, 
$U^{-1}(t,T_s) \neq U^{\dagger}(t,T_s)$\,.

The spectrum of the Hamiltonian $H_s(t)$ can be easily found 
by introducing, as usual, 
a pair of operators,%
\footnote{
They are Schr\"{o}dinger operators 
and the time dependence comes only through the parameters 
$\rho(t)$ and $\omega(t)$.
} 
\begin{align}
 a_s(t) &\equiv \sqrt{\frac{\rho(t)\,\omega(t)}{2}}\,q_s 
               + \ii \sqrt{\frac{1}{2\rho(t)\,\omega(t)}}\,p_s \,,
\label{ann}
\\
 a^\dagger_s(t) &\equiv \sqrt{\frac{\rho(t)\,\omega(t)}{2}}\,q_s 
                      - \ii \sqrt{\frac{1}{2\rho(t)\,\omega(t)}}\,p_s\,.
\label{cre}
\end{align}
We call $a_s(t)$ and $a^\dagger_s(t)$ the annihilation and creation operators 
at time $t$\,. 
Note that $a_s(t)$ and $a^\dagger_s(t)$ are 
hermitian conjugate to each other,
because $\rho(t)\,\omega(t)$ is positive and 
$q^\dagger_s=q_s$ and $p^\dagger_s=p_s$\,.
From the commutation relation \eqref{koukan1}, we have
\begin{align}
 [a_s(t),a^\dagger_s(t)] =1 \,.
\end{align}
The Hamiltonian \eqref{Hamiltonian1} can then be rewritten as
\begin{align}
 H_s(t)&= \frac{\omega(t)}{2}\,\bigl[a^\dagger_s(t)\,a_s(t)+a_s(t)\, a^\dagger_s(t)\bigr] \nn\\
  &=\omega(t)\,\Bigl[a^\dagger_s(t)\,a_s(t)+\frac{1}{2}\Bigr] \,.
\end{align}

We define the state $\ket{0_t, t}$  
as that which vanishes when acted on by $a_s(t)$:
\begin{align}
 a_s(t)\,\ket{0_t , t} = 0\,.
\end{align}
Accordingly, the state 
$\overline{\bra{0_t, t}}\equiv \bra{0_t, t}=\ket{0_t, t}^\dagger$ satisfies
\begin{align}
 \overline{\bra{0_t, t}}\, a_s^\dagger(t) = 0\,.
\end{align}
Then the right and left eigenstates of $H_s(t)$ are given by
\begin{align}
 \ket{n,t}_t&\equiv \frac{1}{n!}\,\bigl[a^\dagger_s(t)\bigr]^n\,\ket{0_t,t}\,,\\
 {}_t\overline{\bra{n,t}}
  &\equiv \frac{1}{n!}\,\overline{\bra{0_t,t}}\,\bigl[a_s(t)\bigr]^n=\ket{n,t}_t^\dagger\,,
\end{align}
which satisfy
\begin{align}
 H_s(t)\,\ket{n,t}_t 
   &= \Bigl(n+\frac{1}{2}\Bigr)\,\omega(t)\,\ket{n,t}_t\,, \\
 {}_t\overline{\bra{n,t}}\,H_s(t)
   &=\Bigl(n+\frac{1}{2}\Bigr)\,\omega(t)\,{}_t\overline{\bra{n,t}}\,.
\end{align}
We call $\ket{0_t,t}=\ket{0,t}_t$ the ground state (or the vacuum) at time $t$, 
since this is the minimum energy state at the moment if $\varepsilon=0$\,.

\begin{figure}[htbp]
\begin{center}
\includegraphics[width=7cm]{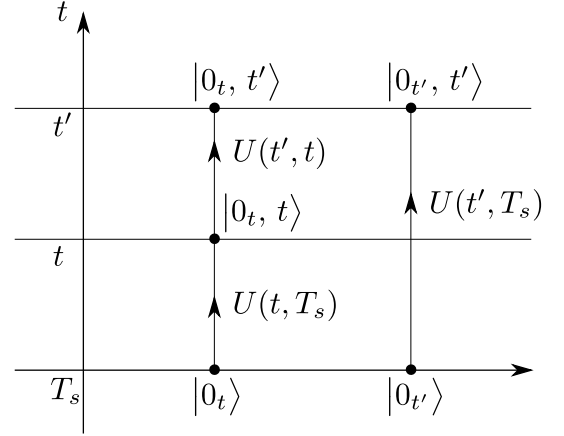}
\begin{quote}
\caption{Time evolution of states. 
The system is quantized in the Schr\"{o}dinger picture 
on the timeslice at $T_s$\,.   
$\ket{0_t,t}$ is the state annihilated by the Schr\"{o}dinger operator $a_s(t)$, 
while $\ket{0_t}$ is the state annihilated by the Heisenberg operator $a(t)$.
\label{fig:state}}
\end{quote}
\end{center}
\vspace{-6ex}
\end{figure}

It is important to note that 
the ground state at time $t'$, $\ket{0_{t'}, t'}$, 
is generically different from the state $\ket{0_t, t'}$; 
the latter is obtained as a time evolution 
of the ground state $\ket{0_t, t}$ at time $t$, 
$\ket{0_t,t'}=U(t',t)\ket{0_t,t}$ 
(see Fig.~\ref{fig:state}). 
Note also that since the Hamiltonian is already specified at each time, 
the vacuum state is uniquely determined, 
and there is no freedom to introduce other vacuum states 
through Bogoliubov transformations. 

\subsection{Heisenberg picture}
\label{sec:Heisenberg}

Now we move from the Schr\"odinger picture to the Heisenberg picture. 
Given a Schr\"odinger operator $O_s(t)$ 
[possibly depending on $t$ through the parameters 
involved when constructing the operator as in Eq.~\eqref{ann}], 
we define the corresponding Heisenberg operator as
\begin{align}
 O(t) \equiv U^{-1}(t,T_s)\, O_s(t)\, U(t,T_s)\,,
\label{heisen}
\end{align}
which satisfies the Heisenberg equation, 
\begin{align}
 &\qquad\dot{O}(t) = \ii\, [H(t),\,O(t)] + \frac{\partial O(t)}{\partial t}
\nn\\
 &\Bigl(\frac{\partial O(t)}{\partial t} 
  \equiv U^{-1}(t,T_s)\,\frac{\partial O_s(t)}{\partial t}\,U(t,T_s)\Bigr)\,.
\end{align}
For our harmonic oscillator, 
the time evolution of the canonical variables is given by
\begin{align}
 \dot{q}(t) &= \ii\, [H(t),\,q(t)] = \frac{p(t)}{\rho(t)} \,, \\
 \dot{p}(t) &= \ii\, [H(t),\,p(t)] = {}- \rho(t)\,\omega^2(t)\, q(t)\,,
\end{align}
and by eliminating $p(t)$, we obtain the differential equation
\begin{align}
 \frac{\rmd}{\rmd t} \Bigl(\rho(t)\,\frac{\rmd}{\rmd t} \, q(t) \Bigr)
 + \rho(t)\,\omega^2(t)\,q(t) =0 \,.
\label{EOM}
\end{align}
This is certainly the equation of motion derived from the Lagrangian 
$L(q,\dot{q},t)=\rho(t)\,\dot{q}^2/2 - \rho(t)\,\omega^2(t)\,q^2/2$\,. 
Note that if we had used $U^{\dagger}(t,T_s)$ in Eq.~\eqref{heisen} 
instead of $U^{-1}(t,T_s)$, 
the equation of motion could not be reproduced correctly 
when $\varepsilon\neq0$\,.

In the Heisenberg picture, the annihilation and creation operators 
become
\begin{align}
 a(t) &\equiv U^{-1}(t,T_s)\,a_s(t)\,U(t,T_s) \nn\\
      &=\sqrt{\frac{\rho(t)\,\omega(t)}{2}}\,q(t) 
       +\ii \sqrt{\frac{1}{2\rho(t)\,\omega(t)}}\,p(t)\,,
\label{ann2}
\\
 \bar{a}(t) &\equiv  U^{-1}(t,T_s)\,a^\dagger_s(t)\,U(t,T_s) \nn\\
      &= \sqrt{\frac{\rho(t)\,\omega(t)}{2}}\,q(t) 
          -\ii \sqrt{\frac{1}{2\rho(t)\,\omega(t)}}\,p(t) \,.
\label{cre2}
\end{align}
They satisfy the commutation relation
\begin{align}
 [a(t), \bar{a}(t)] =1\,, 
\label{koukan2}
\end{align}
but are not hermitian conjugate to each other 
when $\varepsilon\neq0$\,. 
The Hamiltonian is then expressed as
\begin{align}
 &H(t) \equiv U^{-1}(t,T_s)\,H_s(t)\,U(t,T_s)\nn\\
 &=\frac{p^2(t)}{2\rho(t)} + \frac{\rho(t)\,\omega^2(t)\,q^2(t)}{2} 
  =\omega(t)\,\Bigl[\bar{a}(t)\,a(t)+\frac{1}{2}\Bigr] \,.
\end{align}
Note that the states 
$\ket{0_t}\equiv \ket{0_t,T_s}=U^{-1}(t,T_s)\,\ket{0_t,t}$ and 
$\overline{\bra{0_t}}\equiv \overline{\bra{0_t,T_s}}
=\overline{\bra{0_t,t}}\,U(t,T_s)
=\bra{0_t,t}\,U(t,T_s)$ 
(see Fig.~\ref{fig:state}) 
satisfy the equations
\begin{align}
 a(t)\,\ket{0_t} &= 0 = \bra{0_t}\,a^\dagger(t) \,, \\
 \overline{\bra{0_t}}\, \bar{a}(t) &= 0 = \bar{a}^\dagger(t)\,\overline{\ket{0_t}} \,.
\end{align}
$\overline{\bra{0_t}}$ may differ from $\bra{0_t}$ 
since $U(t,T_s)$ is not unitary when $\varepsilon\neq 0$\,.

We denote  by $\{f(t),g(t)\}$ 
a pair of linearly independent c-number solutions of \eqref{EOM}. 
One can easily show that their Wronskian,
\begin{align}
 W[f,g](t) \equiv f(t)\,\dot{g}(t)-\dot{f}(t)\,g(t)\,,
\end{align}
satisfies the equation
\begin{align}
 \frac{\rmd}{\rmd t} (\rho\, W[f,g])
 =f\,\frac{\rmd}{\rmd t}\bigl(\rho\,\dot{g}\bigr)
  -\frac{\rmd}{\rmd t}\bigl(\rho\,\dot{f}\bigr)\,g 
 =0 \,.
\end{align}
Thus the combination (to be called the weighted Wronskian)
\begin{align}
 W_\rho[f,g] \equiv \rho(t)\,W[f,g](t)
\end{align}
does not depend on $t$. 
Since $q(t)$ is also a solution of Eq.~\eqref{EOM}, 
we can expand canonical variables $q(t)$ and $p(t)$ as
\begin{align}
 q(t) &=c_1\, f(t) + c_2\, g(t) \,, 
\label{q_c12}
\\
 p(t) &=\rho(t)\, \dot{q}(t) 
 = \rho(t)\,\bigl[\,c_1\,\dot{f}(t) +c_2\,\dot{g}(t) \,\bigr]\,,
\label{p_c12}
\end{align}
where $c_1$ and $c_2$ are some time-independent quantum operators
living in a space spanned by $q_s$ and $p_s$ with complex coefficients. 
Equations \eqref{q_c12} and \eqref{p_c12} can be solved 
with respect to $c_1$ and $c_2$ as
\begin{align}
 \begin{pmatrix}
  c_1 \\ c_2
 \end{pmatrix}
 = \frac{1}{W_\rho[f,g]}
 \begin{pmatrix}
  \rho\,\dot{g} &-g \cr
 -\rho\,\dot{f} & f 
 \end{pmatrix}(t)\,
 \begin{pmatrix}
  q(t) \\ p(t)
 \end{pmatrix} \,.
\end{align}
This can be further rewritten by using \eqref{ann2} and \eqref{cre2} as
\begin{align}
 \begin{pmatrix}
  c_1 \\ c_2
 \end{pmatrix}
 = C(t)\, 
 \begin{pmatrix}
 a(t) \\ \bar{a}(t)
 \end{pmatrix}\,,
\label{c12-a}
\end{align}
where
\begin{align}
 C(t) &\equiv \frac{1}{W_\rho[f,g]\,\sqrt{2 \rho(t)\,\omega(t)}}
   \begin{pmatrix}
     v(t) &\bar{v}(t)  \cr -u(t)  &-\bar{u}(t) 
   \end{pmatrix} 
\label{Ct}
\end{align}
with
\begin{align}
 \biggl\{\begin{array}{l}
  u(t) \\ \bar{u}(t)
 \end{array}\biggr\}
  &\equiv \rho(t)\,\bigl[\,\dot{f}(t)\pm \ii \omega(t)\,f(t)\,\bigr] \,,
\label{uuvv-1} \\
 \biggl\{\begin{array}{l}
  v(t) \\ \bar{v}(t)
 \end{array}\biggr\} 
  &\equiv \rho(t)\,\bigl[\,\dot{g}(t)\pm \ii \omega(t)\,g(t)\,\bigr] \,. 
\label{uuvv-2}
\end{align}
Note that 
\begin{align}
  \bigl(u\,\bar{v}-v\,\bar{u})(t) &= 2\ii \rho(t)\,\omega(t)\,W_\rho[f,g] \,,
\label{uv}
\\
 \det C(t) &= \frac{\ii}{W_\rho[f,g]}~~\bigl(=\mbox{const})\,,
\label{detC}
\\
 C^{-1}(t) &= \frac{-\ii}{\sqrt{2\rho(t)\,\omega(t)}}
  \begin{pmatrix}
   -\bar{u}(t) &-\bar{v}(t) \cr u(t)  &v(t) 
  \end{pmatrix} \,.
\label{Ct_inv}
\end{align}

\subsection{Bogoliubov coefficients for finite time intervals}
\label{sec:Bogoliubov}

The Bogoliubov coefficients from time $t'$ to time $t$ are defined by
\begin{align}
 \begin{pmatrix}
  a(t) \\ \bar{a}(t)
 \end{pmatrix}
 \equiv 
 \begin{pmatrix}
   \bar{\alpha} &-\bar{\beta} \\
  -\beta &\alpha 
 \end{pmatrix}(t;t')
 \begin{pmatrix}
  a(t') \\ \bar{a}(t')
\end{pmatrix} . 
\label{bogo} 
\end{align}
Since the operators $\{c_1,c_2\}$ in \eqref{c12-a} do not depend on time, 
we have  
\begin{align}
 \begin{pmatrix}
  c_1 \\ c_2
 \end{pmatrix}
 = C(t) \begin{pmatrix}
         a(t) \cr \bar{a}(t)
        \end{pmatrix} 
 = C(t') \begin{pmatrix}
          a(t') \cr \bar{a}(t')
 \end{pmatrix} , 
\label{atime}
\end{align}
from which we find   
\begin{align}
 &\begin{pmatrix}
   \bar{\alpha} &-\bar{\beta} \\
  -\beta &\alpha 
 \end{pmatrix}(t;t')
  = C^{-1}(t)\,C(t') \nn \\
 &= \frac{-\ii}{2W_\rho[f,g]\,
                \sqrt{\rho(t)\,\omega(t)\,\rho(t')\,\omega(t')}}\nn\\
 &\times\begin{pmatrix}
   u(t')\,\bar{v}(t)-v(t')\,\bar{u}(t) &
   \bar{u}(t')\,\bar{v}(t)-\bar{v}(t')\,\bar{u}(t) \\
   v(t')\,u(t)-u(t')\,v(t) &\bar{v}(t')\,u(t)-\bar{u}(t')\,v(t) 
  \end{pmatrix} .
\end{align}
This is the fundamental formula to express the Bogoliubov coefficients 
in terms of a given set of independent solutions $\{f(t),\,g(t)\}$\,.

Due to the commutation relations \eqref{koukan2}, 
the Bogoliubov coefficients should satisfy the relation
\begin{align}
 \bigl(\alpha\, \bar{\alpha} -\beta \,\bar{\beta} \bigr)(t;t') =1 \,.
\label{ab_norm}
\end{align}
This can be directly checked by using the identity \eqref{detC} 
as 
\begin{align}
 (\alpha \bar\alpha-\beta \bar\beta)(t;t')&=\det C^{-1}(t)\,\det C(t') \nn\\
 &=\frac{W_\rho[f,g]}{W_\rho[f,g]}=1\,.
\end{align}
Note that 
$\bar{\alpha} \neq \alpha^{\ast}$ and $\bar{\beta} \neq \beta^{\ast}$ 
when $\varepsilon\neq0$\,. 

\subsection{Wave functions}
\label{sec:wave}

Using the Bogoliubov coefficients, 
we can express the Heisenberg operators $q(t)$ 
with the creation and annihilation operators at a different time $t_I$ as follows:%
\footnote{
In the following, we will use the shorthand notation 
such as $f_I\equiv f(t_I)$ or $\dot{f}_I\equiv \dot{f}(t_I)$ 
when a quantity is evaluated at time $t_I$\,. 
} 
\begin{align}
 q(t)&=\frac{1}{\sqrt{2\rho(t)\,\omega(t)}}\, \bigl(a(t)+\bar{a}(t)\bigr) \nn\\
     &=\frac{1}{\sqrt{2\rho(t)\,\omega(t)}}
       \bigl( \bar{\alpha}(t;t_I)\,a_I
             -\bar{\beta}(t;t_I)\,\bar{a}_I\nn\\
     &\qquad\qquad\qquad\quad +\alpha(t;t_I)\,\bar{a}_I
             -\beta(t;t_I)\,a_I \bigr) 
\nn\\
 &\equiv \varphi(t;t_I)\,a_I +\bar{\varphi}(t;t_I)\,\bar{a}_I \,,
\end{align}
where we have defined functions $\varphi(t;t_I)$ and $\bar{\varphi}(t;t_I)$ 
(to be called {\em wave functions}) as
\begin{align}
 \varphi(t;t_I)
 &\equiv \frac{1}{\sqrt{2\rho(t)\,\omega(t)}}\,
 \bigl(\bar{\alpha}(t;t_I)-\beta(t;t_I)\bigr)\nn\\
 &=\frac{1}{W_\rho[f,g]\, \sqrt{2\rho_I\,\omega_I}}
 \bigl(v_I\, f(t)-u_I\, g(t)\bigr) 
\label{phiin}\,,
\\
 \bar{\varphi}(t;t_I)
 &\equiv \frac{1}{\sqrt{2\rho(t)\,\omega(t)}}\,
 \bigl(\alpha(t;t_I)-\bar{\beta}(t;t_I)\bigr)\nn\\
 &=\frac{1}{W_\rho[f,g]\, \sqrt{2\rho_I\, \omega_I}}
 \bigl(\bar{v}_I\, f(t)-\bar{u}_I\, g(t)\bigr) \,.
\label{phiout}
\end{align}
By using the $t$-independence of $W_\rho[f,g]$ and Eq.~\eqref{uv}, 
we can show that $\varphi(t;t_I)$ and $\bar{\varphi}(t;t_I)$ 
are normalized as
\begin{align}
 W_\rho[\varphi(t;t_I),\bar{\varphi}(t;t_I)] = \ii 
  \quad \bigl(\,\forall t\,,\,\forall t_I \bigr)\,. 
\label{wronskian}
\end{align}
Moreover, the following relation holds for any value of $t$:
\begin{align}
 &\begin{pmatrix}
  \bar{\alpha} &-\bar{\beta} \\ -\beta &\alpha 
 \end{pmatrix}(t_1;t_0)\nn\\
 &= -\ii 
\begin{pmatrix}
 -W_\rho[\bar{\varphi}(t;t_1),\varphi(t;t_0)] 
  &-W_\rho[\bar{\varphi}(t;t_1),\bar{\varphi}(t;t_0)] \\
  W_\rho[\varphi(t;t_1),\varphi(t;t_0)] 
  &W_\rho[\varphi(t;t_1),\bar{\varphi}(t;t_0)] 
\end{pmatrix}
\,. 
\label{albe}
\end{align}

We are now in a position to make a few comments. 

\noindent{\bf basis independence}

We can show that the Bogoliubov coefficients 
and the wave functions $\{\varphi(t;t_I), \bar{\varphi}(t;t_I)\}$ 
do no depend on the choice of a pair of independent solutions $\{f(t),g(t)\}$\,, 
as they should.
In fact, suppose that we take another pair $\{f'(t),g'(t)\}$.  
They should be expressed as linear combinations of $\{f(t),g(t)\}$ of the form
\begin{align}
 \bigl( f'(t) \ \, g'(t) \bigr)
 &=  \bigl( f(t) \ \, g(t) \bigr)\, \Xi \qquad (\Xi \in GL(2,\mathbb{C}))\,,
\end{align} 
from which we have 
\begin{align}
 \begin{pmatrix}
   f' & g' \cr \dot{f}' & \dot{g}' \\
 \end{pmatrix}(t)
   =
 \begin{pmatrix}
   f & g \cr \dot{f} & \dot{g} \\
 \end{pmatrix}\!(t)\,\, \Xi \,,\quad 
 C'(t) = \Xi^{-1}\,C(t) \,.
\end{align}
The new Bogoliubov coefficients associated with the choice $\{f'(t),g'(t)\}$ 
then become
\begin{align}
 &\begin{pmatrix}
  \bar{\alpha}' &-\bar{\beta}' \\ -\beta' &\alpha' 
 \end{pmatrix}(t;t_I)
  = \bigl[C'(t)\bigr]^{-1}\,C'(t_I) 
   \nn\\
 &= C^{-1}(t)\,\Xi\,\Xi^{-1}\,C(t_I) = C^{-1}(t)\,C(t_I) \nn\\
 &= \begin{pmatrix}
      \bar{\alpha} &-\bar{\beta} \\ -\beta &\alpha 
    \end{pmatrix}(t;t_I) \,,
\end{align}
which shows the basis-independence of the Bogoliubov coefficients. 
The wave functions 
$\{\varphi(t;t_I), \bar{\varphi}(t;t_I)\}$ 
are also basis independent 
since they are expressed 
by the basis-independent Bogoliubov coefficients 
[see Eqs.~\eqref{phiin} and \eqref{phiout}]. 

\noindent{\bf lapse independence}

We can show that the Bogoliubov coefficients  
and the wave functions $\varphi(t;t_I)$ behave as scalar functions 
under the temporal reparametrizations preserving the foliation of spacetime. 
In fact, for such reparametrization $t\rightarrow \tilde{t}=\tilde{t}(t)$\,, 
the pull-back of the lapse function $N(t)$ 
[see Eq.~\eqref{parametrization}] is given by 
$N(t) \to \tilde{N}(t)=(\rmd \tilde{t}/\rmd t)\,N\bigl(\tilde{t}(t)\bigr)$\,, 
and we can choose a new pair of solutions $\{\tilde{f}(t),\tilde{g}(t)\}$ 
as $\tilde{f}(t)=f\bigl(\tilde{t}(t)\bigr)$ 
and $\tilde{g}(t)=g\bigl(\tilde{t}(t)\bigr)$. 
Then, we can easily show that the functions $\rho(t)\,\omega(t)$, 
$u(t)$, $v(t)$, and $W_\rho[f(t),g(t)]$ 
transform as scalar functions under the reparametrization. 
Since the Bogoliubov coefficients and the wave function $\varphi(t;t_I)$ 
are written as combinations of these functions, 
they also transform as scalar functions. 
This means that there is no need to care about the temporal reparametrization 
[i.e.~the choice of the lapse function $N(t)$]
when we construct vacua.

\subsection{Feynman propagators}
\label{sec:Feynman}

We consider the region $t_i < t_0 \leq \{t,t'\} \leq t_1 < t_f$\,, 
where $t_f$ and $t_i$ are the future and the past boundaries 
of the spacetime region we consider. 
The in-out and in-in propagators are defined with the following two steps.

\noindent
\underline{\bf step 1}\\[1mm] 
We first introduce the following two-point functions 
from our wave functions $\varphi(t;t_I)$ and $\bar{\varphi}(t;t_I)$:%
\footnote{
If we instead use $G'_{10}(t,t'; t_1,t_0) 
 = \bra{0_{t_1}}\, q^\dagger(t_>)\, q(t_<) \,\ket{0_{t_{0}}}
    /\langle 0_{t_1}\rvert 0_{t_{0}} \rangle$, 
then the corresponding in-out propagator will not coincide with the propagator 
obtained by the standard path integral (see section \ref{sec:path_integral}), 
and thus we do not consider this choice in this paper. 
By contrast, the in-in propagator still has options for its definition 
(e.g.,  $G'_{00}(t,t'; t_0,t_0) 
 = \bra{0_{t_0}}\, {\rm T}\,q(t)\, q(t') \,\ket{0_{t_{0}}}
    /\langle 0_{t_0}\rvert 0_{t_{0}} \rangle$), 
and we leave as a future work a detailed study of such options 
as well as an investigation of the relation to the path integral 
based on the Schwinger-Keldysh formalism 
\cite{Schwinger:1960qe,Keldysh:1964ud}. 
} 
\begin{align}
 &G_{10}(t,t'; t_1,t_0) 
  \equiv \frac{\overline{\bra{0_{t_1}}}\, {\rm T}\, q(t)\, q(t') \,\ket{0_{t_{0}}}}
    {\overline{\langle 0_{t_1}\rvert} 0_{t_{0}} \rangle} \nn \\
 &= \frac{\ii}{W_\rho[\varphi(s;t_1),\bar{\varphi}(s;t_0)]}
    \,\varphi(t_{>};t_1)\,\bar{\varphi}(t_{<};t_0)
\label{G10} 
\nn\\
 &\qquad\qquad\qquad\qquad\qquad\qquad\qquad\mbox{($s$: arbitrary)}\,,
\\[3mm]
 &G_{00}(t,t'; t_0,t_0)
  \equiv \frac{\bra{0_{t_0}}\,q^\dagger(t_>)\, q(t_<)\, \ket{0_{t_{0}}}}
         {\langle 0_{t_0}\rvert 0_{t_{0}} \rangle} \nn \\
 &= \frac{\ii}{V_\rho[\bar{\varphi}^\ast(s;t_0),\,\bar{\varphi}(s;t_0)](T_s)}\,
    \bar{\varphi}^\ast(t_{>};t_0)\,\bar{\varphi}(t_{<};t_0) \,,
\label{G00}
\end{align}
where $t_{>}\equiv {\rm max}(t,t')$\,, $t_{<}\equiv {\rm min}(t,t')$
and $V_\rho[f,g](s)\equiv \rho(s)\,f(s)\,\dot{g}(s)
-\rho^\ast(s)\,\dot{f}(s)\,g(s)$\,.%
\footnote{
Note that when $\varepsilon=0$, 
$V_\rho[f,g](s)$ coincides with $W_\rho[f,g](s)$ 
and thus is constant in $s$\,. 
Otherwise it may depend on $s$\,. } 

\noindent
\underline{\bf step 2}\\[1mm] 
We then define the in-out and in-in propagators  
by sending $t_0$ and $t_1$ to the values at the temporal boundary:
\begin{align}
 G^{\out/\inn}(t,t') 
  &\equiv \lim_{\genfrac{}{}{0pt}{}{t_0 \to t_i}{t_1 \to t_f}}
  G_{10}(t,t'; t_1,t_0)\,,
\label{G_out_in}
\\
 G^{\inn/\inn}(t,t') 
  &\equiv \lim_{t_0 \to t_i}G_{00}(t,t'; t_0,t_0) \,.
\label{G_in_in}
\end{align}

We here make a few comments. 
To obtain the last expression of \eqref{G10}, 
we use the following identities which are direct consequences 
of Eqs.~\eqref{bogo}, \eqref{ab_norm} and \eqref{albe}:
\begin{align}
 a_1&=\bar{\alpha}(t_1;t_0)\, a_0 -\frac{\bar{\beta}(t_1;t_0)}{\alpha(t_1;t_0)}\,
      \bigl(\bar{a}_1 +\beta(t_1;t_0)\, a_0\bigr) \nn\\
    &=\frac{1}{\alpha(t_1;t_0)}
      \bigl(a_0 -\bar{\beta}(t_1;t_0)\, \bar{a}_1 \bigr) \,, \\
 \frac{\overline{\langle 0_{t_1}\rvert} \,a_1\, \bar{a}_0\,\lvert 0_{t_{0}} \rangle}
           {\overline{\langle 0_{t_1}\rvert} 0_{t_{0}} \rangle} 
    &= \frac{1}{\alpha(t_1;t_0)}\,
      \frac{\overline{\langle 0_{t_1}\rvert}
      \bigl(a_0 -\bar{\beta}(t_1;t_0)\, \bar{a}_1 \bigr)\, 
      \bar{a}_0\,\lvert 0_{t_{0}} \rangle}
           {\overline{\langle 0_{t_1}\rvert} 0_{t_{0}} \rangle} \nn\\
  &= \frac{1}{\alpha(t_1;t_0)}
\nn\\
  &= \frac{\ii}{W_\rho[\varphi(s;t_1),\bar{\varphi}(s;t_0)]} 
  \quad \mbox{($s$: arbitrary)}\,. 
\end{align}
We then have
\begin{align}
 &G_{10}(t,t'; t_1,t_0) \nn\\
 &= \frac{1}{\overline{\langle 0_{t_1}\rvert} 0_{t_{0}} \rangle}\,
    \overline{\bra{0_{t_1}}}\,\bigl(\varphi(t_{>};t_1)\,a_1
  +\bar{\varphi}(t_{>};t_1)\,\bar{a}_1 \bigr)\nn\\
 &\qquad\qquad\qquad\times
   \bigl(\varphi(t_{<};t_0)\,a_0 +\bar{\varphi}(t_{<};t_0)\,\bar{a}_0 \bigr)
          \ket{0_{t_{0}}} \nn\\
 &= \varphi(t_{>};t_1)\,\bar{\varphi}(t_{<};t_0)\, 
    \frac{\overline{\bra{0_{t_1}}}\, a_1 \,\bar{a}_0\, \ket{0_{t_{0}}}}
         {\overline{\langle 0_{t_1}\rvert} 0_{t_{0}} \rangle} 
\nn\\
 &= \frac{\ii}{W_\rho[\varphi(s;t_1),\bar{\varphi}(s;t_0)]}
    \,\varphi(t_{>};t_1)\,\bar{\varphi}(t_{<};t_0)\quad\mbox{($s$: arbitrary)}\,.
\end{align}
On the other hand, to obtain the last expression of \eqref{G00}, 
we start from the identities 
\begin{align}
 \bar{a}^\dagger_0 
  &=  {}- \frac{V_\rho[\varphi^\ast(s;t_0),\bar{\varphi}(s;t_0)](T_s)}
       {V_\rho[\bar{\varphi}^\ast(s;t_0),\bar{\varphi}(s;t_0)](T_s)}\,a^\dagger_0\nn\\
  &\quad\ + \frac{\ii}{V_\rho[\bar{\varphi}^\ast(s;t_0),\bar{\varphi}(s;t_0)](T_s)}\,a_0\,,
\label{in-in_a0}
\\
 \frac{\bra{0_{t_0}}\,\bar{a}_0^\dagger\, \bar{a}_0 \,\ket{0_{t_{0}}}}
         {\langle 0_{t_0}\rvert 0_{t_{0}} \rangle} 
  &= \frac{\ii}{V_\rho[\bar{\varphi}^\ast(s;t_0)\,\bar{\varphi}(s;t_0)](T_s)} \,,
\end{align}
which can be shown by using the hermiticity at time $T_s$, 
$q^\dagger(T_s)=q(T_s)$ and $p^\dagger(T_s)=p(T_s)$ 
(see Appendix \ref{appendix:in-in-matrix}). 
We then have
\begin{align}
 &G_{00}(t,t'; t_0,t_0)\nn\\
 &= \frac{1}{\langle 0_{t_0}\rvert 0_{t_{0}} \rangle}\,
    \bra{0_{t_0}}\,\bigl(\varphi(t_{>};t_0)\,a_0
  +\bar{\varphi}(t_{>};t_0)\,\bar{a}_0 \bigr)^\dagger\nn\\
 &\qquad\qquad\qquad \times\bigl(\varphi(t_{<};t_0)\,a_0 +\bar{\varphi}(t_{<};t_0)\,\bar{a}_0 \bigr)\,
    \ket{0_{t_{0}}} \nn\\
 &= \bar{\varphi}^\ast(t_{>};t_0)\,\bar{\varphi}(t_{<};t_0)\,
    \frac{\bra{0_{t_0}}\,\bar{a}_0^\dagger\, \bar{a}_0 \,\ket{0_{t_{0}}}}
         {\langle 0_{t_0}\rvert 0_{t_{0}} \rangle} \nn\\
 &= \frac{\ii}{V_\rho[\bar{\varphi}^\ast(s;t_0),\,\bar{\varphi}(s;t_0)](T_s)}\,
    \bar{\varphi}^\ast(t_{>};t_0)\,\bar{\varphi}(t_{<};t_0) \,.
\end{align}
When we need to specify $T_s$\,, 
we will set $T_s=t_>$ as in \cite{Weinberg:2005vy} 
(see also discussions following Eq.\ \eqref{Minkowski_00}), 
which leads in the Schr\"odinger picture to
\begin{align}
 & G_{00}(t,t';t_0,t_0)
\nn\\
 &=\frac{\bigl( U(t_>,t_0)\,\ket{0_{t_0},t_0}\bigr)^\dagger
   q_s\,U(t_>,t_<) \,q_s \,U(t_<,t_0)\,\ket{0_{t_0},t_0}}
   {||\,U(t_>,t_0)\,\ket{0_{t_0},t_0} \,||^2}\,.
\end{align}

When $\rho(t)$ and $\omega(t)$ are asymptotically constant in the remote past 
[i.e.~$\rho(t) \sim \rho_{\inn}$ 
and $\omega(t) \sim \omega_{\inn}$ as $t \rightarrow t_i$], 
we can choose a pair of independent solutions $\{f(t),g(t)\}$
 as those which behave as
\begin{align}
 f(t) \sim e^{-\ii \omega_{\inn} t} \,, \qquad 
 g(t) \sim e^{+\ii \omega_{\inn} t} \qquad (t \sim t_i)\,.
\end{align}
If we choose such basis, we then have
\begin{align}
 & u_0 \sim 0 \,, \quad 
  \bar{u}_0 \sim -2\ii \rho_{\inn}\,\omega_{\inn}\, e^{-\ii \omega_{\inn}\, t_0}\,, \\
 & v_0 \sim 2\ii \rho_{\inn}\,\omega_{\inn}\, e^{\ii \omega_{\inn}\, t_0} \,, \quad 
  \bar{v}_0 \sim 0 \quad (t_0 \sim t_i)\,,
\end{align}
and from Eqs.~\eqref{phiin} and \eqref{phiout} 
the wave functions at the remote past are found to behave as
\begin{align}
 \varphi(t;t_0) 
  &\sim \frac{1}{\sqrt{2\rho_{\inn}\,\omega_{\inn}}}\,
  e^{-\ii \omega_{\inn}\, (t-t_0)} \,, \\
 \bar{\varphi}(t;t_0) 
  &\sim \frac{1}{\sqrt{2\rho_{\inn}\,\omega_{\inn}}}\,
  e^{+\ii \omega_{\inn}\, (t-t_0)} 
 \quad (t \sim t_i;~t_0 \sim t_i)\,.
\end{align}
A conclusion of the same kind can be obtained for the wave functions 
$\bigl(\varphi(t;t_1),\,\bar\varphi(t;t_1)\bigr)$ 
if $\rho(t)$ and $\omega(t)$ are asymptotically constant at the remote future. 
This behavior of wave functions 
will be directly seen in concrete examples given in section \ref{sec:Minkowski} 
and in Appendix \ref{appendix:asymptotically_Minkowski}.

\section{Simple example: scalar field in Minkowski space}
\label{sec:Minkowski}

In this section, 
to demonstrate how the prescription of the previous section works,  
we consider a free real scalar field $\phi(x)$ 
living in Minkowski space with the metric
\begin{align}
 \rmd s^2 = {}- \rmd t^2+\rmd \bx^2\,.
\end{align}
Another well studied example is investigated within our framework
in Appendix \ref{appendix:asymptotically_Minkowski}. 

\subsection{Setup}

In order to clarify the structure of mode functions, 
we first assume that the spatial part 
is a $(d-1)$-dimensional torus of radius $L/2\pi$\,, 
which we will take infinite afterwards. 
The wave vectors $\bk$ then take the following values:
\begin{align}
 \bk&= \frac{2\pi}{L}\,\bn\quad\bigl( \bn\in\bZZ^{d-1}\bigr)\,.
\end{align}
For $\bk=\bigl(k_1,k_2,\dotsc,k_{d-1}\bigr)$\,, 
we write $\bk>0$ (or $\bk<0$) 
if the first nonvanishing element in the sequence $\{k_1,\,k_2,\,\cdots\}$ 
is positive (or negative). 
Note that $\bk<0$ is equivalent to $-\bk>0$\,.
We write $\bk=0$ if $\bk$ is the zero vector ($\bk={\bf 0}$). 

We introduce a complete set of (real-valued) eigenfunctions 
$\bigl\{Y_{\bk,\,a}(\bx)\bigr\}$ of the spatial Laplacian 
$\Delta_{d-1}=\sum_{i=1}^{d-1}\partial_i^2$ as 
\begin{align}
 \bk=0:\quad&Y_{\bk=0,\,a=1}\equiv\frac{1}{\sqrt{V}}\quad
  \bigl(V\equiv L^{d-1}\bigr)\,,
\label{Yk0}
\\
 \bk>0:\quad&
  Y_{\bk,\, a=1}(\bx)=\sqrt{\dfrac{2}{V}}\,\cos\bk\cdot\bx\,,
\label{Yk1-1}\\
  &Y_{\bk,\, a=2}(\bx)=\sqrt{\dfrac{2}{V}}\,\sin\bk\cdot\bx\,.
\label{Yk1-2}
\end{align}
They satisfy the orthonormal relations,
\begin{align}
 \int\!\!\rmd^{d-1}\bx\,Y_{\bk,\,a}(\bx)\,Y_{\bk',\,a'}(\bx)
  =\delta_{\bk,\bk'}\,\delta_{a,a'}\,,
\end{align}
and we expand the scalar field $\phi(x)$ as
\begin{align}
 \phi(x)=\phi(t,\bx)=\sum_{\bk\geq0}\sum_a\,\phi_{\bk,\,a}(t)\,Y_{\bk,\,a}(\bx)\,.
\end{align}
The action then becomes 
\begin{align}
 &S[\phi(x)]=\int\rmd^d x\,\Bigl[\,
  {}- \frac{1}{2}\,\partial^\mu\phi(x)\,\partial_\mu\phi(x)
  -\frac{m^2}{2}\,\phi^2\,\Bigr]\nn\\
 &=\sum_{\bk\geq0}\sum_a\,\int\rmd t\,\Bigl[\,
   \frac{1}{2}\,\dot{\phi}_{\bk,\,a}^2(t)-\frac{\omega_{k}^2}{2}
   \,\phi_{\bk,\,a}^2(t)
   \,\Bigr]\,,
\end{align}
where 
\begin{align}
 \omega_{k}&\equiv \sqrt{k^2+m^2}
\quad
 \Bigl(k\equiv \abs{\bk}=\Bigl[\sum_{i=1}^{d-1}\,k_i^2\Bigr]^{1/2}\Bigr) \,.
\end{align}
We thus have the following correspondence with the ingredients 
of the previous section:
\begin{align}
 q(t) &=\phi_{\bk,\,a}(t)\,,\quad 
 \rho(t) =e^{\ii\varepsilon}\,,\\
 \omega(t) &=\omega_{k,\,\varepsilon}
\equiv e^{-\ii\varepsilon}\omega_k~~(=\mbox{constant in $t$})\,.
\end{align}

\subsection{Propagator for each mode}
\label{Minkowski_mode}

The equation of motion is given by 
$\ddot{q}+\omega^2 q=0$\,,
and we choose a pair of independent solutions as
\begin{align}
 f(t)=e^{-\ii\omega t}\,,\quad g(t)=e^{\ii\omega t}\,.
\end{align}
Their Wronskian is given 
by $W_\rho[f,g]= 2\,\ii\rho\,\omega=2\,\ii\omega_k$\,, 
which is constant in $t$\,.

The functions $u(t)$ and $v(t)$ are easily found to be%
\footnote{
The Bogoliubov coefficients can be calculated 
by using \eqref{albe} as
\begin{align}
 \alpha(t_1;t_0)&=e^{\ii\omega_k\,(t_1-t_0)}\,, \quad  \beta(t_1;t_0) = 0\,, 
\nn\\
  \bar\alpha(t_1;t_0)&=e^{-\ii\omega_k\,(t_1-t_0)}\,, \quad \bar\beta(t_1;t_0)=0\,, 
\nn
\end{align}
which indicates that the vacuum at a later time, $\ket{0_{t_1}}$, 
coincides with the vacuum at an earlier time, $\ket{0_{t_0}}$\,, up to a phase. 
} 
\begin{align}
 \biggl\{\begin{array}{l}
  u(t) \\ \bar{u}(t)
 \end{array}\biggr\}
  &= \rho\,\bigl[\,\dot{f}(t)\pm \ii\omega\,f(t)\,\bigr]
  = \biggl\{\begin{array}{l}
     0 \\ -2\,\ii\,\omega_k\,e^{-\ii\omega t}
    \end{array}\,,\\
 \biggl\{\begin{array}{l}
  v(t) \\ \bar{v}(t)
 \end{array}\biggr\}
  &= \rho\,\bigl[\,\dot{g}(t)\pm \ii\omega\,g(t)\,\bigr]
  = \biggl\{\begin{array}{l}
     2\,\ii\,\omega_k\,e^{\ii\omega t} \\ 0
    \end{array}\,,
\end{align}
and using \eqref{phiin} and \eqref{phiout} 
we obtain the wave functions as
\begin{align}
 \varphi(t;t_0)&=\frac{1}{W_\rho\,\sqrt{2\rho_0\,\omega_0}}\,
  \bigl[ v_0\,f(t) - u_0\,g(t)\bigr] \nn\\
  &=\frac{1}{\sqrt{2\omega_k}}\,e^{-\ii\omega_{k,\varepsilon}\,(t-t_0)}\,,\\
 \bar\varphi(t;t_0)&=\frac{1}{W_\rho\,\sqrt{2\rho_0\,\omega_0}}\,
  \bigl[ \bar{v}_0\,f(t) - \bar{u}_0\,g(t)\bigr] \nn\\
  &=\frac{1}{\sqrt{2\omega_k}}\,e^{\ii\omega_{k,\varepsilon}\,(t-t_0)}\,,\\
 \varphi(t;t_1)&=\frac{1}{W_\rho\,\sqrt{2\rho_1\,\omega_1}}\,
  \bigl[ v_1\,f(t) - u_1\,g(t)\bigr]\nn\\
  &=\frac{1}{\sqrt{2\omega_k}}\,e^{-\ii\omega_{k,\varepsilon}\,(t-t_1)}\,,\\
 \bar\varphi(t;t_1)&=\frac{1}{W_\rho\,\sqrt{2\rho_1\,\omega_1}}\,
  \bigl[ \bar{v}_1\,f(t) - \bar{u}_1\,g(t)\bigr]\nn\\
  &=\frac{1}{\sqrt{2\omega_k}}\,e^{\ii\omega_{k,\varepsilon}\,(t-t_1)}\,.
\end{align}
From them, we have
\begin{align}
 W_\rho[\varphi(s;t_1),\bar\varphi(s;t_0)]
  &=\ii\,e^{\ii\omega_{k,\varepsilon}(t_1-t_0)}\,,
\\
 V_\rho[\bar\varphi^\ast(s;t_0),\bar\varphi(s;t_0)]
  &=\ii\,e^{\ii(\omega_{k,\varepsilon}-\omega_{k,-\varepsilon})(s-t_0)}\,,
\end{align}
and the two-point functions take the forms
\begin{align}
 &G_{\bk,\,10}(t,t'; t_1,t_0) 
\nn\\
  &=\frac{\ii}{W_\rho[\varphi(s;t_1),\bar\varphi(s;t_0)]}\,
  \varphi(t_>, t_1)\,\bar\varphi(t_<; t_0) \nn\\
  &= \frac{1}{2\,\omega_{k}}\,e^{-\ii\omega_{k,\,\varepsilon}\,(t_>-t_<)}\,, 
\label{Minkowski_10}
\\
 &G_{\bk,\,00}(t,t'; t_0,t_0) 
\nn\\
  &= \frac{\ii}{V_\rho[\bar{\varphi}^\ast(s;t_0)\,
  \bar{\varphi}(s;t_0)](T_s)}\,
  \bar{\varphi}^\ast(t_>;t_0)\,\bar\varphi(t_<; t_0) \nn\\
  &= \frac{1}{2\,\omega_{k}}\,
 e^{- \ii\omega_{k,-\varepsilon}\,(t_>-T_s) 
 -\ii\omega_{k,\varepsilon}(T_s-t_<)} \,.
\label{Minkowski_00}
\end{align}
Note that the dependence on $t_0$ and $t_1$ totally disappear 
in $G_{\bk,\,IJ}(t,t'; t_I,t_J)$ for Minkowski space, 
and thus we need not to take the limit $t_0\to-\infty$\,, $t_1\to+\infty$ 
to obtain the in-out and in-in propagators. 
We see from \eqref{Minkowski_00} 
that the behavior of $G_{\bk,\,00}$ in the region $k\to \infty$ 
gets significantly improved 
if we choose $T_s$ such that $T_s \geq t_>$\,. 
By simply setting $T_s=t_>$\,, 
we obtain
\begin{align}
 G_\bk^{\inn/\inn}(t,t')=G_\bk^{\out/\inn}(t,t')
 =\frac{1}{2\,\omega_{k}}\,
  e^{-\ii\omega_{k,\varepsilon}\,(t_>-t_<)}\,,
\end{align}
which will be denoted by $G_k(t,t')$ 
in the following discussions.

\subsection{Propagator in spacetime}
\label{Minkowski_spacetime}

Once propagators are obtained for each mode $(\bk,a)$\,, 
the propagator in spacetime can be obtained 
by summing them over the modes. 
The manipulation is known very well for Minkowski space, 
but we here review it briefly for later reference.

The in-out or in-in propagator is given by the following summation 
($t_i=-\infty$, $t_f=\infty$): 
\begin{align}
 &G^{\out/\inn}(x,x')\nn\\
 &= \sum_{\bk,\,\bk'\geq0}\sum_{a,\,a'}\,
 \frac{\overline{\bra{0_{t_1}}}\,{\rm T}\,\phi_{\bk,\,a}(t)\,\phi_{\bk',\,a'}(t')\,
      \ket{0_{t_0}}}{\overline{\langle{0_{t_1}} \vert} 0_{t_0}\rangle}
      \Biggr\rvert_{\genfrac{}{}{0pt}{}{t_0\to t_i}{t_1\to t_f}}\,\nn\\
 &\qquad\qquad\qquad\qquad\qquad \times Y_{\bk,\,a}(\bx)\,Y_{\bk',\,a'}(\bx') \,.
\\
 &G^{\inn/\inn}(x,x')\nn\\
 &= \sum_{\bk,\,\bk'\geq0}\sum_{a,\,a'}\,
 \frac{\bra{0_{t_0}}\,\phi^\dagger_{\bk,\,a}(t_>)\,\phi_{\bk',\,a'}(t_<)\,
      \ket{0_{t_0}}}{\langle{0_{t_0}} \vert 0_{t_0}\rangle}
      \Biggr\rvert_{ t_0\to t_i }\,\nn\\
 &\qquad\qquad\qquad\qquad\qquad \times Y_{\bk,\,a}(\bx)\,Y_{\bk',\,a'}(\bx') \,,
\end{align}
Since the two-point functions are diagonalized with respect to the modes,
\begin{align}
 \frac{\overline{\bra{0_{t_1}}}\,{\rm T}\,\phi_{\bk,\,a}(t)\,\phi_{\bk',\,a'}(t')\,
      \ket{0_{t_0}}}{\overline{\langle{0_{t_1}} \vert} 0_{t_0}\rangle}
      \Biggr\rvert_{\genfrac{}{}{0pt}{}{t_0\to t_i}{t_1\to t_f}}
  &= G_k(t,t')\,\delta_{\bk,\bk'}\,\delta_{a,a'}\,,
\\
 \frac{\bra{0_{t_0}}\,\phi^\dagger_{\bk,\,a}(t_>)\,\phi_{\bk',\,a'}(t_<)\,
      \ket{0_{t_0}}}{\langle{0_{t_0}} \vert 0_{t_0}\rangle}\Biggr\rvert_{ t_0\to t_i }
  &= G_k(t,t')\,\delta_{\bk,\bk'}\,\delta_{a,a'}\,,
\end{align}
we have
\begin{align}
 G^{\bigl\{{\genfrac{}{}{0pt}{}{\out/\inn}{\inn/\inn}}\bigr\}}(x,x')
  &= \sum_{\bk\geq 0}\,G_k(t,t')\,\sum_a\,Y_{\bk,a}(\bx)\,Y_{\bk,a}(\bx')\nn\\
  &\equiv \sum_{\bk\geq 0}\,G_k(t,t')\,R_\bk(\bx,\bx')\,.
\end{align}
Here, $R_\bk(\bx,\bx')\equiv \sum_a\,Y_{\bk,a}(\bx)\,Y_{\bk,a}(\bx')$ 
are easily calculated as
\begin{align}
 R_{\bk=0}(\bx,\bx')=\frac{1}{V}\,,\quad
  R_{\bk>0}(\bx,\bx')=\frac{2}{V}\,\cos\bk\cdot(\bx-\bx')\,,
\end{align}
and thus we have
\begin{align}
 &G(x,x')\nn\\
 &=\frac{1}{V}\,\Bigl[
  G_{k=0}(t,t') 
 + 2\,\sum_{\bk>0}\,G_k(t,t')\,\cos\bk\cdot(\bx-\bx')\Bigr]
\nn\\
 &=\frac{1}{V}\,\sum_\bk\,G_k(t,t')\,\cos\bk\cdot(\bx-\bx')\nn\\
 &=\int\frac{\rmd^{d-1}\bk}{(2\pi)^{d-1}}\,
  G_k(t,t')\,\cos\bk\cdot(\bx-\bx')\,,
\label{Minkowski1}
\end{align}
where we have taken the limit $L\to\infty$ in the last equality.

The integration \eqref{Minkowski1} can be performed easily 
(see Appendix \ref{appendix:propagator_Minkowski}), 
and we obtain
\begin{align}
 G(x,x')
 &= \frac{1}{(2\pi)^{\frac{d-1}{2}}\,\abs{\bx-\bx'}^{\frac{d-3}{2}}}\,\nn\\
 &\quad\times  \int_0^\infty\!\! \rmd k\,k^{\frac{d-1}{2}}\,
    G_k(t,t')\,J_{\frac{d-3}{2}}\bigl(k\abs{\bx-\bx'}\bigr)\, 
\nn\\
 &=\frac{m^{(d-2)/2}}{(2\pi)^{d/2}\,
  \bigl(\sigma+\ii0\bigr)^{(d-2)/4}}\,
  K_{\frac{d-2}{2}}\bigl(m\sqrt{\sigma+\ii0}\bigr)
\label{Minkowski_prop}
\end{align}
with $\sigma\equiv (x-x')^2$.  
Here, $J_\nu(z)$ is the Bessel function, and 
$K_\nu(z)$ is the modified Bessel function of the second kind.

\section{Scalar field in de Sitter space}
\label{sec:deSitter}

\subsection{Geometry and definitions}
\label{sec:geometry}

We first recall the geometry of de Sitter space 
and collect the notation and definitions.

$d$-dimensional de Sitter space dS$_d$ 
has the topology $\bRR\times S^{d-1}$ 
and is defined as a hyperboloid
\begin{align}
 &\eta_{MN}\, X^M X^N = \ell^2\quad  (M,N,\cdots =0,\dotsc,d)\,,
\nn\\
 &\bigl(\eta_{MN}\bigr)={\rm diag}(-1,1,\dotsc,1)\,,
\end{align}
in $(d+1)$-dimensional Minkowski space 
with the metric
\begin{align}
 \rmd s^2 = \eta_{MN}\, \rmd X^M \rmd X^N \,.
\end{align}
$\ell$ is called the de Sitter radius. 
The constant Ricci scalar curvature is then given by $R=d\,(d-1)/\ell^2$.

There are several well-known coordinate patches 
which cover the whole or just a part of de Sitter space. 
Among them, we consider 
the global patch and the Poincar\'e (or planer) patch, 
which we will briefly review below. 

\noindent\underline{\bf global patch}: 
This coordinate patch covers the whole region of de Sitter space. 
The embedding of dS$_d$ is given by the functions
\begin{align}
 X^0(\tau,\bOmega) &= \ell\,\sinh\tau \,,
\nn\\
 X^I(\tau,\bOmega) &= \ell\,\cosh\tau \,\,\Omega^I 
\quad  (I=1,\dotsc,d)
\end{align}
with $\bOmega\cdot\bOmega=1$\,.
Here, $\tau$ runs over the range $-\infty<\tau<\infty$ 
and $\bOmega$ is a unit vector in $\bRR^d$ 
spanning a $(d-1)$-dimensional sphere. 
With the coordinates $(\tau,\bOmega)$ the metric has the form 
\begin{align}
 \rmd s^2 &= \ell^2\bigl[ {}- \rmd \tau^2 + \cosh^2\tau \,\rmd \Omega_{d-1}^2 \bigr]\nn\\
          &= \ell^2[{}- \bigl(1-t^2\bigr)^{-2}\rmd t^2 
             +\bigl(1-t^2\bigr)^{-1}\rmd \Omega_{d-1}^2 ]\,.
\end{align}
In the last equality, we have introduced another temporal coordinate $t$ as
\begin{align}
 t \equiv \tanh\tau \qquad (-1<t<1)\,.
\end{align}

\noindent\underline{\bf Poincar\'e patch}: 
This coordinate patch covers only half of de Sitter space. 
The embedding is given by the following functions 
with $\eta<0$ and $\bx\in \mathbb{R}^{d-1}$:
\begin{align}
 X^0(\eta,\bx) &= \frac{\ell^2-\eta^2+\abs{\bx}^2}{-2\eta} \,,\quad 
 X^i(\eta,\bx) = \ell\,\frac{x^i}{-\eta} \,,\nn\\
 X^d(\eta,\bx) &= \frac{\ell^2+\eta^2-\abs{\bx}^2}{-2\eta}\,,
\end{align}
where the spatial norm is defined by 
$\abs{\bx} \equiv \sqrt{\bx\cdot \bx} 
 \equiv \sqrt{\delta_{ij}\,x^i\,x^j}$\,.
Note that this patch only covers the region $X^0+X^d=\ell^2/(-\eta)>0$\,.
In these coordinates, the metric takes the form 
\begin{align}
 \rmd s^2 = \ell^2\,\frac{{}- \rmd \eta^2 + \rmd \bx\cdot \rmd \bx}{\eta^2} \,.
\end{align}
The Poincar\'e patch is not preserved 
under a finite action of de Sitter group ${\rm SO}(1,d)$\,, 
but is still preserved 
under infinitesimal actions of ${\rm SO}(1,d)$\,. 
In fact, the infinitesimal actions are given 
by the Killing vectors $M_{MN}=X_M\partial_N-X_N\partial_M$, 
which take the following forms in the Poincar\'e patch:
\begin{align}
 M_{0d} &=\eta\,\partial_{\eta}+x^i\,\partial_i\,,
\\
 M_{0i}+M_{di} &=\frac{1}{\ell}
 \bigl[2x^i\,\eta\,\partial_\eta+2x^i\,x^j\partial_j \nn\\
       &\qquad\qquad-(-\eta^2+\abs{\bx}^2)\,\partial_i\bigr]\,,
\\
 M_{0i}-M_{di} &=- \ell \,\partial_i\,,
\\
 M_{ij} &=x^i\,\partial_j-x^j\,\partial_i \,.
\end{align}
These Killing vectors do not have the $\partial_{\eta}$ component 
at the boundary of the patch $\eta=0$ as long as $\abs{\bx}<\infty$. 
Similarly, if we define another time coordinate 
$t\equiv 1/(-\eta)$, we find that 
the Killing vectors do not have the $\partial_t$ component 
at another boundary at $t=0$ (i.e., $\eta=-\infty$). 
These results show that the infinitesimal transformations 
map any point inside the Poincar\'e patch 
(i.e., $\abs{\bx}<\infty$ and $-\infty<\eta<0$) 
into the same region.

We define the de Sitter invariant quantity
\begin{align}
 Z(x,x')\equiv \ell^{-2}\,\eta_{MN}\,X^M(x)\,X^N(x') \,,
\end{align}
which is related to the geodesic distance $d(x,x')$ 
between two points $x$ and $x'$ via the relation
\begin{align}
 Z(x,x') = \cos\Bigl(\frac{d(x,x')}{\ell}\Bigr)\,.
\end{align}
It takes the following values depending 
on the positional relation between $x$ and $x'$\,:
\begin{align}
 Z(x,x')\left\{
 \begin{array}{ll}
  >1 & (\text{for $x$ and $x'$ timelikely separated})\cr
  =1 & (\text{for $x$ and $x'$ lightlikely separated})\cr
  <1 & (\text{for $x$ and $x'$ spacelikely separated})
 \end{array}\right. \,,
\end{align}
as can be seen from the identity
\begin{align}
 Z(x,x')
 = 1-\frac{1}{2\,\ell^2}\,\eta_{MN}\,&\bigl(X^M(x)-X^M(x')\bigr)\nn\\
                             &\times\bigl(X^N(x)-X^N(x')\bigr) \,.
\end{align}
In the global and the Poincar\'e coordinates, 
$Z(x,x')$ is written in the form
\begin{align}
 Z(x,x')=\left\{
 \begin{array}{l}
 \displaystyle
  -\sinh\tau\,\sinh\tau' + \cosh\tau\,\cosh\tau'\,\bOmega\cdot\bOmega' \\[1mm]
 \qquad\qquad\qquad\quad(\text{global coords.})\\[2mm]
 \displaystyle 
  \frac{\eta^2 + \eta'^2 - \abs{\bx-\bx'}^2}{2\eta\,\eta'} \\[1mm]
 \qquad\qquad\qquad\quad (\text{Poincar\'e coords.})
 \end{array}\right. \,.
\end{align}

One can easily prove that 
any two-point function $G(x,x')$ that is invariant 
under the infinitesimal actions 
$M_{MN}+M'_{MN}=M^\mu_{MN}(x)\,\partial/\partial x^\mu
+M^\mu_{MN}(x')\,\partial/\partial x^{\prime\mu}$ 
must be a function of the de Sitter invariant $Z(x,x')$\,. 
We will see that all the propagators constructed in this paper 
turn out to be functions of $Z(x,x')$\,.
In what follows (except in subsection \ref{sec:EAdS}), 
we set $\ell=1$\,.

\subsection{Scalar field in the Poincar\'e patch}
\label{sec:Poincare}

We first consider a free real scalar field in the Poincar\'e patch. 
The action takes the form%
\footnote{
In this paper, 
we put a possible curvature-coupling term, 
$(\xi/2)\,R\,\phi^2=\bigl(d(d-1)\xi/2\bigr)\,\phi^2$, 
into the mass term, $(m^2/2)\,\phi^2$\,. 
} 
\begin{align}
 S[\phi(x)]
  &= {}- \frac{1}{2}\,\int\rmd \eta\int \rmd^{d-1}\bx\,\sqrt{-g}\,\nn\\
  &\qquad\quad\times\bigl( g^{\mu\nu}\,\partial_\mu\phi\,\partial_\nu\phi
  + m^2\,\phi^2\bigr)\,.
\end{align}
Using the same eigenfunctions $\{Y_{\bk,\,a}(\bx)\}$ 
as those given in section \ref{sec:Minkowski} 
[Eqs.~\eqref{Yk0}, \eqref{Yk1-1}, and \eqref{Yk1-2}], 
we expand a scalar field $\phi(x)$ as
\begin{align}
 \phi(x)=\phi(\eta,\bx)
        =\sum_{\bk\geq 0}\sum_a\,\phi_{\bk,\,a}(\eta)\,Y_{\bk,\,a}(\bx)\,.
\end{align}
The functions defined in section \ref{sec:general} then take the following form 
(see \eqref{rhomega-1} and \eqref{rhomega-2}):
\begin{align}
 q(\eta) &= \phi_{\bk,a}(\eta)\,,\quad
 \rho(\eta)= e^{\ii\varepsilon}(-\eta)^{-(d-2)}\,,\\
 \omega(\eta) 
 &= e^{-\ii\varepsilon}\sqrt{m^2\,(-\eta)^{-2}+k^2} 
 \equiv e^{-\ii\varepsilon}\,\omega_k(\eta)\,,
\end{align}
from which we introduce%
\footnote{
Our prescription of section \ref{sec:general} cannot be applied directly 
to the exactly massless case, where $\omega(\eta)=0$ for $k=0$\,. 
We actually define the massless theory as the $m\to 0$ limit of a massive theory.
} 
\begin{align}
 m_\varepsilon&\equiv e^{-\ii\varepsilon}\,m \,,\quad
 k_\varepsilon\equiv e^{-\ii\varepsilon}\,k \,.
\end{align}
Note that
\begin{align}
 \omega_{k,0} 
   &\equiv \omega_k(\eta_0) 
   \overset{\eta_0\rightarrow -\infty}{\longrightarrow} k \,,
 \quad 
 \omega_{k,1} 
   \equiv \omega_k(\eta_1)
   \overset{\eta_1\rightarrow -0}{\longrightarrow} 
           \frac{m}{-\eta_1} \,.
\label{poincare-omega}
\end{align}

\subsubsection{Propagators for each mode in the Poincar\'e patch}
\label{sec:mode_Poincare}

The equation of motion \eqref{EOM} takes the form
\begin{align}
 \eta^2\,\ddot{q}(\eta) 
 - (d-2)\,\dot{q}(\eta)
 + \bigl(k_\varepsilon^2\eta^2+m_\varepsilon^2)\,q(\eta) =0\,,
\end{align}
and we choose a set of independent solutions as
\begin{align}
 f(\eta) &= (-\eta)^{\frac{d-1}{2}}\,J_{\nu_\varepsilon}(-k_\varepsilon\eta)\,,\\
 g(\eta) &= (-\eta)^{\frac{d-1}{2}}\,N_{\nu_\varepsilon}(-k_\varepsilon\eta)
\end{align}
with 
\begin{align}
 \nu_\varepsilon \equiv \left\{ 
  \begin{array}{ll}
   \sqrt{\bigl(\frac{d-1}{2}\bigr)^2- m_\varepsilon^2} 
   = \nu + \ii\varepsilon \quad 
     & \bigl(m < \frac{d-1}{2}\bigr) \\ 
   \ii\,\sqrt{m_\varepsilon^2 - \bigl(\frac{d-1}{2}\bigr)^2} 
   = \ii\,\mu + \varepsilon
     & \bigl(m \geq \frac{d-1}{2}\bigr) 
  \end{array}
  \right.\quad\nn\\
 \biggl[\,\nu\equiv \sqrt{\Bigl(\frac{d-1}{2}\Bigr)^2- m^2}\,,\quad
 \mu\equiv \sqrt{m^2 - \Bigl(\frac{d-1}{2}\Bigr)^2}\,\biggr]\,.
\end{align}
Here, $N_\nu(x)$ is the Neumann function. 
Note that $\Ree\nu_\varepsilon > 0$ 
for any positive value of $m$\,. 
The Wronskian is given by $W[f,g](\eta) =-(2/\pi)\,(-\eta)^{d-2}$, 
and thus
\begin{align}
 W_\rho[f,g]=\rho(t)\,W[f,g](t) = {}-e^{\ii\varepsilon} \frac{2}{\pi}\,.
\end{align}
The functions $u(\eta)$ and $v(\eta)$ of \eqref{uuvv-1} and \eqref{uuvv-2}
have the forms
\begin{align}
  \left\{\begin{array}{l}
    u(\eta) \cr \bar{u}(\eta)
  \end{array}\right\} 
  &= -e^{\ii\varepsilon}(-\eta)^{-\frac{d-1}{2}}\nn\\
      &\quad\times\Bigl[\Bigl(\frac{d-1}{2}+\nu_\varepsilon\pm\ii\omega(\eta)\,\eta\Bigr)\,
            J_{\nu_\varepsilon}(-k_\varepsilon\eta)\nn\\
  &\qquad\qquad\qquad\quad +k_\varepsilon\eta\,J_{1+\nu_\varepsilon}(-k_\varepsilon\eta)\Bigr] \,,\\
  \left\{\begin{array}{l}
    v(\eta) \cr \bar{v}(\eta)
  \end{array}\right\} 
  &= -e^{\ii\varepsilon}(-\eta)^{-\frac{d-1}{2}}\nn\\
      &\quad\times\Bigl[\Bigl(\frac{d-1}{2}+\nu_\varepsilon\pm\ii\omega(\eta)\,\eta\Bigr)\,
            N_{\nu_\varepsilon}(-k_\varepsilon\eta)\nn\\
  &\qquad\qquad\qquad\quad +k_\varepsilon\eta\,N_{1+\nu_\varepsilon}(-k_\varepsilon\eta)\Bigr] \,,
\end{align}
where we have used the formulae
\begin{align}
 z\,\frac{\partial J_\nu(z)}{\partial z}&= \nu\, J_{\nu}(z) - z\,J_{\nu+1}(z)\,,\\
 z\,\frac{\partial N_\nu(z)}{\partial z}&= \nu\, N_{\nu}(z) - z\,N_{\nu+1}(z) \,.
\end{align}
The wave functions are then given by
\begin{align}
 \varphi(\eta;\eta_I)
  &= -\frac{\pi\,e^{-\ii\varepsilon}}{2\sqrt{2 \omega_{k,I}}}\,
    (-\eta_I)^{\frac{d-2}{2}} \,
    \bigl[ v_I\,f(\eta) -u_I\,g(\eta)\bigr] \,,\\
 \bar{\varphi}(\eta;\eta_I)
  &= -\frac{\pi\,e^{-\ii\varepsilon}}{2\sqrt{2 \omega_{k,I}}}\,
    (-\eta_I)^{\frac{d-2}{2}} \,
    \bigl[ \bar{v}_I\,f(\eta) -\bar{u}_I\,g(\eta)\bigr] \,,
\end{align}
from which the two-point functions are obtained as
\begin{align}
 &G_{10}(\eta,\eta'; \eta_1,\eta_0) 
\nn\\
 &= \frac{\ii}{W_\rho[\varphi(s;\eta_1),\bar{\varphi}(s;\eta_0)]}
    \,\varphi(\eta_{>};\eta_1)\,\bar{\varphi}(\eta_{<};\eta_0)
\label{Poin_G10}
\nn\\
 &\qquad\qquad\qquad\qquad\qquad\qquad\qquad\mbox{($s$: arbitrary)}\,,
\\[3mm]
 &G_{00}(\eta,\eta'; \eta_0,\eta_0)
\nn\\
 &= \frac{\ii}{V_\rho[\bar{\varphi}^\ast(s;\eta_0),
 \,\bar{\varphi}(s;\eta_0)](\eta_s)}\,
    \bar{\varphi}^\ast(\eta_{>};\eta_0)\,\bar{\varphi}(\eta_{<};\eta_0) \,.
\label{Poin_G00}
\end{align}

We now send $\eta_0$\,, $\eta_1$ to the boundary of the Poincar\'{e} patch; 
$\eta_0\rightarrow \eta_i=-\infty$ 
and $\eta_1\rightarrow \eta_f=0$\,. 
By using the asymptotic forms of the Bessel functions,%
\footnote{
We have used the inequalities $\Ree \nu_\varepsilon >0$ and 
$\Ree (-\ii k_\varepsilon \eta)>0$\,.
} 
\begin{align}
 J_{\nu_\varepsilon}(-k_\varepsilon\eta)&\overset{\eta\rightarrow -\infty}{\sim} 
    \frac{1}{\sqrt{2\pi}}\,(-k_\varepsilon\eta)^{-1/2}
          \biggl[1+\ii\,\frac{\nu_\varepsilon^2-(1/4)}{2(-k_\varepsilon\eta)}\biggr]\nn\\
    &\qquad\qquad\times
    e^{-\ii\bigl(k_\varepsilon\eta +\frac{\pi (2\nu_\varepsilon +1)}{4}\bigr)} \,,\\
 N_{\nu_\varepsilon}(-k_\varepsilon\eta)&\overset{\eta\rightarrow -\infty}{\sim} 
    \frac{-\ii}{\sqrt{2\pi}}\,(-k_\varepsilon\eta)^{-1/2}
          \biggl[1+\ii\,\frac{\nu_\varepsilon^2-(1/4)}{2(-k_\varepsilon\eta)}\biggr]\nn\\
     &\qquad\qquad\times
     e^{-\ii\bigl(k_\varepsilon\eta +\frac{\pi (2\nu_\varepsilon +1)}{4}\bigr)} \,,
\\
 J_{\nu_\varepsilon}(-k_\varepsilon\eta)&\overset{\eta\rightarrow 0}{\sim} 
   \frac{1}{\Gamma(1+\nu_\varepsilon)}\,
   \Bigl(-\frac{k_\varepsilon\eta}{2}\Bigr)^{\nu_\varepsilon} \,,\\
 N_{\nu_\varepsilon}(-k_\varepsilon\eta)&\overset{\eta\rightarrow 0}{\sim} 
   -\frac{\Gamma(\nu_\varepsilon)}{\pi}\,
    \Bigl(-\frac{k_\varepsilon\eta}{2}\Bigr)^{-\nu_\varepsilon} \,,
\end{align}
one can easily show that $u_I$, $\bar{u}_I$, $v_I$, and $\bar{v}_I$ ($I=0,1$) 
have the asymptotic forms
\begin{align}
    u_0
    &\sim \ii v_0 
\nn\\
    &\sim -\frac{e^{\ii\varepsilon}}{\sqrt{2\pi k_\varepsilon}}
    \Bigl(\frac{d-2}{2}\Bigr)\, (-\eta_0)^{-\frac{d}{2}}\, 
     e^{-\ii\bigl(k_\varepsilon\eta_0+\frac{\pi (2\nu_\varepsilon +1)}{4}\bigr)}
    \,,\\
    \bar{u}_0 
   &\sim \ii \bar{v}_0
\nn\\
   &\sim -\ii e^{\ii\varepsilon}\sqrt{\frac{2k_\varepsilon}{\pi}}\, (-\eta_0)^{-\frac{d-2}{2}}
 e^{-\ii\bigl(k_\varepsilon\eta_0+\frac{\pi (2\nu_\varepsilon +1)}{4}\bigr)}\,,\\
  \left\{\begin{array}{l}
    u_1 \cr \bar{u}_1
  \end{array}\right\} 
   &\sim
  {}- \frac{e^{\ii\varepsilon}\,(k_\varepsilon/2)^{\nu_\varepsilon}}{\Gamma(1+\nu_\varepsilon)} 
  (-\eta_1)^{-\frac{d-1}{2}+\nu_\varepsilon}\nn\\
  &\qquad\times \Bigl(\frac{d-1}{2}+ \nu_\varepsilon \mp \ii m_\varepsilon\Bigr) \,,\\
  \left\{\begin{array}{l}
    v_1 \cr \bar{v}_1
  \end{array}\right\} 
   &\sim
 \frac{e^{\ii\varepsilon}\,\Gamma(\nu_\varepsilon)\,(k_\varepsilon/2)^{-\nu_\varepsilon}}{\pi}\, 
  (-\eta_1)^{-\frac{d-1}{2}-\nu_\varepsilon} \nn\\
 &\qquad\times \Bigl(\frac{d-1}{2}- \nu_\varepsilon \mp 
 \ii m_\varepsilon\Bigr) \,.
\end{align}
Since $\nu_\varepsilon$ always has a positive real part, 
we obtain the relation 
$(-\eta_1)^{-\nu_\varepsilon}\gg (-\eta_1)^{\nu_\varepsilon}$ 
in the limit $\eta_1\to 0$\,, 
from which we find 
\begin{align}
 \lvert u_1\rvert \ll  \lvert v_1 \rvert \,,\qquad  
 \lvert \bar{u}_1 \rvert \ll \lvert \bar{v}_1 \rvert \,.
\end{align}
Thus, we find that the wave functions behave as 
\begin{align}
 \varphi(\eta;\eta_0) 
  &\sim -\frac{\pi \,e^{-\ii\varepsilon}}{2\sqrt{2k}}\,
  (-\eta_0)^{\frac{d-2}{2}}\,v_0\,
  \bigl[f(\eta)-\ii g(\eta)\bigr]\,,
\label{in-phi}
\\ 
 \bar\varphi(\eta;\eta_0) 
  &\sim -\frac{\pi \,e^{-\ii\varepsilon}}{2\sqrt{2k}}\,
  (-\eta_0)^{\frac{d-2}{2}}\,\bar{v}_0\,
  \bigl[f(\eta)-\ii g(\eta)\bigr]\,,
\label{in-bar-phi}
\\
 \varphi(\eta;\eta_1) 
  &\sim -\frac{\pi \,e^{-\ii\varepsilon}}{2\sqrt{2m}}\,
  (-\eta_1)^{\frac{d-1}{2}}\,v_1\,f(\eta)\,,
\label{out-phi}
\\
 \varphi(\eta;\eta_1) 
  &\sim -\frac{\pi \,e^{-\ii\varepsilon}}{2\sqrt{2m}}\,
  (-\eta_1)^{\frac{d-1}{2}}\,\bar{v}_1\,f(\eta)\,,
\label{out-bar-phi}
\end{align}
with which the weighted Wronskian becomes
\begin{align}
 &W_\rho[\varphi(\eta;\eta_1),\bar\varphi(\eta;\eta_0)]
\nn\\
 &\sim \frac{\pi^2\,e^{-2\ii\varepsilon}}{8\sqrt{km}}\,
  (-\eta_0)^{\frac{d-2}{2}} (-\eta_1)^{\frac{d-1}{2}}\,v_1\,\bar{v}_0\,
  W_\rho\bigl[f(\eta),\,f(\eta)-\ii g(\eta)\bigr]
\nn\\
 &=\frac{\ii\pi}{4\sqrt{km}}\,e^{-\ii\varepsilon}\,v_1\,\bar{v}_0\,
  (-\eta_0)^{\frac{d-2}{2}} (-\eta_1)^{\frac{d-1}{2}} \,,
\label{Poin_W}
\\
 &V_\rho[\bar\varphi^\ast(\eta;\eta_0),\bar\varphi(\eta;\eta_0)]
\nn\\
 &\sim \frac{\pi^2}{8k}\,
  (-\eta_0)^{d-2}\,\lvert \bar{v}_0 \rvert^2\,
  V_\rho\bigl[\bigl(f(\eta)-\ii g(\eta)\bigr)^\ast,\,
  f(\eta)-\ii g(\eta)\bigr] \,.
\label{Poin_V}
\end{align}

The in-out propagator can be readily obtained 
by substituting \eqref{in-bar-phi}, \eqref{out-phi} 
and \eqref{Poin_W} into \eqref{Poin_G10} as
\begin{align}
 G^{\out/\inn}_\bk(\eta,\eta') 
 &=\frac{\pi}{2}\,e^{-\ii\varepsilon}\,f(\eta_>)\,
 \bigl[f(\eta_<)-\ii g(\eta_<)\bigr]
\nn\\
 &= \frac{\pi}{2}\, [(-\eta)(-\eta')]^{\frac{d-1}{2}} \,
       J_{\nu}\bigl(-k_\varepsilon\eta_>\bigr) \,
       H^{(2)}_{\nu}\bigl(-k_\varepsilon\eta_<\bigr) \,.
\end{align}
Here, $H_\nu^{(1,2)}(x)$  are the Hankel functions 
defined by $H_\nu^{(1,2)}(x)\equiv J_\nu(x)\pm\ii N_\nu(x)$\,, 
and we have set $\varepsilon=0$ in the last expression 
as far as it does not change the analytic property of the propagator. 
One the other hand, 
in order to calculate the in-in propagator, 
\begin{align}
 &G^{\inn/\inn}_\bk(\eta,\eta') 
  =\frac{\ii}{V_\rho}\,\bigl[ f(\eta_>)-\ii g(\eta_>)\bigr]^\ast\,
  \bigl[ f(\eta_<)-\ii g(\eta_<) \bigr]
\nn\\
 &\Bigl( V_\rho=V_\rho\bigl[\bigl(f(s)-\ii g(s)\bigr)^\ast,\,
  f(s)-\ii g(s)\bigr](\eta_s) \Bigr)\,,
\label{Poin_in_in}
\end{align}
we first notice 
that the complex conjugate of 
$f(\eta)-\ii g(\eta)=(-\eta)^{\frac{d-1}{2}}\,
H^{(2)}_{\nu_\varepsilon}(-k_\varepsilon\eta)$ 
is given by  
$\bigl[f(\eta)-\ii g(\eta)\bigr]^\ast
=(-\eta)^{\frac{d-1}{2}}\,
H^{(1)}_{\nu_{-\varepsilon}}(-k_{-\varepsilon}\eta)$\,. 
Thus, for the small mass case, $m<(d-1)/2$ ($\nu\in\bRR$)\,,  
we have $\bigl[H^{(2)}_{\nu}(z)\bigr]^\ast
=H^{(1)}_{\nu}(z^\ast)$\,, 
so that we obtain 
\begin{align}
 f(\eta)-\ii g(\eta)
 &=(-\eta)^{\frac{d-1}{2}}\,
 H^{(2)}_{\nu}(-k_\varepsilon\eta)+O(\varepsilon)\,,
\label{Poin_in}
\\
 \bigl[  f(\eta)-\ii g(\eta)\bigr]^\ast
 &=(-\eta)^{\frac{d-1}{2}}\,
 H^{(1)}_{\nu}(-k_{-\varepsilon}\eta)+O(\varepsilon)\,,
\label{Poin_in-ast}
\end{align}
which lead to 
\begin{align}
 V_\rho = \frac{4\ii}{\pi} + O(\varepsilon)\,.
\label{Poin_V_small}
\end{align}
For the large mass case, 
$m\geq (d-1)/2$ ($\nu=\ii\mu\in \ii\bRR$)\,,  
we have $\bigl[H^{(2)}_{\nu}(z)\bigr]^\ast
=\bigl[H^{(2)}_{\ii\mu}(z)\bigr]^\ast
=H^{(1)}_{-\ii\mu}(z^\ast)
=e^{-\pi\mu}\,H^{(1)}_{\ii\mu}(z^\ast)$\,, 
so that we obtain 
\begin{align}
 f(\eta)-\ii g(\eta)
 &=(-\eta)^{\frac{d-1}{2}}\,
 H^{(2)}_{\nu}(-k_\varepsilon\eta)+O(\varepsilon)\,,
\\
 \bigl[  f(\eta)-\ii g(\eta)\bigr]^\ast
 &= e^{-\pi\mu}\,(-\eta)^{\frac{d-1}{2}}\,
 H^{(1)}_{\nu}(-k_{-\varepsilon}\eta)+O(\varepsilon)\,,
\end{align}
which lead to
\begin{align}
 V_\rho=
  e^{-\pi\mu} \,\frac{4\ii}{\pi} + O(\varepsilon)\,.
\label{Poin_V_large}
\end{align}
Substituting Eqs.\ \eqref{Poin_in}--\eqref{Poin_V_large} 
to \eqref{Poin_in_in}, 
we obtain
\begin{align}
  &G^{\rm in/in}(\eta,\eta')
\nn\\
  &= \frac{\pi}{4}\, [(-\eta)(-\eta')]^{\frac{d-1}{2}} \,
   H^{(1)}_{\nu}\bigl(-k_{-\varepsilon}\eta_>\bigr) \,
   H^{(2)}_{\nu}\bigl(-k_\varepsilon\eta_<\bigr) \,.
\label{Poin_Ginin}
\end{align}

We here make a few comments. 
With $\varepsilon$ set to zero, 
the wave function $\bar\varphi(\eta;\eta_0)$ 
converges in the limit $\eta_0\rightarrow -\infty$ 
to a finite function 
\begin{align}
 \varphi^\ast_{\inn}(\eta)\equiv 
 \frac{\sqrt{\pi}}{2}\,(-\eta)^{\frac{d-1}{2}}\,
 H^{(2)}_{\nu}(-k\,\eta)
\end{align} 
up to an oscillatory phase, 
while $\varphi(\eta;\eta_1)$ diverges 
as $(-\eta_1)^{-\nu}$ in the limit $\eta_1\rightarrow 0$\,. 
This difference can be attributed to the fact 
that the timelike vector $\partial_\eta$ becomes asymptotically 
a Killing vector in the remote past, 
but does not in the remote future.%
\footnote{
In fact, the Lie derivative of $g_{\mu\nu}$ 
with respect to the vector $\xi=\partial_\eta$ 
is $\pounds_\xi\,g_{\mu\nu}\propto (-\eta)^{-3}$\,. 
} 
The finite asymptotic function $\varphi_{\inn}(\eta)$ 
coincides with the positive-mode wave function 
associated with the Euclidean vacuum up to a phase.
One should note that, in the limit $\eta_1\to 0$\,, 
the divergence in $\varphi(\eta;\eta_1)$ 
is canceled out with that in $W_\rho[\varphi(\eta,\eta_1),\bar{\varphi}(\eta,\eta_0)]$ 
and the in-out propagator is obtained with a finite value.

For completeness, we show the asymptotic forms the Bogoliubov coefficients
\begin{align}
 \alpha(\eta_1;\eta_0)
 &\sim {}- \frac{\Gamma(\nu)}{2\sqrt{2\pi\,m}}\, 
     (-k\,\eta_1/2)^{-\nu_\varepsilon} \nn\\
 &\times\Bigl(\frac{d-1}{2}- \nu - \ii m \Bigr)\,
     e^{- \ii \bigl(k_\varepsilon\eta_0 +\frac{\pi(2 \nu +1)}{4}\bigr)}\,,\\
 \beta(\eta_1;\eta_0)
 &\sim {}- \ii \frac{(d-2)\Gamma(\nu)}{8\sqrt{2\pi\,m}}\, 
     (-k\,\eta_1/2)^{-\nu_\varepsilon} (-k\,\eta_0)^{-1}\nn\\
 &\times\Bigl(\frac{d-1}{2}- \nu - \ii m \Bigr)\,
     e^{- \ii \bigl(k_\varepsilon\eta_0 +\frac{\pi(2 \nu +1)}{4}\bigr)}\,,
\end{align}
which diverges in either of the limits 
$\eta_0\to -\infty$ and $\eta_1\to 0$\,.

\subsubsection{Propagators in the Poincar\'e patch}
\label{sec:spacetime_Poincare}

Since the eigenfunctions $\{Y_{\bk,\,a}(\bx)\}$ are the same as 
those given in section \ref{sec:Minkowski}, 
the propagators in spacetime can be written in the form
\begin{align}
 &G^{\{\genfrac{}{}{0pt}{}{\out/\inn}{\inn/\inn}\}}(x,x') \nn\\
 &= \sum_{\bk\geq 0}\,G^{\{\genfrac{}{}{0pt}{}{\out/\inn}{\inn/\inn}\}}_\bk(\eta,\eta')\,
    \sum_a\,Y_{\bk,a}(\bx)\,Y_{\bk,a}(\bx')\nn\\
 &= \frac{1}{(2\pi)^{\frac{d-1}{2}}\,\abs{\bx-\bx'}^{\frac{d-3}{2}}}\nn\\
 &\times  \int_0^\infty \rmd k\,k^{\frac{d-1}{2}}\,
             J_{\frac{d-3}{2}}\bigl(k\abs{\bx-\bx'}\bigr)\, 
    G^{\{\genfrac{}{}{0pt}{}{\out/\inn}{\inn/\inn}\}}_\bk(\eta,\eta') \,,
\end{align}
as in the case of Minkowski space (see Appendix \ref{appendix:propagator_Minkowski}).

For the in-out propagator, we have
\begin{align}
 &G^{\out/\inn}(x,x') \nn\\
 &=\frac{e^{-\ii\varepsilon}\,\pi}{2}\, 
  \frac{[(-\eta)(-\eta')]^{\frac{d-1}{2}}}
       {(2\pi)^{\frac{d-1}{2}}\,\abs{\bx-\bx'}^{\frac{d-3}{2}}}\,\nn\\
 &\times\int_0^\infty \rmd k\,k^{\frac{d-1}{2}}\,
    J_{\frac{d-3}{2}}\bigl(k\abs{\bx-\bx'}\bigr) \, 
    J_{\nu}(-k_\varepsilon\eta_>) \,
    H^{(2)}_{\nu}(-k_\varepsilon\eta_<) \,.
\end{align}
As is proved in Appendix \ref{appendix:Poincare}, 
this can be integrated to the form%
\footnote{
Here we have taken the limit $\varepsilon\to 0$\,. 
$\LP_\nu^\mu(z)$ and $\LQ_\nu^\mu(z)$ 
denote the associated Legendre functions of the first and second kind 
that are defined on the complex $z$-plane other than the cut 
along the real axis to the left of the point $z=1$\,. 
There are another type of associated Legendre functions 
that are defined on the interval $(-1,1)$, 
which we denote by $\LPc(x)$ and $\LQc(x)$\,. 
See Appendix \ref{appendix:Legendre_addition} for their definitions 
and several useful identities. 
} 
\begin{align}
 G^{\out/\inn}(x,x') 
   = \frac{e^{-\ii\pi(d-2)}}{(2\pi)^{d/2}}\,(u^2-1)^{-\frac{d-2}{4}}\,
     \LQ_{\nu-1/2}^{\frac{d-2}{2}}(u) 
\label{Poincare_in_out_full}
\end{align}
with $u= Z(x,x')-\ii 0$\,.

The in-in propagator 
\begin{align}
 &G^{\inn/\inn}(x,x') 
\nn\\
 &=\frac{\pi}{4}\, 
  \frac{[(-\eta)(-\eta')]^{\frac{d-1}{2}}}
       {(2\pi)^{\frac{d-1}{2}}\,\abs{\bx-\bx'}^{\frac{d-3}{2}}}
\nn\\
 &\times\int_0^\infty \rmd k\,k^{\frac{d-1}{2}}\,
    J_{\frac{d-3}{2}}\bigl(k\abs{\bx-\bx'}\bigr) \, 
    H^{(1)}_{\nu}(-k_{-\varepsilon}\eta_>) \,
    H^{(2)}_{\nu}(-k_\varepsilon\eta_<) 
\end{align}
can be rewritten in a similar manner to the form
\begin{align}
 &G^{\inn/\inn}(x,x') 
\nn\\
 &= \frac{\Gamma\bigl(\frac{d-1}{2}+\nu\bigr)\,\Gamma\bigl(\frac{d-1}{2}-\nu\bigr)}
         {2\,(2\pi)^{d/2}}\,(u^2-1)^{-\frac{d-2}{4}}\,
  \LP_{\nu-1/2}^{-\frac{d-2}{2}}(u) 
\label{Poincare_in_in_full}
\end{align}
with $u\equiv -Z(x,x')+\ii\,0$\,. 
A proof is given also in Appendix \ref{appendix:Poincare}.
This can be further rewritten as
\begin{align}
 &G^{\inn/\inn}(x,x') \nn\\
 &= \frac{\Gamma\bigl(\frac{d-1}{2}+\nu\bigr)\,\Gamma\bigl(\frac{d-1}{2}-\nu\bigr)}
        {(4\pi)^{d/2}\,\Gamma\bigl(d/2\bigr)}\,\nn\\
 &\quad\times F\Bigl(\frac{d-1}{2}+\nu\,,\,\frac{d-1}{2}-\nu\,;\,\frac{d}{2}\,;\,\frac{1-u}{2}\Bigr)\\
 &= \frac{\Gamma(\frac{d-1}{2})}{4 \pi^{(d-1)/2}\,
    \sin\bigl[\pi\bigl(\frac{d-1}{2}-\nu\bigr)\bigr]}\,
    C_{\nu-\frac{d-1}{2}}^{\frac{d-1}{2}}(u)\,,
\end{align}
where Kummer's relation \eqref{Kummers_relation} 
has been used in the first equality 
and $C_{\nu}^{\mu}(x)$ is the Gegenbauer function. 
This propagator is the same as the well-known in-in propagator 
associated with the Euclidean vacuum.

If we consider the massless limit where $\nu\rightarrow(d-1)/2$, 
the in-in propagator diverges, as was pointed out in \cite{Allen:1985ux}. 
In contrast, we find that the in-out propagator has a finite massless limit:
\begin{align}
 G^{\out/\inn}(x,x') 
   = \frac{e^{-\ii\pi(d-2)}}{(2\pi)^{d/2}}\,(u^2-1)^{-\frac{d-2}{4}}\,
     \LQ_{\frac{d-2}{2}}^{\frac{d-2}{2}}(u) 
\nn\\
 \mbox{(massless)}\,.
\end{align}

\subsection{Scalar field in the global patch}
\label{sec:global}

In the global patch, as a complete set of 
eigenfunctions of the spatial Laplacian $\Delta_{d-1}$ on $S^{d-1}$\,, 
we take  (real-valued) spherical harmonics 
$\{Y_{LM}(\bOmega)\}$\,.
They satisfy $\Delta_{d-1}\,Y_{LM} = -L(L+d-2)\,Y_{LM}$\ $(M=1,\dotsc,N_L^{(d)})$\,, 
and the degeneracy $N_L^{(d)}$ is given by
\begin{align}
 N_L^{(d)} = \frac{(L+d-3)!}{(d-2)!\,L!}\,(2L+d-2) 
\end{align}
with the exceptional case $d=2$ and $L=0$ where $N_0^{(2)}\equiv 1$. 
We choose them such that they are orthonormal: 
\begin{align}
 \int\rmd \bOmega \,Y_{LM}(\bOmega)\,Y_{L'M'}(\bOmega) 
  = \delta_{LL'}\,\delta_{MM'} \,.
\end{align}
Then, by expanding $\phi(x)$ as 
\begin{align}
 \phi(x) = \sum_{L,M}\phi_{LM}(t)\,Y_{LM}(\bOmega) \,, 
\end{align}
the mode function $q(t)\equiv \phi_{LM}(t)$ describes a harmonic oscillator 
with time-dependent mass and frequency of the following form 
[see \eqref{rhomega-1} and \eqref{rhomega-2}]:
\begin{align}
 \rho(t) &= e^{\ii\varepsilon}\bigl(1-t^2\bigr)^{-\frac{d-3}{2}}  \,, \\
 \omega(t) &=\bigl(1-t^2\bigr)^{-1}\,
               e^{-\ii\varepsilon}\sqrt{L\,(L+d-2)\,\bigl(1-t^2\bigr)+m^2}\nn\\
 &\equiv \bigl(1-t^2\bigr)^{-1} \,\bar{m}(t)\,.
\end{align}

\subsubsection{Propagators for each mode in the global patch}
\label{sec:mode_global}

The equation of motion takes the form
\begin{align}
 \ddot{q}(t)+(d-3)\,\frac{t}{1-t^2}\,\dot{q}(t)+\omega^2(t)\,q(t)=0\,.
\end{align}
We choose a pair of independent solutions as
\begin{align}
 f(t) = \bigl(1-t^2\bigr)^{\frac{d-1}{4}}\,
  \LPc^{\nu_\varepsilon}_{k_\varepsilon}(t) \,, \quad 
 g(t) = \bigl(1-t^2\bigr)^{\frac{d-1}{4}}\,
  \LQc^{\nu_\varepsilon}_{k_\varepsilon}(t) \,,
\end{align}
where 
\begin{align}
 k_\varepsilon &=-\frac{1}{2}
 +\sqrt{\Bigl(\frac{d-2}{2}\Bigr)^2 +e^{-2\ii\varepsilon}L(L+d-2)} \nn\\
&=k-\ii\varepsilon \quad \Bigl(k\equiv L+\frac{d-3}{2}\Bigr)\,,
\end{align}
and $\LPc_\nu^\mu(t)$ and $\LQc_\nu^\mu(t)$ are the associated Legendre functions 
defined on the interval $(-1,1)$ (see Appendix \ref{appendix:Legendre_addition}). 
The weighted Wronskian (constant in $t$) then has the form
\begin{align}
 W_\rho[f,g] = \rho(t)\,W[f,g](t)
  = e^{\ii\varepsilon}\,\frac{\Gamma(k_\varepsilon+\nu_{\varepsilon}+1)}{\Gamma(k_\varepsilon-\nu_{\varepsilon}+1)}\,.
\end{align}
The functions $u(t)$ and $v(t)$ defined in \eqref{uuvv-1} and \eqref{uuvv-2} are then given by
\begin{align}
 \biggl\{\begin{array}{l}
  u(t) \\ \bar{u}(t)
 \end{array}\biggr\}
  &=e^{\ii\varepsilon}\bigl(1-t^2\bigr)^{-\frac{d-1}{4}}\nn\\
  &\times\Bigl\{\Bigl[\Bigl(k_\varepsilon-\frac{d-3}{2}\Bigr)\,t\pm\ii\bar{m}(t)\Bigr]\LPc^{\nu_\varepsilon}_{k_\varepsilon}(t)\nn\\
  &\qquad\qquad\quad -\bigl(k_\varepsilon-\nu_{\varepsilon}+1\bigr)\LPc^{\nu_\varepsilon}_{k_\varepsilon+1}(t)\Bigr\}  \,, \\
 \biggl\{\begin{array}{l}
  v(t) \\ \bar{v}(t)
 \end{array}\biggr\}
  &=e^{\ii\varepsilon}\bigl(1-t^2\bigr)^{-\frac{d-1}{4}}\nn\\
  &\times\Bigl\{\Bigl[\Bigl(k_\varepsilon-\frac{d-3}{2}\Bigr)\,t\pm\ii\bar{m}(t)\Bigr]\LQc^{\nu_\varepsilon}_{k_\varepsilon}(t)\nn\\
  &\qquad\qquad\quad -\bigl(k_\varepsilon-\nu_{\varepsilon}+1\bigr)\LQc^{\nu_\varepsilon}_{k_\varepsilon+1}(t)\Bigr\}  \,, 
 \end{align}
and the wave functions $(\varphi,\,\bar{\varphi})$ take the form
\begin{align}
 \varphi(t;t_I) 
  &=  \frac{e^{-\ii\varepsilon}\,\Gamma(k_\varepsilon-\nu_{\varepsilon}+1)}
           {\sqrt{2\abs{\bar{m}_{I}}}\,\Gamma(k_\varepsilon+\nu_{\varepsilon}+1)}\,\nn\\
  &\qquad \times(1-t_I^2)^{\frac{d-1}{4}} 
      \bigl(v_I \, f(t)-u_I \, g(t) \bigr) \,, \\
 \bar{\varphi}(t;t_I) 
  &= \frac{e^{-\ii\varepsilon}\,\Gamma(k_\varepsilon-\nu_{\varepsilon}+1)}
          {\sqrt{2\abs{\bar{m}_{I}}}\, 
     \Gamma(k_\varepsilon+\nu_{\varepsilon}+1)}\,\nn\\
  &\qquad \times(1-t_I^2)^{\frac{d-1}{4}} 
     \bigl(\bar{v}_I \, f(t)-\bar{u}_I \, g(t) \bigr)\,.
\end{align}

We now send $t_0$, $t_1$ to the boundary of the global patch; 
$t_0 \rightarrow t_i=-1$ and $t_1 \rightarrow t_f=1$\,. 
Using \eqref{p1} and \eqref{q1}, 
we see that $u_1=u(t_1)$ and $v_1=v(t_1)$ take the following asymptotic forms 
in the limit $t_1 \rightarrow 1$\,:
\begin{align}
 \biggl\{\begin{array}{l}
  u_1 \\ \bar{u}_1
 \end{array}\biggr\}
  &\sim 
   \frac{2^{\nu_{\varepsilon}}\,e^{\ii\varepsilon}\sin(\pi\nu_{\varepsilon})\,
   \Gamma(\nu_{\varepsilon})}{\pi}\nn\\
  &\times\biggl(-\frac{d-1}{2}+\nu_{\varepsilon}\pm\ii\bar{m}_1\biggr)\,
   \bigl(1-t_1^2\bigr)^{-\frac{d-1}{4}-\frac{\nu_{\varepsilon}}{2}} \,, \\
 \biggl\{\begin{array}{l}
  v_1 \\ \bar{v}_1
 \end{array}\biggr\}
  &\sim 
   2^{\nu_{\varepsilon}-1}\,e^{\ii\varepsilon}\cos(\pi\nu_{\varepsilon})\,
  \Gamma(\nu_{\varepsilon})\nn\\
  &\times\biggl(-\frac{d-1}{2}+\nu_{\varepsilon}\pm\ii\bar{m}_1\biggr)\,
   \bigl(1-t_1^2\bigr)^{-\frac{d-1}{4}-\frac{\nu_{\varepsilon}}{2}}
\nn\\
 &\sim \frac{\pi \cos\pi\nu_\varepsilon}{2\sin\pi\nu_\varepsilon}
  \biggl\{\begin{array}{l}
  u_1 \\ \bar{u}_1
 \end{array}\biggr\}
\,,
\end{align}
and thus we find that the wave functions behave as%
\footnote{
The asymptotic forms given in Eq.~\eqref{global_out} 
do not satisfy the normalization condition \eqref{wronskian}. 
Actually, to ensure this normalization, 
we need to add a subleading term proportional to 
$\bigl(1-t_1^2\bigr)^{\nu_{\varepsilon}/2}$, 
which is omitted from the above asymptotic forms. 
The asymptotic forms, however, are yet sufficient 
for calculating various propagators.
\label{footnote:global}
} 
\begin{align}
 \biggl\{\begin{array}{l}
  \varphi(t;t_1) \\ \bar{\varphi}(t;t_1)
 \end{array}\biggr\}
 &\sim\frac{e^{-\ii\varepsilon}\,\Gamma(k_\varepsilon-\nu_{\varepsilon}+1)}
           {\sqrt{2m}\,\Gamma(k_\varepsilon+\nu_{\varepsilon}+1)}\,
 (1-t_1^2)^{\frac{d-1}{4}} 
\nn\\
  &\quad \times
  \biggl\{\begin{array}{l}
  v_1 \\ \bar{v}_1
 \end{array}\biggr\}\,
 \Bigl[
  f(t)-\frac{2\sin\pi\nu_\varepsilon}{\pi\cos\pi\nu_\varepsilon}\,g(t)
 \Bigr]
\nn\\
 &= \gamma_1\,
 \biggl\{\begin{array}{l}
  v_1 \\ \bar{v}_1
 \end{array}\biggr\}\,
 (1-t^2)^{\frac{d-1}{4}}\,\LPc^{-\nu_\varepsilon}_{k_\varepsilon}(t)\,.
\label{global_out}
\end{align}
Here, $\gamma_1\equiv e^{-\ii\varepsilon} 
(1-t_1^2)^{(d-1)/4}/(\sqrt{2m}\,\cos\pi\nu_\varepsilon)$\,,
and we have used the formula \eqref{LPc_mmu} 
and the fact that $\bar{m}_1 \to e^{-\ii\varepsilon}m$ ($t\to +1$)\,.

Similarly, we see that 
$u_0=u(t_0)$ and $v_0=v(t_0)$ have the following asymptotic forms 
in the limit $t_0 \rightarrow -1$\,:
\begin{align}
 \biggl\{\begin{array}{l}
  u_0 \\ \bar{u}_0
 \end{array}\biggr\}
  &\sim 
  \left\{\begin{array}{l}
   \displaystyle  
     \frac{e^{\ii\varepsilon}\,2^{-\nu_\varepsilon}\,
           \cos(\pi k_\varepsilon)\,\Gamma(k_\varepsilon+\nu_\varepsilon+1)}
          {\Gamma(\nu_\varepsilon+1)\,
            \Gamma(k_\varepsilon-\nu_\varepsilon+1)} \\[3mm]
   \displaystyle  
     \times \Bigl(\frac{d-1}{2}+\nu_\varepsilon\pm\ii\bar{m}_0\Bigr)\,
        \Bigl(1-t_0^2\bigr)^{-\frac{d-1}{4}+\frac{\nu_{\varepsilon}}{2}} \\[2mm]
    \qquad\qquad\qquad\qquad (d:\odd)\\[3mm]
   \displaystyle  
    {}- \frac{e^{\ii\varepsilon}\,2^{\nu_\varepsilon}\,
     \sin(\pi k_\varepsilon)\,\Gamma(\nu_\varepsilon)}{\pi}\\[2mm]
   \displaystyle  
     \times \Bigl(\frac{d-1}{2}-\nu_\varepsilon\pm\ii\bar{m}_0\Bigr)\,
        \bigl(1-t_0^2\bigr)^{-\frac{d-1}{4}-\frac{\nu_{\varepsilon}}{2}}\\[2mm]
    \qquad\qquad\qquad\qquad (d:\even)\\
  \end{array}\right. \,, \\
 \biggl\{\begin{array}{l}
  v_0 \\ \bar{v}_0
 \end{array}\biggr\}
  &\sim 
  \left\{\begin{array}{ll}
    \displaystyle
       {}- e^{\ii\varepsilon}\,2^{\nu_\varepsilon-1}\,
       \cos(\pi k_\varepsilon)\,\Gamma(\nu_\varepsilon) \\[2mm]
   \displaystyle  
       \times\Bigl(\frac{d-1}{2}-\nu_\varepsilon\pm\ii\bar{m}_0\Bigr)\,
       \bigl(1-t_0^2\bigr)^{-\frac{d-1}{4}-\frac{\nu_{\varepsilon}}{2}} \\[2mm]
    \qquad\qquad\qquad\qquad (d:\odd)\\[3mm]
    \displaystyle
       {}-\frac{e^{\ii\varepsilon}\,2^{-\nu_\varepsilon-1} \, 
         \pi \, \Gamma(k_\varepsilon+\nu_\varepsilon+1)}
            {\sin(\pi k_\varepsilon)\,
             \Gamma(\nu_\varepsilon+1)\,\Gamma(k_\varepsilon-\nu_\varepsilon+1)}\\[3mm]
   \displaystyle  
       \times\Bigl(\frac{d-1}{2}+\nu_\varepsilon\pm\ii\bar{m}_0\Bigr)\,
       \bigl(1-t_0^2\bigr)^{-\frac{d-1}{4}+\frac{\nu_{\varepsilon}}{2}} \\
    \qquad\qquad\qquad\qquad (d:\even)
  \end{array}\right. \,.
\end{align}
Here, in adopting the asymptotic forms of 
$\LPc_{k_\varepsilon}^{\nu_\varepsilon}(t_0)$ 
and $\LQc_{k_\varepsilon}^{\nu_\varepsilon}(t_0)$ 
for $t_0\to-1$ [see \eqref{pm1} and \eqref{qm1}]\,, 
we have used the fact that $\Ree \nu_{\varepsilon} >0$\,, 
which particularly means that  
$\bigl(1-t_0^2\bigr)^{-\frac{\nu_{\varepsilon}}{2}}
 \gg\bigl(1-t_0^2\bigr)^{\frac{\nu_{\varepsilon}}{2}}$\,.
The same inequality can be used to show that
\begin{align}
 \biggl\{\begin{array}{l}
  u_0 \\ \bar{u}_0
 \end{array}\biggr\}
&\ll
 \biggl\{\begin{array}{l}
  v_0 \\ \bar{v}_0
 \end{array}\biggr\}
 \quad \mbox{($d$\,: odd)}\,,
\\
 \biggl\{\begin{array}{l}
  u_0 \\ \bar{u}_0
 \end{array}\biggr\}
&\gg
 \biggl\{\begin{array}{l}
  v_0 \\ \bar{v}_0
 \end{array}\biggr\}
 \quad \mbox{($d$\,: even)}\,,
\end{align} 
from which we find that
\begin{align}
 \biggl\{\begin{array}{l}
  \varphi(t;t_0) \\ \bar{\varphi}(t;t_0)
 \end{array}\biggr\}
& \sim 
 \gamma_0\,\biggl\{\begin{array}{l}
  v_0 \\ \bar{v}_0
 \end{array}\biggr\}
 (1-t^2)^{\frac{d-1}{4}}\,\LPc^{\nu_\varepsilon}_{k_\varepsilon}(t)
\nn\\
 & \qquad\qquad\qquad \mbox{($d$\,: odd)}\,,
\label{global_in_odd}
\\
 \biggl\{\begin{array}{l}
  \varphi(t;t_0) \\ \bar{\varphi}(t;t_0)
 \end{array}\biggr\}
 &\sim 
 -\gamma_0\,\biggl\{\begin{array}{l}
  u_0 \\ \bar{u}_0
 \end{array}\biggr\}
 (1-t^2)^{\frac{d-1}{4}}\,\LQc^{\nu_\varepsilon}_{k_\varepsilon}(t)
\nn\\
 &\qquad\qquad\qquad \mbox{($d$\,: even)}\,,
\label{global_in_even}
\end{align}
where $\gamma_0\equiv e^{-\ii\varepsilon}\,
 \Gamma(k_\varepsilon-\nu_{\varepsilon}+1)\,
 (1-t_0^2)^{\frac{d-1}{4}}/
 \bigl(\sqrt{2m}\,
 \Gamma(k_\varepsilon+\nu_{\varepsilon}+1)\bigr)$\,.

With the wave functions $\varphi(t;t_I)$ at hand 
[see \eqref{global_out}, \eqref{global_in_odd} and \eqref{global_in_even}], 
the weighted Wronskian can be readily obtained as
\begin{align}
 &W_\rho[\varphi(t;t_1),\,\bar\varphi(t;t_0)]
\nn\\
 &\sim
 \Biggl\{
 \begin{array}{ll}
  +\gamma_1 \gamma_0\,v_1 \bar{v}_0\,(1-t^2)\,
  W_\rho[\LPc^{-\nu_\varepsilon}_{k_\varepsilon}(t),\,
  \LPc^{\nu_\varepsilon}_{k_\varepsilon}(t)]
  & \mbox{($d$\,: odd)} \\
  -\gamma_1\gamma_0\,v_1\bar{u}_0\,(1-t^2)\,
  W_\rho[\LPc^{-\nu_\varepsilon}_{k_\varepsilon}(t),\,
  \LQc^{\nu_\varepsilon}_{k_\varepsilon}(t)]
  & \mbox{($d$\,: even)}
 \end{array}
\nn\\
 &= 
 \Biggl\{
 \begin{array}{ll}
  +e^{\ii\varepsilon}\,\gamma_1\gamma_0\,v_1\bar{v}_0\,
  \dfrac{2}{\pi}\,\sin\pi\nu_\varepsilon 
  & \mbox{($d$\,: odd)} \\
  -e^{\ii\varepsilon}\,\gamma_1\gamma_0\,v_1\bar{u}_0\,
  \cos\pi\nu_\varepsilon
  & \mbox{($d$\,: even)} 
 \end{array}\,.
\end{align} 
We then obtain the in-out propagator 
\begin{align}
  &G^{\out/\inn}_{L}(t,t') \nn\\
  &=\lim_{\genfrac{}{}{0pt}{}{t_0 \rightarrow -1}{t_1 \rightarrow +1}}
    \frac{\ii}{W_\rho[\varphi(t;t_1),\,\bar\varphi(t;t_0)]}\,
    \varphi(t_{>};t_1)\,\bar{\varphi}(t_{<};t_0) \nn\\[2mm]
  &=\left\{\begin{array}{ll}
    \displaystyle
     \frac{\ii\pi}{2\sin\pi\nu}\,
     \bigl[(1-t_{>}^2)\,(1-t_{<}^2)\bigr]^{(d-1)/4}\,
     \LPc^{-\nu}_{k}(t_>)\,
     \LPc^{\nu}_{k}(t_<) \\
     \qquad\qquad\qquad\qquad\qquad\qquad \mbox{($d$\,: odd)}\,,\\[3mm]
    \displaystyle
     \frac{\ii}{\cos\pi\nu}\,
     \bigl[(1-t_{>}^2)\,(1-t_{<}^2)\bigr]^{(d-1)/4}\,
     \LPc^{-\nu}_{k}(t_>)\,
     \LQc^{\nu}_{k}(t_<) \\
     \qquad\qquad\qquad\qquad\qquad\qquad \mbox{($d$\,: even)}\,.
 \end{array}\right.
\label{in-out_global}
\end{align}
In the last expression, we have set $\varepsilon=0$\,.

To obtain the in-in propagator, 
we first notice that 
\begin{align}
 V_\rho &\equiv V_\rho[\bar{\varphi}^\ast(t;t_0),\bar\varphi(t;t_0)]
\nn\\
 &=\Biggl\{
 \begin{array}{ll}
  \lvert \gamma_0\rvert^2\, \lvert \bar{v}_0\rvert^2\, V_\rho[f^\ast,f](T_s)
  & \mbox{($d$\,: odd)} \\
  \lvert \gamma_0\rvert^2\, \lvert \bar{u}_0\rvert^2\, V_\rho[g^\ast,g](T_s)
  & \mbox{($d$\,: even)}
 \end{array}\,,
\end{align}
from which we have
\begin{align}
&G^{\inn/\inn}_{L}(t,t') \nn\\
  &=\lim_{t_0 \rightarrow -1}
    \frac{\ii}{V_\rho[\bar{\varphi}^\ast(t;t_0),\,\bar{\varphi}(t;t_0)](T_s)}\,
    \bar{\varphi}^\ast(t_{>};t_0)\,\bar{\varphi}(t_{<};t_0) \nn\\[2mm]
  &=\left\{\begin{array}{ll}
    \displaystyle
     \frac{\ii}
          {V_\rho[f^\ast,f](T_s)}\,
          f^\ast(t_>) \, f(t_<)
     & \mbox{($d$\,: odd)}\,,\\[5mm]
    \displaystyle
     \frac{\ii}
          {V_\rho[g^\ast,g](T_s)}\,
          g^\ast(t_>) \, g(t_<)
     & \mbox{($d$\,: even)}\,.
 \end{array}\right.
\end{align}
When mass is large [$m\geq (d-1)/2$ and thus $\nu=\ii\mu\in\ii\bRR$]\,, 
we have
\begin{align}
 [f(t)]^\ast
 &=(1-t^2)^{\frac{d-1}{4}}\,
 \bigl[\LPc^{\nu_\varepsilon}_{k_\varepsilon}(t)\bigr]^\ast
\nn\\
 &=(1-t^2)^{\frac{d-1}{4}}\,
 \LPc^{-\nu_\varepsilon}_{k_\varepsilon}(t) + {\cal O}(\varepsilon)\,,
\\
 [g(t)]^\ast
 &=(1-t^2)^{\frac{d-1}{4}}\,
 \bigl[\LQc^{\nu_\varepsilon}_{k_\varepsilon}(t)\bigr]^\ast
\nn\\
 &=(1-t^2)^{\frac{d-1}{4}}\,
 \LQc^{-\nu_\varepsilon}_{k_\varepsilon}(t) + {\cal O}(\varepsilon)\,,
\end{align}
and thus 
\begin{align}
 V_\rho[f^\ast,f](T_s)&=\frac{2\ii\sinh\pi\mu}{\pi}+\mathcal{O}(\varepsilon)\,, 
\\
 V_\rho[g^\ast,g](T_s)&=\frac{\ii\pi\sinh\pi\mu}{2}+\mathcal{O}(\varepsilon)\,.
\end{align}
The in-in propagator is then obtained as
\begin{align}
 &G^{\inn/\inn}_{L}(t,t') \nn\\
  &=\left\{\begin{array}{l}
    \displaystyle
     \frac{\pi}{2\sinh(\pi\mu)}\,
     \bigl[(1-t_{>}^2)\,(1-t_{<}^2)\bigr]^{\frac{d-1}{4}} \\
     \qquad\qquad\qquad\times \LPc^{-\ii\mu}_{k}(t_>)\,
     \LPc^{\ii\mu}_{k}(t_<) \qquad (d:\odd)\,,\\
    \displaystyle
     \frac{2}{\pi\sinh(\pi\mu)}\,
     \bigl[(1-t_{>}^2)\,(1-t_{<}^2)\bigr]^{\frac{d-1}{4}}\,\\
     \qquad\qquad\qquad\times \LQc^{-\ii\mu}_{k}(t_>)\,
     \LQc^{\ii\mu}_{k}(t_<) \qquad (d:\even)\,.
 \end{array}\right.
\label{in-in_global}
\end{align}
On the other hand, 
when mass is small 
[$m<(d-1)/2$ and thus $\nu\in\bRR$]\,, 
we have $f^\ast(t)=f(t)+{\cal O}(\varepsilon)$\,,  
$g^\ast(t)=g(t)+{\cal O}(\varepsilon)$\,, 
and thus ${V_\rho[f^\ast,f](T_s)}={\cal O}(\varepsilon)$\,, 
${V_\rho[g^\ast,g](T_s)}={\cal O}(\varepsilon)$\,. 
This means that 
$\lim_{t_0\to -1}G_{00}(t,t';\,t_0,t_0)$ 
has the singularity of the form ${\cal O}(\varepsilon^{-1})$ 
and we cannot set $\varepsilon=0$\,.

The wave functions at the remote past and future had been obtained 
for the heavy mass case ($m\geq (d-1)/2$)  
in \cite{Mottola:1984ar,Bousso:2001mw} 
as%
\footnote{
We here give the following identities 
which are useful in comparing our results 
with the literature:
\begin{align}
 &\frac{2^{\frac{d-1}{2}+L}\,\cosh^{L}\tau\,e^{(-\frac{d-1}{2}-L-\nu)\,\tau}}
         {\Gamma(1+\nu)} \nn\\
 &\times F\Bigl(\frac{d-1}{2}+L,\frac{d-1}{2}+L+\nu;1+\nu;-e^{-2\tau}\Bigr) 
  =\bigl(1-t^2\bigr)^{\frac{d-1}{4}}\,\LPc^{-\nu}_{k}(t) \,,\nn\\
 &\frac{2^{\frac{d-1}{2}+L}\,\cosh^{L}\tau\,e^{(\frac{d-1}{2}+L-\nu)\,\tau}}
         {\Gamma(1-\nu)}\nn\\
 &\times F\Bigl(\frac{d-1}{2}+L,\frac{d-1}{2}+L-\nu;1-\nu;-e^{+2\tau}\Bigr) 
  =\bigl(1-t^2\bigr)^{\frac{d-1}{4}}\,\LPc^{\nu}_{k}(-t) \nn\\
 &= \frac{\Gamma(k+1+\nu)}{\Gamma(k+1-\nu)}\,(1-t^2)^{\frac{d-1}{4}}
 \left\{\begin{array}{ll}
    (-1)^k\,\LPc_k^{-\nu}(t) & (d:\odd)\cr 
    (-1)^{k+1/2}\,(2/\pi)\,\LQc_k^{-\nu}(t) &(d:\even)
          \end{array}\right. \,.\nn
\end{align}
} 
\begin{align}
 \varphi_{\out}(t)&\propto (1-t^2)^{\frac{d-1}{4}}\,\LPc_k^{-\ii\mu}(t)\,,
\\
 \varphi_{\inn}^{\ast}(t)&\propto\biggl\{\begin{array}{l}
 (1-t^2)^{\frac{d-1}{4}}\,\LPc_k^{\ii\mu}(t) \quad(d:\odd)\\
 (1-t^2)^{\frac{d-1}{4}}\,\LQc_k^{\ii\mu}(t) \quad(d:\even)
 \end{array} \,,
\end{align}
by requiring that $\varphi_\inn^\ast(t)$ $(\varphi_\out(t))$ 
be regular for $t\to -1$ ($t\to +1$) 
and an analytic function in the lower half of complex $m^2$ plane 
(see also \cite{Krotov:2010ma} 
where they are obtained by suitably choosing the Jost functions). 
Our propagators \eqref{in-out_global} and \eqref{in-in_global}
for $m \geq (d-1)/2$ are consistent with these wave functions.

From \eqref{albe}, the Bogoliubov coefficient $\alpha(t_1;t_0)$ 
can be found to have the asymptotic form
\begin{align}
 &\alpha(t_1;t_0) \nn\\
 &\textstyle 
  \sim {}- \frac{2\ii\sin(\pi\nu)}{\pi}
  \biggl[\frac{2^{\nu-1}\, \Gamma(\nu)}{\sqrt{2\bar{m}_{1}}}
         \Bigl(-\frac{d-1}{2}+\nu+\ii\bar{m}_1\Bigr)\,
         \bigl(1-t_1^2\bigr)^{-\frac{\nu_{\varepsilon}}{2}}\biggr] \nn\\
  &\textstyle 
  \quad\times 
    \biggl[-\frac{2^{\nu-1}\,\cos(\pi k)\,
                  \Gamma(\nu)\,\Gamma(k-\nu+1)}
                 {\sqrt{2\bar{m}_{0}}\Gamma(k+\nu+1)}\nn\\
  &\textstyle 
  \qquad\qquad\times 
            \Bigl(\frac{d-1}{2}-\nu-\ii\bar{m}_0\Bigr)
            \bigl(1-t_0^2\bigr)^{-\frac{\nu_{\varepsilon}}{2}}\biggr] \,,\\
 &\beta(t_1;t_0) \nn\\
 &\textstyle 
  \sim {}+ \frac{2\ii\sin(\pi\nu)}{\pi}
  \biggl[\frac{2^{\nu-1}\,\Gamma(\nu)}{\sqrt{2\bar{m}_{1}}}
  \Bigl(-\frac{d-1}{2}+\nu+\ii\bar{m}_1\Bigr)
  \bigl(1-t_1^2\bigr)^{-\frac{\nu_{\varepsilon}}{2}}\biggr] \nn\\
  &\textstyle 
  \quad\times 
  \biggl[-\frac{2^{\nu-1}\cos(\pi k)\,\Gamma(\nu)\,
  \Gamma(k-\nu+1)}
               {\sqrt{2\bar{m}_{0}}\,\Gamma(k+\nu+1)}\nn\\
  &\textstyle 
  \qquad\qquad\times 
  \Bigl(\frac{d-1}{2}-\nu+\ii\bar{m}_0\Bigr)
  \bigl(1-t_0^2\bigr)^{-\frac{\nu_{\varepsilon}}{2}}\biggr]
\end{align}
in odd dimensions, and
\begin{align}
 &\alpha(t_1;t_0) \nn\\
 &\textstyle\sim 
  {}- \ii\cos(\pi\nu)\,
  \biggl[\frac{2^{\nu-1}\, \Gamma(\nu)}{\sqrt{2\bar{m}_{1}}}
         \Bigl(-\frac{d-1}{2}+\nu+\ii\bar{m}_1\Bigr)\,
         \bigl(1-t_1^2\bigr)^{-\frac{\nu_{\varepsilon}}{2}}\biggr] \nn \\
 &\textstyle \times 
  \biggl[\frac{2^{\nu}\,\sin(\pi k)\,
               \Gamma(\nu)\,\Gamma(k-\nu+1)}
              {\pi\,\sqrt{2\bar{m}_{0}}\,\Gamma(k+\nu+1)}\,
         \Bigl(\frac{d-1}{2}-\nu-\ii\bar{m}_0\Bigr)\,
         \bigl(1-t_0^2\bigr)^{-\frac{\nu_{\varepsilon}}{2}}\biggr]\,, \\
 &\beta(t_1;t_0) \nn\\
 &\textstyle\sim {}+ \ii\cos(\pi\nu)
  \biggl[\frac{2^{\nu-1} \, \Gamma(\nu)}{\sqrt{2\bar{m}_{1}}}
  \Bigl(-\frac{d-1}{2}+\nu+\ii\bar{m}_1\Bigr)
  \bigl(1-t_1^2\bigr)^{-\frac{\nu_{\varepsilon}}{2}}\biggr] \nn \\
 &\textstyle \times 
   \biggl[\frac{2^{\nu} \sin (\pi k)\,
   \Gamma(\nu)\,\Gamma(k-\nu+1)}
               {\pi\sqrt{2\bar{m}_{0}}\,\Gamma(k+\nu+1)}
   \Bigl(\frac{d-1}{2}-\nu+\ii\bar{m}_0\Bigr)
   \bigl(1-t_0^2\bigr)^{-\frac{\nu_{\varepsilon}}{2}}\biggr]
\end{align}
in even dimensions.

\subsubsection{Propagators in the global patch}
\label{sec:spacetime_global}

We now make a sum over all modes 
to obtain the propagators, 
$G^{\out/\inn}(x,x')$ and $G^{\inn/\inn}(x,x')$.
For $d\geq 3$, the summation over $M$ can be written 
with the Gegenbauer polynomials as
\begin{align}
 \sum_{M=1}^{N_L^{(d)}} Y_{LM}(\bOmega)\, Y_{LM}(\bOmega') 
 &= \frac{2L+d-2}{(d-2)\,\abs{\Omega_{d-1}}}\,
   C_L^{\frac{d-2}{2}}(\bOmega \cdot \bOmega') \nn\\
 & \bigl(\,\abs{\Omega_{d-1}} = 2\pi^{\frac{d}{2}}/{\Gamma(d/2)}\,\bigr)\,.
\label{gegen}
\end{align}
As for $d=2$, the sum has the form
\begin{align}
 \sum_{M=1}^{N_L^{(2)}}Y_{LM}(\bOmega)\,Y_{LM}(\bOmega')
 =\left\{\begin{array}{ll}
   \displaystyle\frac{1}{2\pi} & (L=0)\,,\\
   \displaystyle\frac{\cos(L \bOmega\cdot\bOmega)}{\pi} & (L\geq 1)\,,
  \end{array}\right.
\end{align}
which is the same as the $d\rightarrow2$ limit 
of the expression \eqref{gegen}. 
Thus Eq.~\eqref{gegen} can be understood to hold for any dimensionality $d\geq 2$.
The in-out propagator in spacetime then takes the form
\begin{align}
 &G^{\out/\inn}(x,x') \nn\\
 &=\sum_{L=0}^{\infty} G^{\out/\inn}_{L}(t,t')\,
                       \sum_{M=1}^{N_L^{(d)}} Y_{LM}(\bOmega)\,Y_{LM}(\bOmega') \nn\\
 &=\sum_{L=0}^{\infty} \frac{2L+d-2}{(d-2)\,\abs{\Omega_{d-1}}}\,
                  G^{\out/\inn}_{L}(t,t')\,C_L^{\frac{d-2}{2}}(\bOmega\cdot\bOmega')\,,
\end{align}
which becomes
\begin{align}
 &G^{\out/\inn}_{\odd}(x,x') \nn\\
 &= \frac{\ii\pi}{2\sin(\pi\nu)\,(d-2)\,\abs{\Omega_{d-1}}}\,
    \bigl[ (1-t_{>}^2)\,(1-t_{<}^2)\bigr]^{(d-1)/4} \nn\\
 & \times \sum_{L=0}^{\infty}(2L+d-2)\, 
     \LPc^{-\nu}_{k}(t_>)\,\LPc^{\nu}_{k}(t_<) \, C_L^{\frac{d-2}{2}}(\cos\theta) 
\end{align}
in odd dimensions, and 
\begin{align}
 &G^{\out/\inn}_{\even}(x,x') \nn\\
 &= \frac{\ii}{\cos(\pi\nu)\,(d-2)\,\abs{\Omega_{d-1}}}\,
    \bigl[ (1-t_{>}^2)\,(1-t_{<}^2)\bigr]^{(d-1)/4} \nn\\
 & \times \sum_{L=0}^{\infty}(2L+d-2)\, 
     \LPc^{-\nu}_{k}(t_>)\,\LQc^{\nu}_{k}(t_<)\, C_L^{\frac{d-2}{2}}(\cos\theta) 
\end{align}
in even dimensions. 
Here, we have defined $\theta$ via the relation $\bOmega \cdot \bOmega'\equiv \cos\theta$.

Using Eqs.~\eqref{formulaodd} and \eqref{formulaeven} 
and introducing
\begin{align}
 u_{\pm}(x,x')
  &\equiv {}- Z(x,x') \pm \ii 0\nn\\
  &= {}-  \frac{-t\,t'+\cos\theta}{(1-t^2)^{\frac{1}{2}}\,(1-t'^2)^{\frac{1}{2}}} 
  \pm \ii 0\,,
\end{align}
we can rewrite the in-out propagator in a de Sitter invariant form:
\begin{align}
 &G^{\out/\inn}_{\odd}(x,x') \nn\\
 &=\frac{-e^{-\ii\pi\frac{d-2}{2}}}{2\,(2\pi)^{\frac{d}{2}}\,\sin(\pi\nu)}\nn\\
  &\times \biggl[(u_{+}^2-1)^{-\frac{d-2}{4}}\,
             \LQ^{\frac{d-2}{2}}_{\nu-\frac{1}{2}}(u_{+})
           -(u_{-}^2-1)^{-\frac{d-2}{4}}\,
             \LQ^{\frac{d-2}{2}}_{\nu-\frac{1}{2}}(u_{-}) \biggr] \,,
\label{globalodd}
\\
 &G^{\out/\inn}_{\even}(x,x') \nn\\
 &=\frac{\ii e^{-\ii\pi\frac{d-2}{2}}}{2\,(2\pi)^{\frac{d}{2}}\,\cos(\pi\nu)}\nn\\
  &\times \biggl[(u_{+}^2-1)^{-\frac{d-2}{4}}\,
             \LQ^{\frac{d-2}{2}}_{\nu-\frac{1}{2}}(u_{+})
           +(u_{-}^2-1)^{-\frac{d-2}{4}}\,
             \LQ^{\frac{d-2}{2}}_{\nu-\frac{1}{2}}(u_{-}) \biggr] \,.
\label{globaleven}
\end{align}

In the massless limit $m\to 0$ (or $\nu\rightarrow(d-1)/2$), we have
\begin{align}
 \sin\pi\nu\rightarrow 0 \quad (d:\odd)\,, \qquad 
 \cos\pi\nu\rightarrow 0 \quad (d:\even)\,,
\end{align}
and thus the propagators \eqref{globalodd} and \eqref{globaleven} diverge. 
We thus conclude that there exists no finite massless limit in the global patch, 
as opposed to the case of the Poincar\'e patch.

On the other hand, the in-in propagator 
in the heavy mass case ($m>(d-1)/2$) takes the form
\begin{align}
 &G^{\inn/\inn}(x,x') \nn\\
 &=\sum_{L=0}^{\infty} \frac{2L+d-2}{(d-2)\,\abs{\Omega_{d-1}}}\,
                  G^{\inn/\inn}_{L}(t,t')\,C_L^{\frac{d-2}{2}}(\bOmega\cdot\bOmega')\,,
\end{align}
which becomes
\begin{align}
 &G^{\inn/\inn}_{\odd}(x,x') \nn\\
 &= \frac{\ii\pi}{2\sin(\pi\nu)\,(d-2)\,\abs{\Omega_{d-1}}}\,
    \bigl[ (1-t_{>}^2)\,(1-t_{<}^2)\bigr]^{\frac{d-1}{4}} \nn\\
 &\quad \times \sum_{L=0}^{\infty}(2L+d-2)\, 
     \LPc^{-\nu}_{k}(t_>)\,\LPc^{\nu}_{k}(t_<) \, C_L^{\frac{d-2}{2}}(\cos\theta)
\label{global_in-in_odd}
\end{align}
in odd dimensions, and 
\begin{align}
 &G^{\inn/\inn}_{\even}(x,x') \nn\\
 &= \frac{2\ii}{\pi\sin(\pi\nu)\,(d-2)\,\abs{\Omega_{d-1}}}\,
    \bigl[ (1-t_{>}^2)\,(1-t_{<}^2)\bigr]^{\frac{d-1}{4}} \nn\\
 &\quad \times \sum_{L=0}^{\infty}(2L+d-2)\, 
     \LQc^{-\nu}_{k}(t_>)\,\LQc^{\nu}_{k}(t_<)\, C_L^{\frac{d-2}{2}}(\cos\theta) 
\label{global_in-in_even}
\end{align}
in even dimensions. 
Here, $\nu=\ii\mu=\ii\sqrt{m^2-(d-1)^2/4}$ ~$(\mu\in\bRR_{+})$\,. 
Note that $G^{\inn/\inn}_{\odd}(x,x') =G^{\out/\inn}_{\odd}(x,x') $, 
which is consistent with a well-known fact 
that the in-vacuum equals the out-vacuum up to a phase in odd dimensions 
(see \cite{Bousso:2001mw}). 
The summations in \eqref{global_in-in_odd} and \eqref{global_in-in_even} 
can be carried out analytically 
by using \eqref{formulaodd} and \eqref{formulaeven3}, 
and are again expressed in de Sitter invariant forms,
\begin{align}
 &G^{\inn/\inn}_{\odd}(x,x') =G^{\out/\inn}_{\odd}(x,x') 
\nn\\
 &=\frac{{}-e^{-\ii\pi\frac{d-2}{2}}}{2\,(2\pi)^{\frac{d}{2}}\,\sin(\pi\nu)}
     \Bigl[(u_{+}^2-1)^{-\frac{d-2}{4}}\,
             \LQ^{\frac{d-2}{2}}_{\nu-\frac{1}{2}}(u_{+})\nn\\
 &\qquad\qquad\qquad\qquad -(u_{-}^2-1)^{-\frac{d-2}{4}}\,
             \LQ^{\frac{d-2}{2}}_{\nu-\frac{1}{2}}(u_{-}) \Bigr] \,,
\\
 &G^{\inn/\inn}_{\even}(x,x') \nn\\
 &= \frac{e^{-\ii\pi\frac{d-2}{2}}}{2\,(2\pi)^{\frac{d}{2}}\,\sin\pi\nu}\,
    \Bigl\{
     (u_+^2-1)^{-\frac{d-2}{4}}\,\LQ_{\nu-\frac{1}{2}}^{\frac{d-2}{2}}(u_+)\nn\\
 &\qquad\qquad\qquad\qquad 
 - (u_-^2-1)^{-\frac{d-2}{4}}\,\LQ_{\nu-\frac{1}{2}}^{\frac{d-2}{2}}(u_-) \nn\\
 &\qquad + \frac{\ii\,\pi}{\cos\pi\nu}\,
     \bigl[e^{-\ii\pi\nu}\,(u_+^2-1)^{-\frac{d-2}{4}}\,
                             \LP_{\nu-\frac{1}{2}}^{\frac{d-2}{2}}(u_+)\nn\\
 &\qquad\qquad\qquad\qquad
         + e^{\ii\pi\nu}\,(u_-^2-1)^{-\frac{d-2}{4}}\, 
                             \LP_{\nu-\frac{1}{2}}^{\frac{d-2}{2}}(u_-)
     \bigr]\Bigr\}  \nn\\
 &= {}-\frac{e^{-\ii\pi\frac{d-2}{2}}}{2\,(2\pi)^{\frac{d}{2}}\,\sin\pi\nu}\,
    \Bigl\{
     (u_+^2-1)^{-\frac{d-2}{4}}\,\LQ_{\nu-\frac{1}{2}}^{\frac{d-2}{2}}(u_+) \nn\\
 &\qquad\qquad\qquad\qquad\qquad 
 - (u_-^2-1)^{-\frac{d-2}{4}}\,\LQ_{\nu-\frac{1}{2}}^{\frac{d-2}{2}}(u_-) \nn\\
 &\qquad {}- \frac{\pi}{\cos(\pi\nu)}\,
     \bigl[(u_+^2-1)^{-\frac{d-2}{4}}\,
                             \LP_{\nu-\frac{1}{2}}^{\frac{d-2}{2}}(-u_+)\nn\\
 &\qquad\qquad\qquad  - (u_-^2-1)^{-\frac{d-2}{4}}\, 
                             \LP_{\nu-\frac{1}{2}}^{\frac{d-2}{2}}(-u_-)
     \bigr]\Bigr\}\,.
\end{align}

\section{Feynman path integral in de Sitter space}
\label{sec:path_integral}

\begin{figure}[htbp]
\begin{center}
\includegraphics[width=8cm]{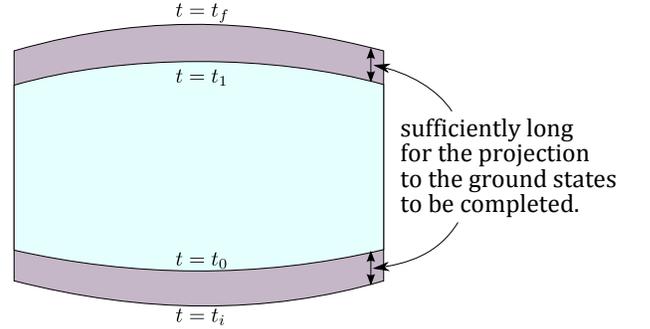}
\begin{quote}
\caption{The spacetime region where the path integral is performed. 
\label{fig:region}}
\end{quote}
\end{center}
\vspace{-6ex}
\end{figure}

In this section, we consider the Feynman propagator 
obtained by a path-integral in curved spacetime 
with the background metric \eqref{parametrization}, 
\begin{align}
 &\langle \phi(x)\,\phi(x') \rangle 
  \equiv \frac{\int[\rmd\phi]\,\phi(x)\,\phi(x')\,e^{\ii S_{\varepsilon}[\phi]}} 
              {\int[\rmd\phi]\,e^{\ii S_{\varepsilon}[\phi]}}\,, 
\label{propagator-path-int}
\\
  &S_\varepsilon[\phi] \equiv \frac{1}{2}\,\int\rmd t\int\rmd^{d-1}x\,\sqrt{h}\, 
  e^{\ii\varepsilon} N^{-1}\, A^{d-1} 
\nn\\
 & \times\Bigl(\partial_t\phi\,\partial_t\phi 
          +e^{-2\ii\varepsilon} N^2\,\bigl[A^{-2}\,\phi\,\Delta_{d-1}\phi
  -m^2\phi^2\bigr]\Bigr)\,.
\end{align}
This action gives a Hamiltonian of the form 
$H_s(t)=e^{-\ii\varepsilon}\,[H_s(t)\rvert_{\varepsilon=0}]$ in the Schr\"odinger picture.
We expect that the propagator defined by the path integral 
agrees with the in-out propagator obtained in the preceding sections.  
In fact, suppose that the base spacetime where $\phi(x)$ lives has a sufficiently large 
noncompact region in the temporal direction near the future and past boundaries 
at $t=t_f$ and $t=t_i$, respectively (see Fig.~\ref{fig:region}). 
Then, due to the existence of $\ii\varepsilon$\,, 
if we first define the path integral 
for a finite interval $(t_0,t_1)$ 
and send the initial time $t_0$ and the final time $t_1$ 
to the infinite past $t_i$ and the infinite future $t_f$, respectively, 
then the initial and final states would be well kept subject 
to the projection to the instantaneous ground state at each moment $t_0$ or $t_1$\,, 
and the dominant contribution to the path integral 
will be only from the configurations that are in the instantaneous ground states 
near the temporary boundaries.  
In this section, we first identify the (sufficient) condition 
under which such projection onto the instantaneous ground states can happen, 
and show that both the Poincar\'e and the global patches 
indeed satisfy this condition.

\subsection{Effective noncompactness in the temporal direction}
\label{iepsilon}

We first expand $\phi(x)$ as in \eqref{mode_expn}:
\begin{align}
 \phi(x) = \sum_n \phi_n(t)\, Y_n(\bx)\,.
\end{align}
The propagator is then written as a sum of the propagators over the mode $n$\,:
\begin{align}
 \langle \phi(x)\,\phi(x') \rangle 
 &= \sum_{n}\langle \phi_n(t)\,\phi_n(t') \rangle\, Y_{n}(\bx)\,Y_{n}(\bx') \,, 
\end{align}
where
\begin{align}
 &\langle \phi_n(t)\,\phi_n(t') \rangle 
 \equiv \lim_{{t_1\to t_f \atop t_0\to t_i}}
 \frac{\int[\rmd\phi_n]^{t_1}_{t_0}\,
  e^{\ii S_{n,\varepsilon}[\phi_n;\,t_1,t_0]}\,\phi_n(t)\,\phi_n(t')}
              {\int[\rmd\phi_n]^{t_1}_{t_0}\,
  e^{\ii S_{n,\varepsilon}[\phi_n;\,t_1,t_0]}} \,,
\label{PI_mode}
\\ 
 &S_{n,\varepsilon}[\phi_n;\,t_1,t_0] 
\nn\\
 &\equiv \int_{t_0}^{t_1}\!\!\rmd t\,\frac{1}{2}\,
  \bigl[e^{\ii\varepsilon}\,\abs{\rho(t)}\,\dot\phi_n^2(t)
        -e^{-\ii\varepsilon}\abs{\rho(t)}\,\abs{\omega_{n}(t)}^2\phi_n^2(t)\bigr] \,.
\label{PI_mode2}
\end{align}
For a fixed mode $n$\,, 
the propagator $ \langle \phi_n(t)\,\phi_n(t') \rangle $ 
can be given the following operator representation in the Schr\"odinger picture 
(we assume $t>t'$ in what follows): 
\begin{align}
 \lim_{\genfrac{}{}{0pt}{}{t_0 \to t_i}{t_1 \to t_f}}
 \frac{\bra{\psi_1,t_1}\,U(t_1,t)\,\phi_{n,s}\,U(t,t')\,\phi_{n,s}\,
  U(t',t_0)\,\ket{\psi_0,t_0}}
      {\bra{\psi_1,t_1}\,U(t_1,t_0)\,\ket{\psi_0,t_0}}\,.
\label{amplitude1}
\end{align}
Here, $\ket{\psi_1,t_1}$ and $\ket{\psi_0,t_0}$ are the final and the initial states 
to be specified as boundary conditions when performing a path integral,  
and are formally taken to be $\bra{\phi_n}\psi_1,t_1\rangle=\bra{\phi_n}\psi_0,t_0\rangle=1$ 
for the path integral \eqref{PI_mode}. 
In the following, we show that the amplitude \eqref{amplitude1} 
can be replaced by the in-out propagator 
\begin{align}
 &\lim_{\genfrac{}{}{0pt}{}{t_0 \to t_i}{t_1 \to t_f}}
 \frac{\bra{0_{t_1},t_1}\,U(t_1,t)\,\phi_{n,s}\,
 U(t,t')\,\phi_{n,s}\,U(t',t_0)\,\ket{0_{t_0},t_0}}
      {\bra{0_{t_1},t_1}\,U(t_1,t_0)\,\ket{0_{t_0},t_0}} 
\label{amplitude2}
\end{align}
for arbitrary $\ket{\psi_1,t_1}$ and $\ket{\psi_0,t_0}$\,, 
provided that the change of time variable 
$t\to\sigma(t)$ such that $\lvert\omega_n(t)\rvert\,\rmd t = \rmd\sigma$ 
maps the region $(t_i,t_f)$ onto a noncompact region for both sides, 
(i.e., $\sigma_i\equiv \sigma(t_i)=-\infty$ 
and $\sigma_f\equiv \sigma(t_f)=+\infty$)\,. 
When this condition is met, the foliation under consideration 
will be said to be 
{\em effectively noncompact in the temporal direction} for the mode $n$\,.

\begin{figure}[htbp]
\vspace{3ex}
\begin{center}
\includegraphics[width=7.5cm]{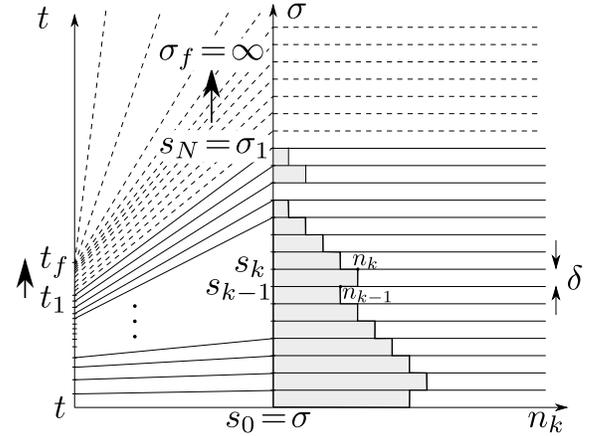}
\begin{quote}
\vspace{-3mm}
\caption{A path labeled by $\{n_k,s_k\}$\,. 
The amplitude suffers from a suppression 
proportional to the area of the shaded region. 
If $\sigma_1\to\sigma_f=\infty$ as $t_1\to t_f$\,, 
only such paths survive 
that are the instantaneous ground states in the far future 
(i.e., $n_k=0$ for large enough $k$)\,.
\label{fig:NewFIG3}}
\end{quote}
\end{center}
\vspace{-9ex}
\end{figure}

It is enough to show that the following equalities hold
for an arbitrary state $\ket{\psi}$:
\begin{align}
 &\lim_{t_1\to t_f}\bra{\psi_1,t_1}\,U(t_1,t)\,\ket{\psi}
\nn\\
 &=\lim_{t_1\to t_f}\bra{\psi_1,t_1} 0_{t_1},t_1\rangle\,
 \bra{0_{t_1},t_1}\,U(t_1,t)\,\ket{\psi}\,,
\label{amp_t1t}
\\
 &\lim_{t_0\to t_i}\bra{\psi}\,U(t,t_0)\, \ket{\psi_0,t_0}
\nn\\
 &=\lim_{t_0\to t_i}\bra{\psi}\,U(t,t_0)\, \ket{0_{t_0},t_0}\,
 \bra{0_{t_0},t_0} \psi_0,t_0\rangle\,.
\label{amp_tt0}
\end{align}
To show the first equality, we first introduce a new time coordinate $\sigma$ 
such that $\lvert\omega_n(t)\rvert\,\rmd t=\rmd \sigma$\,, 
which maps the time interval $(t,t_1)$ to a new interval $(\sigma,\sigma_1)$\,.
Then the Hamiltonian for the mode $n$ becomes (we will omit the index $n$ for brevity)%
\footnote{
We will discard the zero-point energy in the following discussions.
} 
\begin{align}
 H_{s}(t)\,\rmd t 
 &= e^{-\ii\varepsilon}\,\lvert\omega(t)\rvert\,
 a_{s}^\dagger(t)\,a_{s}(t)\,\rmd t
\nn\\
 &= e^{-\ii\varepsilon}\,b_{s}^\dagger(\sigma)\,b_{s}(\sigma)\,\rmd \sigma
\end{align}
with $b_{s}(\sigma)\equiv a_{s}(t(\sigma))$\,, 
and the time evolution operator becomes
\begin{align}
 U(t_1,t)={\rm T}\exp\Bigl[-\ii e^{-\ii\varepsilon}
 \int_\sigma^{\sigma_1}\!\!\rmd\sigma\,b_s^\dagger(\sigma)\,b_s(\sigma)\Bigr]\,.
\end{align} 
We then introduce a small interval $\delta$ 
and divide the new interval into $N$ segments (see Fig.~\ref{fig:NewFIG3}), 
\begin{align}
 N=N(\delta,\sigma_1-\sigma)\equiv \frac{\sigma_1-\sigma}{\delta}\,.
\end{align}
By introducing $s_k\equiv \sigma + k\,\delta$ $(k=0,1,\ldots,N)$ 
with $s_0=\sigma$ and $s_N=\sigma_1$, 
the time evolution operator becomes 
\begin{align}
 U(t_1,t)
  =\prod_{k=1}^N\, 
  \exp\Bigl[-\ii \delta e^{-\ii\varepsilon}
  b_s^\dagger(s_k)\,b_s(s_{k-1})\Bigr]  \,.
\label{psi_evolve}
\end{align}
Substituting this to the amplitude $\bra{\psi_1,t_1}\,U(t_1,t)\,\ket{\psi}$ 
and inserting the identity $\displaystyle\sum_{n_k=0}^\infty \ket{n_k,s_k}\bra{n_k,s_k}=1$ 
at each time $s_k$\,, 
we obtain 
\begin{align}
 &\bra{\psi,t_1}\,U(t_1,t)\,\ket\psi 
\nn\\
 &= \sum_{\{n_k\} }  e^{- \varepsilon \,\delta\,\sum_k\,n_k}\,
  \langle \psi_1,t_1 \ket{n_N,s_N}\,\bra{n_0,s_0} \psi \rangle
\nn\\
 &\quad\times \prod_{k=1}^N \,
               \bra{n_{k},s_{k}} \,
  e^{-\ii \delta\,b_s^\dagger(s_k)\,b_s(s_k)}\,\ket{n_{k-1},s_{k-1}}\,.
\label{U-psi}
\end{align}
We thus find that the amplitude $\bra{\psi_1,t_1}\,U(t_1,t)\,\ket{\psi}$ 
is expressed as a sum over the paths, 
each path corresponding to an evolution of energy levels (not of ``positions'') 
and represented by a sequence $\{n_k\}$ $(k=0,1,\ldots,N)$\,. 
We see that each path receives a suppression factor 
$\exp\bigl[-\varepsilon\,\delta\sum_{k=1}^N n_k\bigr]
=\exp\bigl[-\varepsilon\times\mbox{(area)}\bigr]$\,, 
where $\mbox{(area)}$ is the area of the shaded region 
in Fig.~\ref{fig:NewFIG3}.

We now take the limit $t_1\to t_f$\,. 
If the foliation is effectively noncompact in the temporal direction 
(i.e., if $\sigma_1$ approaches $\sigma_f=\infty$), 
then $N(\delta,\sigma_1-\sigma)$ goes to infinity as $t_1\to t_f$ 
for the fixed small number $\delta$\,, 
and thus the suppression factor removes the contribution 
from any path having a nonvanishing tail for large $t$ 
and projects onto a set of paths 
satisfying the condition $n_k\to 0$ $(k\to\infty)$\,.
This proves the equality \eqref{amp_t1t}. 
Equality \eqref{amp_tt0} can also be proved in the same way.

We can easily show that 
both the Poincar\'e and the global patches are 
effectively noncompact in the temporal direction for any mode. 
As for the Poincar\'e patch, 
the frequency for the mode $\bk$ is given by 
$\lvert\omega_\bk(\eta)\rvert=\sqrt{m^2(-\eta)^{-2}+k^2}$\,, 
and thus it behaves as 
$\lvert\omega_\bk(\eta)\rvert\sim k$ $(\eta\sim\eta_i=-\infty)$ 
and 
$\lvert\omega_\bk(\eta)\rvert\sim m/(-\eta)$ $(\eta\sim\eta_f=0)$\,. 
We thus have 
\begin{align}
 -\sigma_i\sim \int_{\eta_i}\!\rmd \eta\,\lvert\omega_\bk(\eta)\rvert
  &= k\,(-\eta_i)+{\rm const}=+\infty\,,
\nn\\
 \sigma_f\sim \int^{\eta_f}\!\!\rmd \eta\,\lvert\omega_\bk(\eta)\rvert
  &= m\,\log\frac{1}{-\eta_f}+{\rm const}=+\infty\,.
\end{align}
This shows that the Poincar\'e patch is effectively noncompact 
for nonvanishing modes $\bk$\,.%
\footnote{
The zero mode $\bk=0$ does not satisfy the condition. 
However, since the mode belongs to a continuous spectrum, 
this does not give rise to a problem. 
} 
As for the global patch,  
the frequency for the mode $(L,M)$ is given by 
$\lvert\omega_L(t)\rvert=(1-t^2)^{-1}\,\sqrt{m^2+L(L+d-2)\,(1-t^2)}$\,, 
and thus it behaves as 
$\lvert\omega_L(t)\rvert\sim (m/2)\,(1+t)^{-1}$ $(t\sim t_i=-1)$ 
and 
$\lvert\omega_L(t)\rvert\sim (m/2)\,(1-t)^{-1}$ $(t\sim t_f=+1)$\,. 
We thus have 
\begin{align}
 -\sigma_i\sim \int_{t_i}\!\rmd t\,\lvert\omega_L(t)\rvert
  &= \frac{m}{2}\,\log\frac{1}{1+t_i}+{\rm const}=+\infty\,,
\nn\\
 \sigma_f\sim \int^{t_f}\!\!\rmd t\,\lvert\omega_L(t)\rvert
  &= \frac{m}{2}\,\log\frac{1}{1-t_f}+{\rm const}=+\infty\,.
\end{align}
This shows that the global patch is effectively noncompact 
for any mode $(L,M)$\,.

Once the equivalence is established, 
we can easily understand 
why the obtained in-out propagators 
are written with the de Sitter invariant $Z(x,x')$\,. 
In fact, since the patches we consider are preserved 
under infinitesimal actions of ${\rm SO}(1,d)$,  
and since a path integral (for a free scalar field) 
can be defined as respecting the symmetry 
under the infinitesimal actions of ${\rm SO}(1,d)$ 
(which is indeed the case only after we take the limit $\varepsilon\to 0$), 
the propagator obtained by such path integral 
(and thus the in-out propagator) 
must be invariant under the infinitesimal actions 
of ${\rm SO}(1,d)$\,. 
As was mentioned in the last paragraph 
of subsection \ref{sec:geometry}, 
this invariance is sufficient to ensure 
that the propagator can be written with the de Sitter invariant $Z(x,x')$\,.

\subsection{Numerical check}
\label{numerical}

In this subsection, 
we numerically demonstrate 
that the equivalence between the two propagators certainly holds, 
one obtained by a path integral with the $\ii\varepsilon$ prescription 
and another obtained as the in-out propagator  
using the instantaneous ground states.

We first rewrite the action \eqref{PI_mode2} 
with a new variable $\chi_n(t)=\abs{\rho(t)}^{1/2}\,\phi_n(t)
\equiv e^{\sigma(t)}\,\phi_n(t)$   
(we will omit the mode label $n$ for simplicity). 
Then, the action for each mode has the form
\begin{align}
 S_{\varepsilon}[\chi;\,t_1,t_0] &= \int_{t_0}^{t_1}\!\!\rmd t\,\frac{1}{2}\,\chi(t)\,
  \bigl[-e^{\ii\varepsilon}\,\partial_t^2-\Omega^2_\varepsilon(t)\bigr]\,\chi(t) \,,\\
  \Omega^2_\varepsilon(t) &\equiv e^{-\ii\varepsilon}\,\abs{\omega(t)}^2
                                 -e^{\ii\varepsilon}\,\bigl(\dot{\sigma}(t)\bigr)^2
                                 -e^{\ii\varepsilon}\,\ddot{\sigma}(t) \,,
\end{align}
and the propagator for each mode is given by
\begin{align}
 \langle \phi(t)\,\phi(t') \rangle 
 &= \abs{\rho(t)\,\rho(t')}^{-1/2}\,
   \langle \chi(t)\,\chi(t') \rangle \nn\\
 &= \abs{\rho(t)\,\rho(t')}^{-1/2}\,
   \bigbra{t}\, \frac{\ii}{-e^{\ii\varepsilon}\partial_t^2-\Omega_\varepsilon^2}\,\bigket{t'}\,.
\label{propagator_num}
\end{align}
We numerically evaluate the propagator \eqref{propagator_num} 
by dividing the interval $(t_i,t_f)$  into $N$ parts 
and by calculating the inverse of the matrix 
corresponding to $\ii^{-1}\,(-e^{\ii\varepsilon}\partial_t^2-\Omega^2_\varepsilon)$. 
We take a uniform spacing $a\equiv(t_1-t_0)/N$ for brevity  
and write the time variable as
$t=a\,r$ with $r$ an integer in the region $r_0 < r < r_1$ 
($r_0\equiv t_0/a$ and $r_1\equiv t_1/a$). 
We then introduce dimensionless variables $\chi_r$ as
\begin{align}
 \chi(t) = a^{1/2}\,\chi_r\,,
\end{align}
with which the action becomes 
\begin{align}
 S_{\varepsilon}[\chi] 
 &= \frac{1}{2}\,\sum_{r,s}S_{\varepsilon,rs}\,\chi_r\,\chi_s \,,
\\
 S_{\varepsilon,rs}
 &\equiv \bigl[2e^{\ii\varepsilon}-a^2\,\Omega_\varepsilon(ar)\bigr]\,\delta_{r,s}-e^{\ii\varepsilon}\delta_{r,\,s+1}
 -e^{\ii\varepsilon}\delta_{r,\,s-1} \,.
\end{align}
Since 
\begin{align}
 \langle \chi_r\chi_{r'}\rangle
 &=\frac{\int\rmd^N \!\chi\,\,\chi_r\chi_{r'}
 \exp\bigl(\frac{\ii}{2}\,\sum S_{\varepsilon,\,ss'}\,\chi_s\,\chi_{s'}\bigr)}
 {\int \rmd^N \!\chi\,\, 
 \exp\bigl(\frac{\ii}{2}\,\sum S_{\varepsilon,\,ss'}\,\chi_s\,\chi_{s'}\bigr)} \nn\\
 &=\ii (S_\varepsilon^{-1})_{rr'}\,,
\end{align}
the propagator is obtained as
\begin{align}
 \langle \phi(t)\,\phi(t') \rangle 
 = \ii a\,\abs{\rho(t)\,\rho(t')}^{-1/2}\,
  \bigl(S^{-1}_{\varepsilon}\bigr)_{r\,r'}
\label{propagator_discrete}\nn\\
 \bigl(t= a\,r\,,\ t'=a\,r'\big) \,.
\end{align}
We numerically calculate the inverse matrix \eqref{propagator_discrete}
for both of the Poincar\'{e} and global patches 
and compare the result with our in-out propagators 
obtained in section \ref{sec:deSitter}.

\begin{figure}[htbp]
\begin{center}
\includegraphics[width=5.8cm]{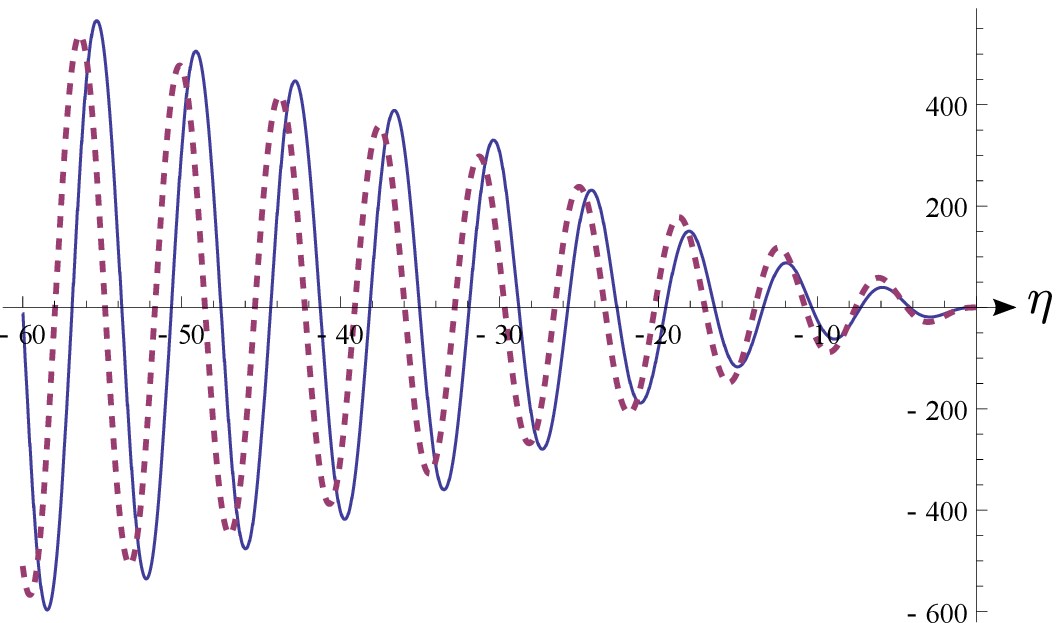}
\hspace{1cm}
\includegraphics[width=5.8cm]{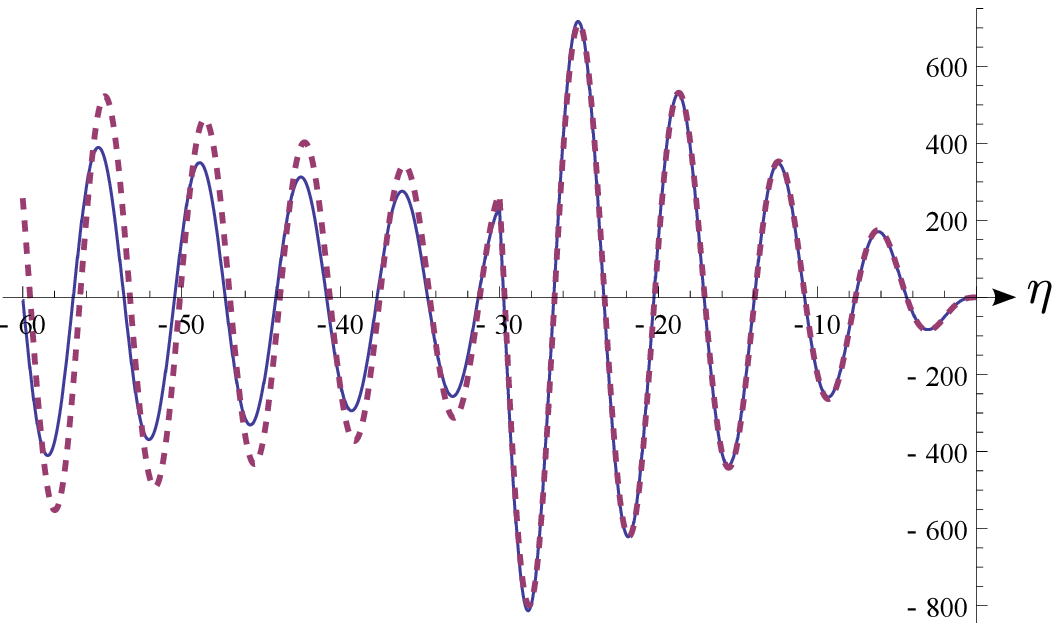}
\begin{quote}
\caption{Poincar\'e patch with light mass: 
the real part (upper) and the imaginary part (lower) 
of the in-out propagator $G^{\out/\inn}_{\bk}(\eta,\eta')$ (dashed curve) 
and the propagator $\langle \phi_\bk(\eta)\,\phi_\bk(\eta') \rangle$ (solid curve)
are shown for $d=4$, $m=0.5$, $k=1$, $a=0.02$, $\varepsilon = 0.01$, 
$\eta_0=-60$, $\eta_1=-0.01$, and $\eta'=-30.005$. 
Recall that $\eta_i=-\infty$ and $\eta_f=0$\,. 
\label{fig:poincare_ie-light}}
\end{quote}
\end{center}
\vspace{-6ex}
\end{figure}

\begin{figure}[htbp]
\begin{center}
\includegraphics[width=5.8cm]{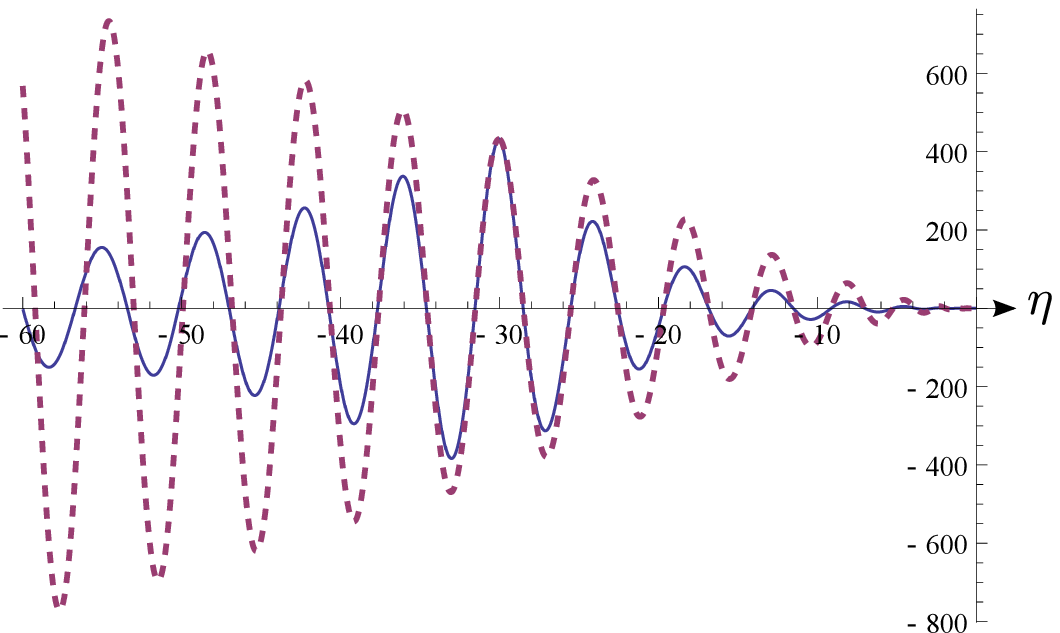}
\hspace{1cm}
\includegraphics[width=5.8cm]{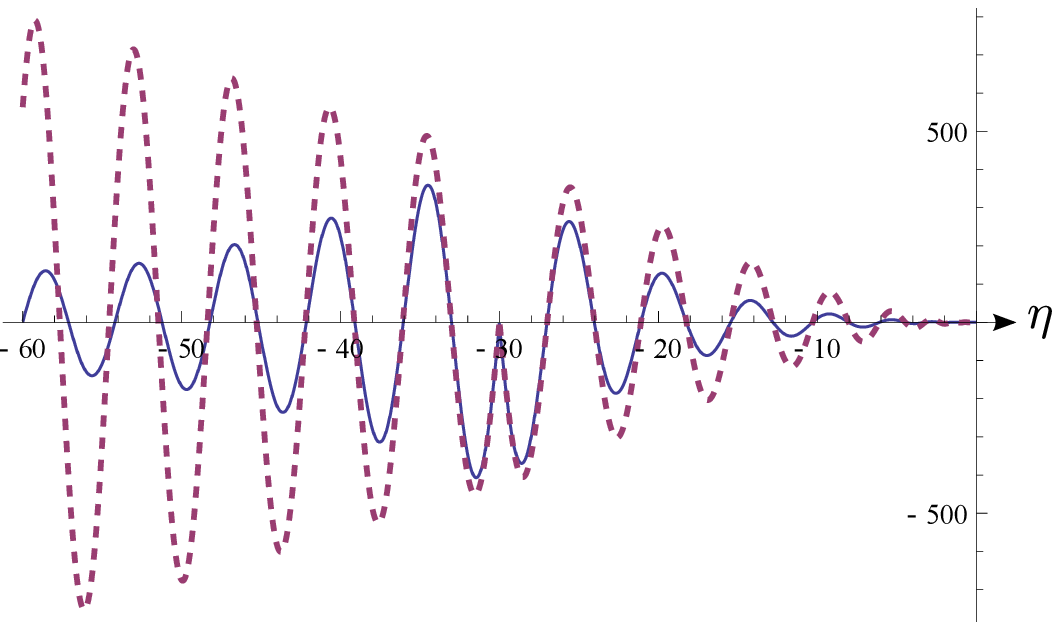}
\begin{quote}
\caption{Poincar\'e patch with heavy mass: 
the real part (upper) and the imaginary part (lower) 
of the in-out propagator $G^{\out/\inn}_{\bk}(\eta,\eta')$ (dashed curve) 
and the propagator $\langle \phi_\bk(\eta)\,\phi_\bk(\eta') \rangle$ (solid curve)
are shown for $d=4$, $m=9$, $k=1$, $a=0.02$, $\varepsilon = 0.07$, 
$\eta_0=-60$, $\eta_1=-0.01$, and $\eta'=-30.005$. 
Recall that $\eta_i=-\infty$ and $\eta_f=0$\,. 
\label{fig:poincare_ie-heavy}}
\end{quote}
\end{center}
\vspace{-6ex}
\end{figure}

\begin{figure}[htbp]
\vspace{3ex}
\begin{center}
\includegraphics[width=5.8cm]{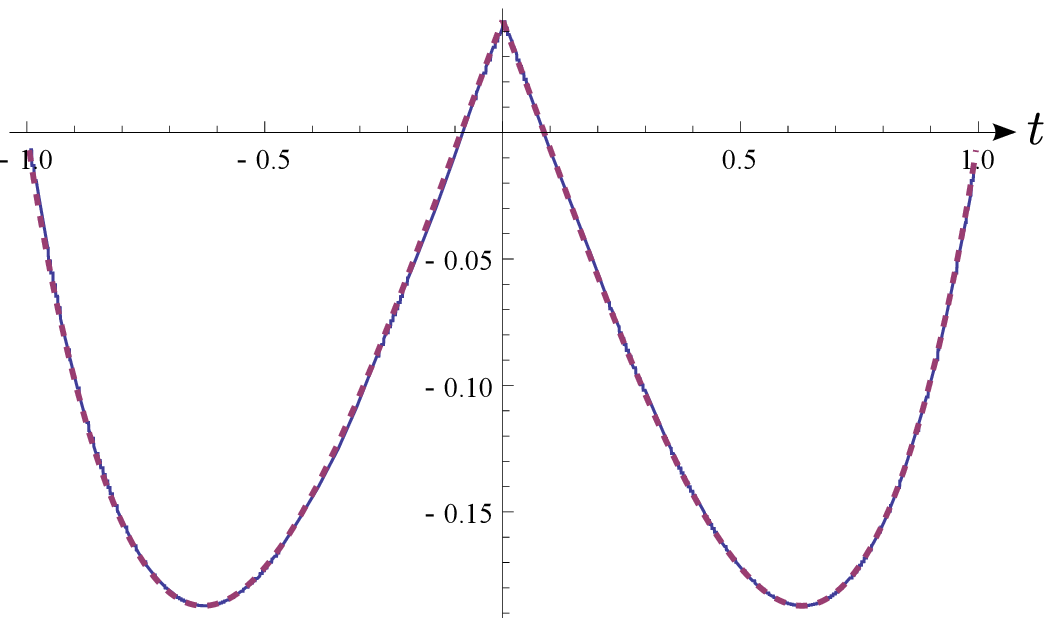}
\hspace{1cm}
\includegraphics[width=5.8cm]{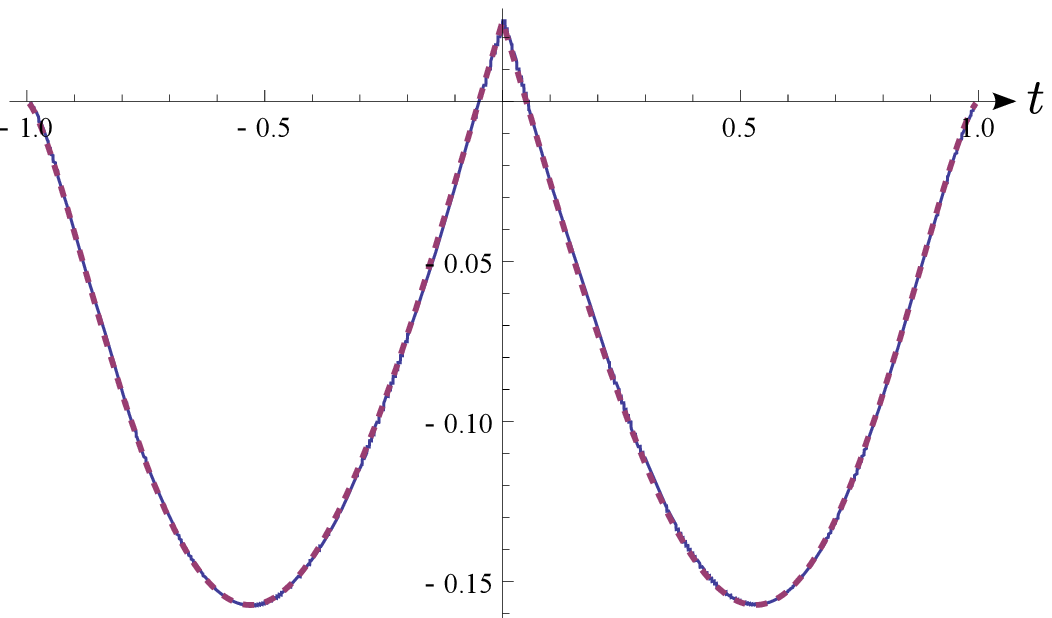}
\begin{quote}
\vspace{-5mm}
\caption{Global patch with light mass ($d$\,: odd/even): 
the imaginary part of the in-out propagator 
$G^{\out/\inn}_{L}(t,0)$ (dashed curve) 
and the propagator $\langle \phi_L(t)\,\phi_L(0) \rangle$ (solid curve)
are shown for $d=3$ (upper) [$d=4$ (lower)], $m=0.5$, $L=2$, $a=0.005$, $\varepsilon = 10^{-10}$, 
$t_0=-0.995$, and $t_1= 0.995$. 
Recall that $t_i=-1$ and $t_f=+1$\,. 
The real part is zero. 
\label{fig:global_ie-light}}
\end{quote}
\end{center}
\vspace{-6ex}
\end{figure}

\begin{figure}[htbp]
\vspace{3ex}
\begin{center}
\includegraphics[width=5.8cm]{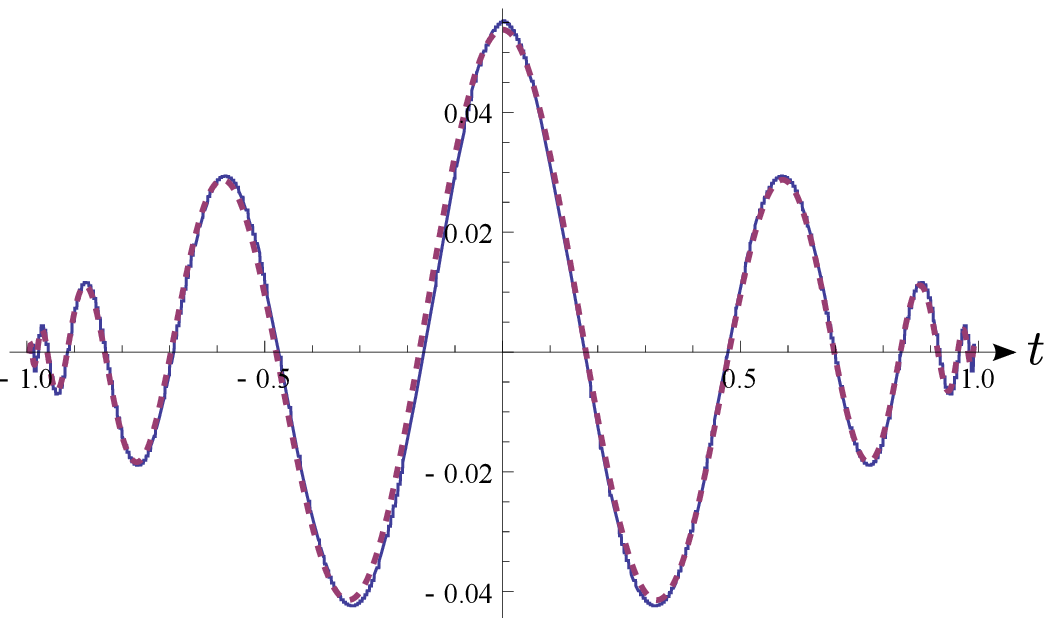}
\hspace{1cm}
\includegraphics[width=5.8cm]{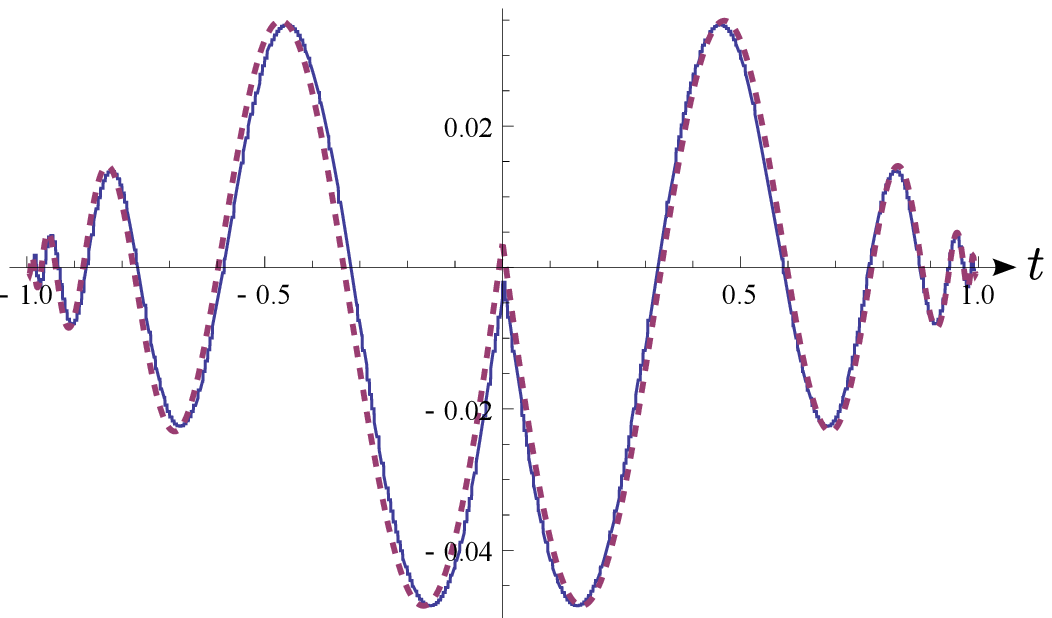}
\begin{quote}
\vspace{-5mm}
\caption{Global patch with heavy mass ($d$\,: odd): 
the real part (upper) and the imaginary part (lower) 
of the in-out propagator $G^{\out/\inn}_{L}(t,0)$ (dashed curve) 
and the propagator $\langle \phi_L(t)\,\phi_L(0) \rangle$ (solid curve)
are shown for $d=3$, $m=9$, $L=2$, $a=0.005$, $\varepsilon = 0.07$, 
$t_0=-0.995$, and $t_1= 0.995$. 
Recall that $t_i=-1$ and $t_f=+1$\,. 
\label{fig:global_ie-heavy-d3}}
\end{quote}
\end{center}
\vspace{-6ex}
\end{figure}

\begin{figure}[htbp]
\vspace{3ex}
\begin{center}
\includegraphics[width=5.8cm]{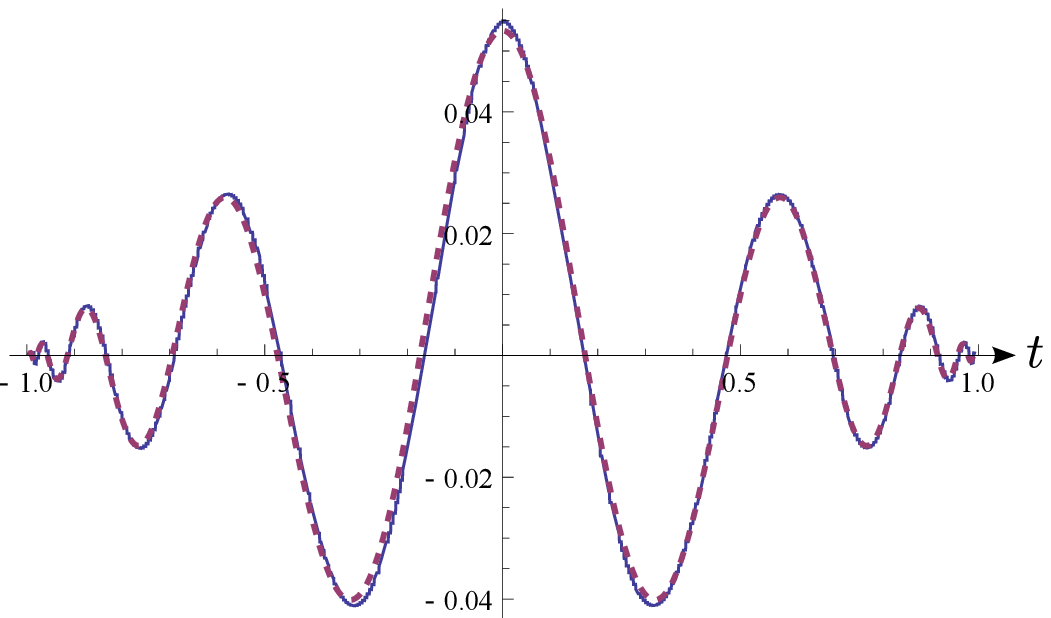}
\hspace{1cm}
\includegraphics[width=5.8cm]{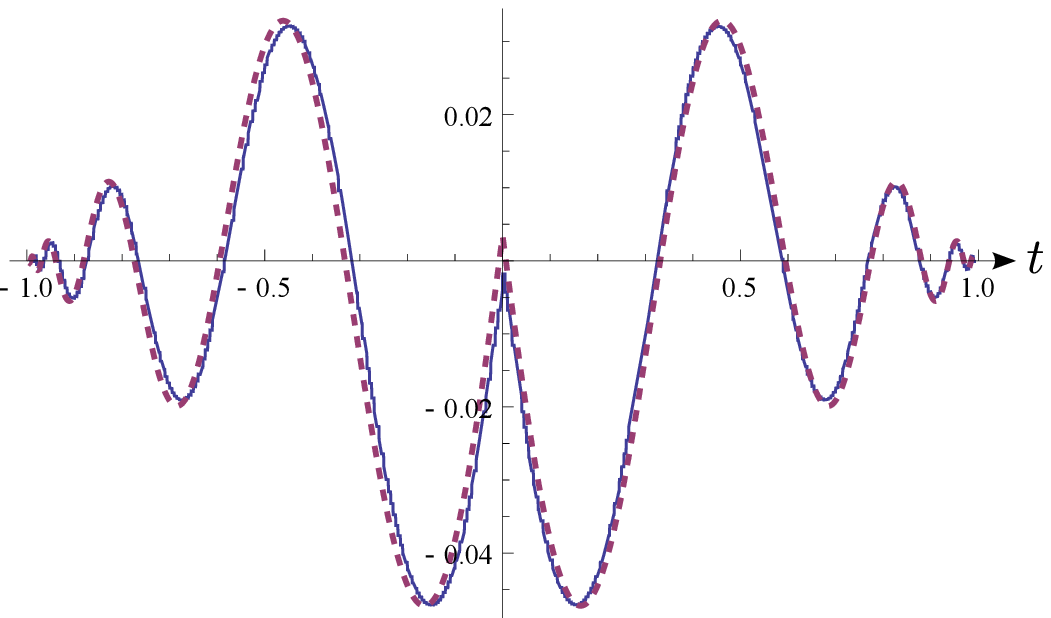}
\begin{quote}
\vspace{-5mm}
\caption{Global patch with heavy mass ($d$\,: even): 
the real part (upper) and the imaginary part (lower) 
of the in-out propagator $G^{\out/\inn}_{L}(t,0)$ (dashed curve) 
and the propagator $\langle \phi_L(t)\,\phi_L(0) \rangle$ (solid curve)
are shown for $d=4$, $m=9$, $L=1$, $a=0.005$, $\varepsilon = 0.07$, 
$t_0=-0.995$, and $t_1= 0.995$. 
Recall that $t_i=-1$ and $t_f=+1$\,. 
\label{fig:global_ie-heavy-d4}}
\end{quote}
\end{center}
\vspace{-6ex}
\end{figure}

The results in the Poincar\'e case for $\{d=4,\,m=0.5\}$ and $\{d=4,\,m=9\}$ 
are depicted in Figs.~\ref{fig:poincare_ie-light} and \ref{fig:poincare_ie-heavy}, 
while those in the global patch for $\{d=3,\,4,\,m=0.5\}$, 
$\{d=3,\,m=9\}$, and $\{d=4,\,m=9\}$ 
are in Figs.~\ref{fig:global_ie-light}, \ref{fig:global_ie-heavy-d3}, 
and \ref{fig:global_ie-heavy-d4}. 
We find that there is a perfect agreement for the global patch, 
while there exists a small discrepancy for the Poincar\'{e} patch. 
We observe that the discrepancy gets reduced as one takes a finer mesh near $\eta=0$ 
and a larger value for $\abs{\eta_0}$\,,
and expect that it will disappear eventually.  
We thus are almost convinced that 
the equivalence between the two propagators 
is confirmed numerically. 

\section{Heat kernel representation and the composition principle}
\label{sec:heat_kernel}

\subsection{General theory}
\label{sec:heat_general}

We consider the random walk of a relativistic particle 
moving in a Lorentzian manifold with the metric
\begin{align}
 \rmd s^2 = g_{\mu\nu}(x)\,\rmd x^\mu\,\rmd x^\nu\,.
\end{align}
Its trajectory is uniquely specified by the functions $X^\mu(\lambda)$ 
$(0\leq \lambda \leq 1)$\,, 
up to reparametrizations $\lambda\to f(\lambda)$  
such that $\rmd f(\lambda)/\rmd \lambda>0$, $f(0)=0$, $f(1)=1$\,.  
The amplitude connecting two points $x$ and $x'$ 
is then given by the Feynman path integral:
\begin{align}
 A(x,x') = \int_{X(0)=x'}^{X(1)=x}\!\!
  \frac{[\rmd X^\mu(\lambda)]}{{\rm Vol}({\rm Diff}_1)}
  \,e^{\,\ii I_0[X(\lambda)]}\,,
\end{align}
where  ${\rm Vol}({\rm Diff}_1)$ is the gauge volume 
of one-dimensional diffeomorphisms 
\begin{align}
 X^\mu(\lambda) \to \tilde{X}^\mu(\lambda)=X^\mu(f(\lambda))\,,
\end{align}
and we propose to set the action $I_0[X]$ for the random walk 
in a Lorentzian manifold as 
\begin{align}
 &I_0\bigl[ X(\lambda)\bigr] \nn\\
 &\equiv {}- (m-\ii\varepsilon)\,\int_0^1\!\!\rmd\lambda\,
  \sqrt{-g_{\mu\nu}\bigl(X(\lambda)\bigr)\,
  \dot{X}^\mu(\lambda)\,\dot{X}^\nu(\lambda)-\ii \varepsilon'}\,.
\label{I_NG}
\end{align} 
Note the presence of two infinitesimal imaginary parts, 
$\ii \varepsilon$ and $\ii \varepsilon'$\,, in $I_0[X(\lambda)]$ 
($\varepsilon,\,\varepsilon'>0$). 
The first $(\ii\varepsilon)$ is the standard one, 
which manifestly suppresses the contribution from such paths  
that are prolonged in the timelike direction. 
We further have introduced the second one $(\ii\varepsilon')$  
in order to define the path integral for any shape of path 
in a Lorentzian manifold. 
In fact, for a timelike segment ($\dot{X}^2<0$)  
we can neglect $\varepsilon'$ 
and the action becomes the standard action for a timelike path, 
while for a spacelike segment ($\dot{X}^2>0$) 
we can rewrite the square root as
\begin{align}
 \sqrt{-\dot{X}^2-\ii\varepsilon'}
  =\sqrt{e^{-\ii(\pi-0)}\dot{X}^2}
  ={}- \ii\sqrt{\dot{X}^2}
\end{align}
and the action gives the path-integral weight 
which suppresses the contribution from such paths 
that are stretched largely in the spacelike direction.

The action \eqref{I_NG} of Nambu-Goto type is equivalent to the following 
action of Polyakov type:
\begin{align}
 I[X(\lambda),e(\lambda)]
  =\int_0^1\!\!\rmd\lambda\,
  \Bigl[\,\frac{\dot{X}^2+\ii\varepsilon'}{2\,e}
  - \frac{m^2-\ii\varepsilon}{2}\,e\,\Bigr]\,,
\end{align}
where $e(\lambda)>0$ is the einbein defined on the one-dimensional manifold. 
We can see from this expression that $\ii\varepsilon$ and $\ii\varepsilon'$ 
give imaginary parts of the same sign. 
The new action $I[X(\lambda),e(\lambda)]$ also has the invariance 
under the one-dimensional diffeomorphisms 
\begin{align}
 X^\mu(\lambda) &\to \tilde{X}^\mu(\lambda)=X^\mu(f(\lambda))\,,\\
  e(\lambda) &\to \tilde{e}(\lambda)=\frac{\rmd f(\lambda)}{\rmd\lambda}\,
  e(f(\lambda))\,,
\end{align}
and we can take a gauge fixing where $e(\lambda)={\rm constant}\equiv T$\,. 
However, as is discussed in detail in \cite{Polyakov_gauge_strings}, 
such constant $T=\int_0^1\!\rmd\lambda\,e(\lambda)$ 
is actually ${\rm Diff}_1$-invariant and needs to be further integrated, 
so that we obtain%
\footnote{
There may arise a divergence when calculating the Jacobian 
to obtain the second line, 
but such divergence should be ultra-local in quantum mechanics 
(i.e., one-dimensional field theory with a coordinate $\lambda$) 
and can be simply dealt with by an additive renormalization of mass $m$, 
as in the Euclidean space considered in \cite{Polyakov_gauge_strings}.
} 
\begin{align}
 A(x,x')&= \int_{X(0)=x'}^{X(1)=x}
  \frac{[\rmd X^\mu(\lambda)\,\rmd e(\lambda)]}{{\rm Vol}({\rm Diff}_1)}
  \,e^{\,\ii I_0[X(\lambda)]}
\nn\\
 &= \int_0^\infty\!\!\rmd T\,\int_{X(0)=x'}^{X(1)=x}\![\rmd X(\lambda)]\nn\\
 &\quad\times\exp\Bigl[\,\ii\int_0^1\!\!\rmd\lambda\Bigl(
  \frac{\dot{X}^2+\ii\varepsilon'}{2\,T}-\frac{m^2-\ii\varepsilon}{2}\,T
  \Bigr)\Bigr]
\nn\\
 &= \int_0^\infty\!\!\rmd T\,e^{-\varepsilon'/(2T)}\,
  \int_{X(0)=x'}^{X(T)=x}\![\rmd X(t)]\nn\\
 &\quad\times\exp\Bigl[\,\ii\int_0^T\!\!\rmd t\Bigl(
  \frac{\dot{X}^2(t)+\ii\varepsilon'}{2}-\frac{m^2-\ii\varepsilon}{2}
  \Bigr)\Bigr]\,,
\label{A_ampl}
\end{align}
where in the last line we have rewritten the expression 
with $t\equiv T\,\lambda$\,. 
The path integral is nothing but that for the quantum mechanical amplitude 
from the state $\ket{x'}$ to the state $\ket{x}$ 
with the Hamiltonian 
\begin{align}
 H&=\frac{1}{2}\bigl(-\Box + m^2 - \ii\varepsilon \bigr) \nn\\
  &\equiv \frac{1}{2}\Bigl[{}- \frac{1}{\sqrt{-g}}\,\partial_\mu\,
  \bigl(\sqrt{-g}\,g^{\mu\nu}\,
  \partial_\nu) + m^2 -\ii\varepsilon\Bigr]\,, 
\end{align}
and thus we obtain the expression 
\begin{align}
 A(x,x')&= \int_0^\infty\!\!\rmd T\,e^{{}- \varepsilon'/(2T)}\,
  \bra{x} e^{{}- \ii (T/2)\,(-\Box+m^2-\ii\varepsilon)} \ket{x'}
\nn\\
  &= \int_0^\infty\!\!\rmd T\,e^{{}- \ii (T/2)\,(m^2-\ii\varepsilon)
   -\varepsilon'/(2T)}\,K(x,x';\,T)\,.
\end{align}
Here, $K(x,x';\,T)$ is the heat kernel of the d'Alembertian $\Box$, 
\begin{align}
 K(x,x';\,T)\equiv \bra{x} e^{\ii (T/2)\,\Box} \ket{x'}\,,
\end{align}
which satisfies the following equations:
\begin{align}
 \ii\,\frac{\partial}{\partial T}\,K(x,x';\,T) 
  &= -\frac{1}{2}\,\Box_x\,K(x,x';\,T)\,, \label{heat_equation}\\
 K(x,x';\,T=0) &= \delta^d(x,x')\equiv \frac{1}{\sqrt{-g}}\,\delta^d(x-x')
 \label{heat_initial}\,.
\end{align}
Note that we need to multiply \eqref{A_ampl} by $1/2$ 
to obtain the propagator $G(x,x')$ of a neutral particle 
(i.e., particle\,$=$\,anti-particle): 
\begin{align}
 &G(x,x') =\frac{1}{2}\,A(x,x')\nn\\
 &=\frac{1}{2}\,\int_0^\infty\!\!\rmd T\,e^{{}- \ii (T/2)\,(m^2-\ii\varepsilon)
   -\varepsilon'/(2T)}\,K(x,x';\,T)\,.
\label{propagator_heat_kernel}
\end{align}

Since the propagator $G(x,x')$ can formally  be written as 
$G(x,x')=\ii\,\bra{x}\,(\Box-m^2+\ii\varepsilon)^{-1}\,\ket{x'}$\,, 
one can easily show that $G(x,x')$ satisfies the following 
{\em composition law} \cite{Polyakov:2007mm}:
\begin{align}
 \frac{\partial}{\partial m^2}\,G(x,x')
  =\ii\,\int\!\!\sqrt{-g(y)}\, \rmd^d y\, G(x,y)\,G(y,x')\,,
\label{composition-law}
\end{align}
which is consistent with the asymptotic form of $G(x,x')$ 
for large timelike separation with large mass:
\begin{align}
 G(x,x') \sim e^{{}- \ii m\,L(x,x')}\,,
\end{align}
where $L(x,x')$ is the timelike geodesic distance 
between $x$ and $x'$\,. 
The relation \eqref{composition-law} has been proposed by Polyakov 
as a principle to be satisfied by quantum field theory 
in curved spacetime 
in order for the propagator to be interpreted 
as representing a sum over paths of a relativistic particle 
in the spacetime. 
If the spacetime has a global timelike Killing vector 
(as does Minkowski space), 
one can define a common vacuum of scalar field 
from the past through the future, 
and the relativistic particle corresponds to a one-particle state. 
Note that such interpretation is not always possible 
when spacetime has no global timelike Killing vector \cite{BD} 
(see also \cite{Hollands:2008vx} for a recent discussion). 

As a simple example, we consider a neutral particle 
propagating in a $d$-dimensional Minkowski space with the metric 
\begin{align}
 \rmd s^2 = {}- \rmd t^2 + \rmd \bx^2\,.
\end{align}
Then the heat kernel can be calculated with the momentum representation as
\begin{align}
 K(x,x';\,T)&= \int\frac{\rmd^d p}{(2\pi)^d}\,
  e^{\ii p(x-x')}\,e^{{}- \ii (T/2)\, p^2}
\nn\\
 &=\ii\,\Bigl(\frac{1}{2\pi\ii T}\Bigr)^{d/2}\,
  \exp\Bigl[\ii \frac{(x-x')^2}{2T}\Bigr]\,.
\end{align}
Here, the first $\ii$ reflects the fact 
that the Gaussian integral over $p_0$ has the opposite sign of quadratic term 
to that for the other variables $p_i$ $(i=1,\dotsc,d-1)$. 
Substituting this to \eqref{propagator_heat_kernel}, 
we obtain the following expression: 
\begin{align}
 G(x,x')&=\frac{\ii^{-(d-2)/2}}{(4\pi)^{d/2}}\,\int_0^\infty\!\!\rmd s\,s^{-\frac{d}{2}}\,
  e^{-\frac{z^2}{4\,s}-a\,s} 
\end{align}
with $z^2=\varepsilon'-\ii(x-x')^2$ and $a=\ii(m^2-\ii\varepsilon)$\,. 
This integration can be easily performed, 
and we obtain 
\begin{align}
 &G(x,x') = \frac{\ii^{-(d-2)/2}}{(4\pi)^{d/2}}\,a^{\frac{d-2}{2}}\,2^{\frac{d}{2}}\,
  \bigl(\sqrt{a}\,z\bigr)^{-\frac{d-2}{2}}\,K_{\frac{d-2}{2}}\bigl(\sqrt{a}\,z\bigr)
\nn\\
 &=\frac{m^{d-2}}{(2\pi)^{d/2}}\,
  \bigl(m\sqrt{\sigma+\ii\varepsilon'}\bigr)^{-\frac{d-2}{2}}\,
  K_{\frac{d-2}{2}}\bigl(m\sqrt{\sigma+\ii\varepsilon'}\bigr)\,,
\end{align}
where $\sigma\equiv (x-x')^2$\,. 
This certainly agrees with the propagator \eqref{Minkowski_prop} of a real scalar field.

\subsection{de Sitter case}
\label{heat_dS}

In this subsection, we check that 
the in-out propagators \eqref{Poincare_in_out_full}, 
\eqref{globalodd}, and \eqref{globaleven} 
indeed satisfies the composition law 
by giving their heat kernel representations.%
\footnote{ 
See \cite{Akhmedov:2009ta} (also \cite{Polyakov:2007mm}) 
for the direct evaluation of the random walk 
in de Sitter space, 
which is based on the heat kernel for Euclidean AdS space 
obtained in \cite{Grosche:1987de}.  
} 

\subsubsection{Poincar\'{e} patch}

We start from the following integral representation 
of the associated Legendre functions  
(a proof is given in Appendix \ref{appendix:Legendre_integral}):
\begin{align}
 &\LQ_{\nu-1/2}^{\frac{d-2}{2}}(u)\nn\\
 &= e^{\ii\pi\frac{d-2}{2}} \,
 \int_0^\infty\rmd\lambda\, 
 \frac{\Gamma\bigl(\frac{d-1}{2}+\ii\lambda\bigr)\,
       \Gamma\bigl(\frac{d-1}{2}-\ii\lambda\bigr)}
      {\Gamma(\ii\lambda)\,\Gamma(-\ii\lambda)}\,
 \frac{\LP_{\ii\lambda-1/2}^{-\frac{d-2}{2}}(u)}{\nu^2+\lambda^2} \nn\\
 &\qquad\qquad\qquad\qquad\qquad\qquad\bigl[d\in\mathbb{Z}\,,\quad \Ree\nu>0\bigr]\,.
\label{Legendre_integral_rep}
\end{align}
Since $\nu_\varepsilon^2$ has a positive imaginary part, 
we have
\begin{align}
 &e^{-\ii\pi\frac{d-2}{2}}\,\bigl(u^2-1\bigr)^{-\frac{d-2}{4}}\,
  \LQ_{\nu_\varepsilon-1/2}^{\frac{d-2}{2}}(u)\nn\\
 &= \frac{1}{2\ii}\int_0^\infty \rmd T\, 
 \int_0^\infty\rmd\lambda\, 
 \frac{\Gamma\bigl(\frac{d-1}{2}+\ii\lambda\bigr)\,
       \Gamma\bigl(\frac{d-1}{2}-\ii\lambda\bigr)}
      {\Gamma(\ii\lambda)\,\Gamma(-\ii\lambda)}\nn\\
 &\qquad\times\bigl(u^2-1\bigr)^{-\frac{d-2}{4}}\,
 \LP_{\ii\lambda-1/2}^{-\frac{d-2}{2}}(u)\,
 e^{\ii\frac{T}{2}\,(\nu_\varepsilon^2+\lambda^2)} \nn\\
 &= \frac{\ii}{2^{2-\frac{d}{2}}}\int_0^\infty \rmd T\, 
 \int_0^\infty\rmd\lambda\, 
  \frac{\lambda\,\Gamma\left(\frac{d-1}{2}\right) \sinh(\pi\lambda)}
       {\sqrt{\pi}\,\cos\bigl[\pi\bigl(\frac{d}{2}-\ii\lambda\bigr)\bigr]}\nn\\
 &\qquad \times C_{\ii\lambda-\frac{d-1}{2}}^{\frac{d-1}{2}}(u)\,
 e^{\ii\frac{T}{2}\,(\nu_\varepsilon^2+\lambda^2)} \,.
\label{eQ_sinhC}
\end{align}
Here, $u=Z-\ii 0$, 
and to obtain the second line we have used the identity
\begin{align}
 \LP_{\ii\lambda-\frac{1}{2}}^{-\frac{d-2}{2}}(u) 
 &=  2^{-\frac{d-2}{2}}\,\bigl(u^2-1\bigr)^{\frac{d-2}{4}}\nn\\
 &\times\frac{\Gamma(d-1)\,\Gamma\bigl(\ii\lambda-\frac{d-3}{2}\bigr)}
      {\Gamma(d/2)\,\Gamma\bigl(\ii\lambda+\frac{d-1}{2}\bigr)}\,
 C_{\ii\lambda-\frac{d-1}{2}}^{\frac{d-1}{2}}(u) \,.
\end{align}
We thus find that the in-out propagator \eqref{Poincare_in_out_full} 
in the Poincar\'{e} patch has the heat kernel representation of the form
\begin{align}
 &G(x,x') = \frac{1}{2}\int_0^\infty\rmd T\,
 e^{-\ii \frac{m^2-\ii\varepsilon}{2}\,T}\, K(x,x';\,T)\,,\\
 &K(x,x';\,T) \nn\\
 &= -\frac{e^{-\ii\pi\frac{d-3}{2}}\,\bigl(u^2-1\bigr)^{-\frac{d-2}{4}}}{(2\pi)^{d/2}} \nn\\
 &\quad \times \int_0^\infty \rmd \lambda\,
 \frac{\Gamma\bigl(\frac{d-1}{2}+\ii\lambda\bigr)\,\Gamma\bigl(\frac{d-1}{2}-\ii\lambda\bigr)}
      {\Gamma(\ii\lambda)\,\Gamma(-\ii\lambda)}\nn\\
 &\qquad\times\LP_{\ii\lambda-1/2}^{-\frac{d-2}{2}}(u)\, 
         e^{\ii\frac{T}{2}\,\bigl(\lambda^2+(\frac{d-1}{2})^2\bigr)} \nn\\
 &= \frac{e^{-\ii\pi\frac{d-3}{2}}\,\Gamma\bigl(\frac{d-1}{2}\bigr)}{2\pi^{\frac{d+1}{2}}}\,
    \int_0^\infty \rmd \lambda\,
 \frac{\lambda\,\sinh(\pi\lambda)}{\cos\bigl[\pi\bigl(\frac{d}{2}-\ii\lambda\bigr)\bigr]}\nn\\
 &\qquad\qquad\qquad\times C_{\ii\lambda-\frac{d-1}{2}}^{\frac{d-1}{2}}(u)\,
 e^{\ii\frac{T}{2}\,\bigl(\lambda^2+(\frac{d-1}{2})^2\bigr)} \,.
\end{align}

The above heat kernel certainly satisfies Eqs.~\eqref{heat_equation} and \eqref{heat_initial}.
In fact, Eq.~\eqref{heat_equation} can be shown in the following way:
\begin{align}
 &\ii\,\frac{\partial}{\partial T}\,K(x,x';\,T)\nn\\
 &=-\frac{e^{-\ii\pi\frac{d-3}{2}}\,\Gamma\bigl(\frac{d-1}{2}\bigr)}
         {4\pi^{\frac{d+1}{2}}}\,
    \int_0^\infty \rmd \lambda\,
 \frac{\lambda\,\sinh(\pi\lambda)\,\bigl(\lambda^2+\bigl(\frac{d-1}{2}\bigr)^2\bigr)}
      {\cos\bigl[\pi\bigl(\frac{d}{2}-\ii\lambda\bigr)\bigr]}\nn\\
 &\qquad\qquad\qquad\qquad\qquad\times C_{\ii\lambda-\frac{d-1}{2}}^{\frac{d-1}{2}}(u)\,
 e^{\ii\frac{T}{2}\,\bigl(\lambda^2+(\frac{d-1}{2})^2\bigr)} \nn\\
 &= -\frac{1}{2}\,\bigl[-\bigl(Z^2-1\bigr)\, \partial_Z^2 K(x,x';\,T)
                        -d\,Z\,\partial_Z K(x,x';\,T)\bigr] \nn\\
 &=-\frac{1}{2}\,\Box_x K(x,x';\,T) \,,
\end{align}
where we have used the Gegenbauer differential equation
\begin{align}
 &\bigl(1-u^2\bigr)\, \partial_u^2 
  C_{\ii\lambda-\frac{d-1}{2}}^{\frac{d-1}{2}}(u)
 -d\,u\,\partial_u C_{\ii\lambda-\frac{d-1}{2}}^{\frac{d-1}{2}}(u)\nn\\
 &-\Bigl[\lambda^2+\Bigl(\frac{d-1}{2}\Bigr)^2\Bigr]\, 
  C_{\ii\lambda-\frac{d-1}{2}}^{\frac{d-1}{2}}(u) =0 \,,
\end{align}
and the fact that the Klein-Gordon operator for functions 
of the de Sitter invariant $f(Z)$ can be written as
\begin{align}
 \Box f(Z) = \bigl(1-Z^2\bigr)\, \partial_Z^2 f(Z) -d\,Z\,\partial_Z f(Z)\,.
\end{align}
The initial condition \eqref{heat_initial} can be shown to hold 
by using the heat kernel equation and the equality
\begin{align}
 (\Box_x-m^2+\ii\varepsilon)G^{\out/\inn}(x,x') =\frac{\ii}{\sqrt{-g}}\,\delta^d(x-x')
\end{align}
as follows:
\begin{align}
 &\frac{\ii}{\sqrt{-g}}\,\delta^d(x-x') \nn\\
 &=\frac{1}{2}\int_0^\infty\rmd T\,e^{-\ii \frac{m^2-\ii\varepsilon}{2}\,T}\,
 (\Box_x-m^2+\ii\varepsilon)K(x,x';\,T)\nn\\
 &=-\ii\int_0^\infty\rmd T\,\frac{\partial}{\partial T}
    \bigl[e^{-\ii \frac{m^2-\ii\varepsilon}{2}\,T}\,K(x,x';\,T)\Bigr]\nn\\
 &=\ii K(x,x';\,0)\,.
\end{align}

\subsubsection{Global patch}

In a similar way, 
using Eq.~\eqref{eQ_sinhC}, 
we can show that the in-out propagators \eqref{globalodd} and \eqref{globaleven} 
have the heat kernel representation of the form
\begin{align}
 G^{\out/\inn}_{\{\genfrac{}{}{0pt}{}{\odd}{\even}\}}(x,x') 
   = \frac{1}{2}\int_0^\infty\rmd T\,e^{-\ii \frac{m^2-\ii\varepsilon}{2}\,T}\,
      K^{\out/\inn}_{\{\genfrac{}{}{0pt}{}{\odd}{\even}\}}(x,x';\,T)
\end{align}
with
\begin{align}
 &K^{\out/\inn}_{\odd}(x,x';\,T) \nn\\
  &= \frac{(-1)^{\frac{d+1}{2}}\,\Gamma\bigl(\frac{d-1}{2}\bigr)}
          {4\pi^{\frac{d+1}{2}}\,\sin(\pi\nu)}\,
     \int_0^\infty \rmd \lambda\, \lambda \nn\\
 &\quad\times 
 \bigl[C_{\ii\lambda-\frac{d-1}{2}}^{\frac{d-1}{2}}(u_+)
      -C_{\ii\lambda-\frac{d-1}{2}}^{\frac{d-1}{2}}(u_-)\bigr]\,
 e^{\ii\frac{T}{2}\,\bigl(\lambda^2+(\frac{d-1}{2})^2\bigr)} \,,\\
 &K^{\out/\inn}_{\even}(x,x';\,T) \nn\\
  &= \frac{(-1)^{\frac{d+2}{2}}\,\Gamma\bigl(\frac{d-1}{2}\bigr)}
          {4\pi^{\frac{d+1}{2}}\,\cos(\pi\nu)}\,
    \int_0^\infty \rmd \lambda\, \lambda\,\tanh(\pi\lambda) \nn\\
 &\quad\times 
 \bigl[C_{\ii\lambda-\frac{d-1}{2}}^{\frac{d-1}{2}}(u_+)
       +C_{\ii\lambda-\frac{d-1}{2}}^{\frac{d-1}{2}}(u_-)\bigr] \,
    e^{\ii\frac{T}{2}\,\bigl(\lambda^2+(\frac{d-1}{2})^2\bigr)}\,. 
\end{align}
Equations \eqref{heat_equation} and \eqref{heat_initial} 
also hold for these heat kernels, 
as can be shown in the same way as above. 

\subsection{Relation to the Green function in Euclidean AdS space}
\label{sec:EAdS}

As has been pointed out in \cite{Akhmedov:2009ta}, 
the in-out propagator in the Poincar\'e patch 
is directly related to the Green function in Euclidean AdS space 
through an analytic continuation. 
In this subsection, we demonstrate this equivalence 
with precise numerical constants.

$d$-dimensional Euclidean AdS space (EAdS$_{d}$)
is defined as the hypersurface in a $(d+1)$-dimensional Minkowski space
with the relation 
\begin{align}
 \eta_{MN}\, Y^M\, Y^N = {}- \ell^{\prime 2} \quad 
 (M,N=0,\dotsc,d) \,,
\end{align}
where $\ell^{\prime}$ is called the AdS radius. 
EAdS$_{d}$ has two connected components. 
A frequently used coordinate system 
which covers only a single connected component is 
the Poincar\'e coordinates $(z,y^i)$ $(i=1,\dotsc,d-1)$ 
that are defined by the following embedding: 
\begin{align}
 Y^0 &= \frac{\ell^{\prime 2} +z^2+{\abs{\by}}^2}{2 z}\,,\quad
 Y^i = \ell^{\prime}\,\frac{y^i}{z}\,,\nn\\
 Y^d &= \frac{\ell^{\prime 2} -z^2-{\abs{\by}}^2}{2 z}\,.
\label{EAdSPoemb}
\end{align}
We here have chosen the component with $z>0$. 
The metric then takes the form
\begin{align}
\rmd s^2 = \ell^{\prime 2}\,\frac{\rmd z^2 + \rmd \by\cdot \rmd \by}{z^2}\,. 
\end{align}

The Green function in EAdS space is known to have the following form 
(see, e.g., \cite{Klebanov:1999tb}):%
\footnote{
In fact, solving the Klein-Gordon equation with a delta function source 
using  the EAdS invariant $Z'=Z'(y,y')$\,, 
we see that the Green function is a linear combination of 
$(Z'^2-1)^{-(d-2)/4} \LP_{\nu'-1/2}^{(d-2)/2}(Z')$
and $(Z'^2-1)^{-(d-2)/4} \LQ_{\nu'-1/2}^{(d-2)/2}(Z')$\,. 
By requiring that the Green function damps at large separation (cluster property), 
only the latter solution is selected, as can be seen from the asymptotic forms 
of the associated Legendre functions [see \eqref{LP_as} and \eqref{LQ_as}]. 
The normalization is then determined by requiring  
that the Green function coincides with that in Euclidean space 
for infinitesimal separation of $y$ and $y'$ [or $Z'(y,y')\to 1$]\,. 
} 
\begin{align}
 &G_{\text{EAdS}}(y,y')\nn\\
 &=\frac{e^{-\ii\pi\,(d-2)/2}}{(2\pi)^{d/2}\,\ell^{\prime\, d-2}}\,
  (Z^{\prime 2}(y,y')-1)^{-\frac{d-2}{4}}\,
     \LQ_{\nu'-1/2}^{(d-2)/2}(Z'(y,y'))\,, 
\end{align}
where $Z'(y,y')$ is the invariant of EAdS space,  
\begin{align}
 Z'(y,y') &\equiv {}- \ell^{\prime -2}\,\eta_{MN}\, Y^M(y)\, Y^N(y')\nn\\
  &=1+\,\frac{(z-z')^2+\abs{\by-\by'}^2}{2\,z\,z'}\,,  
\end{align}
and
\begin{align}
 \nu' \equiv \sqrt{\Bigl(\frac{d-1}{2}\Bigr)^2+m^2\,\ell^{\prime 2}}\,.
\end{align}
Note that $Z'$ is always larger than unity.

The coordinate system $(z,y^i)$ is related to the Poincar\'e coordinates 
$(\eta,x^i)$ of $d$-dimensional de Sitter space 
through the analytic continuation
\begin{align}
 z=e^{\ii\frac{\pi-0}{2}}(-\eta) \,,\quad y^i=x^i\,,\quad \ell^{\prime}
  =e^{\ii\frac{\pi-0}{2}} \ell\,,
\label{EAdS_trsf}
\end{align}
or equivalently, 
\begin{align}
 Y^0=\ii X^d\,,\quad Y^i = X^i\,,\quad Y^d=\ii X^0\,. 
\end{align}
In fact, one can easily show that the metrics of EAdS$_d$ and dS$_d$ 
transform to each other.  
One also finds the relations
\begin{align}
 \nu'&=\sqrt{ \Bigl( \frac{d-1}{2} \Bigr)^2+m^2\,\ell^{\prime 2}}\nn\\
 &=\sqrt{ \Bigl( \frac{d-1}{2} \Bigr)^2-m^2\,\ell^{2}+\ii 0} \nn\\
 &=\nu_{\varepsilon}\,, 
 \\
 Z'(y,y')&=\frac{z^2+z^{\prime 2}+\abs{\by-\by'}^2}{2\,z\,z'}\nn\\
 &=\frac{(-\eta)^2 + (-\eta')^2 - \abs{\bx-\bx'}^2}{2\eta\,\eta'}-\ii 0\nn\\
 &=Z(x,x')-\ii 0\,,
\end{align}
with which the Green function on EAdS space can be rewritten as 
\begin{align}
 G_{\text{EAdS}}(y,y')
 &=\frac{e^{-\ii\pi\,(d-2)}}{(2\pi)^{d/2}\,\ell^{d-2}}\,(u^2-1)^{-\frac{d-2}{4}}\,
     \LQ_{\nu_{\varepsilon}-1/2}^{(d-2)/2}(u) \nn\\ 
 &\qquad\qquad(u=Z(x,x')-\ii 0)\,. 
\end{align}
This agrees with the in-out propagator \eqref{Poincare_in_out_full} 
in the Poincar\'{e} patch of de Sitter space: 
\begin{align}
 G^{\rm in/out}(x,x') 
   &= \frac{e^{-\ii\pi(d-2)}}{(2\pi)^{d/2}\,\ell^{d-2}}\,(u^2-1)^{-\frac{d-2}{4}}\,
     \LQ_{\nu-1/2}^{(d-2)/2}(u)\nn\\
   &\qquad\qquad(u= Z(x,x')-\ii 0)\,. 
\end{align}

As pointed out in \cite{Bousso:2001mw}, 
the Green function of EAdS space has no direct relation with 
the in-in propagators associated with $\alpha$-vacuum of de Sitter space for any $\alpha$. 
We see that it is the in-out propagator (in the Poincar\'{e} patch) 
which is actually related to the Green function of EAdS space. 
We thus expect that we can obtain a deep insight on the dS/CFT correspondence 
\cite{Strominger:2001pn} 
by analytically continuing the Euclidean AdS/CFT correspondence 
and by interpreting the result in terms of the in-out propagators 
(not of the in-in propagators).  

\section{Discussions and conclusion}
\label{sec:conclusion}

In this paper, 
we have considered quantum theory of a free scalar field 
in nonstatic spacetime. 
We first developed a framework 
to treat a harmonic oscillator with time-dependent parameters, 
and then applied it to investigate a free scalar field in de Sitter space, 
both in the Poincar\'{e} and the global patches.

We have taken the vacuum state at each moment $t_I$  
to be the instantaneous ground state of the Hamiltonian at the moment. 
We developed a calculation method 
to obtain the wave function $\varphi(t;t_I)$ associated with the vacuum. 
The in-out and in-in propagators are then obtained 
from the wave functions 
by sending the initial and final times 
to the past and future infinities.

A major advantage of our prescription in defining the vacuum  
is that we do not need to introduce 
``positive-energy wave functions'' 
that cannot be defined in a definite way 
for a spacetime with no asymptotic timelike Killing vector.

We have applied our method 
to calculate the in-out and in-in propagators in de Sitter space. 
The obtained propagators take de Sitter invariant forms,  
and are consistent with the results known in the literature. 
What actually happens is that, 
when the time $t_I$ is sent to a temporal boundary, 
our wave function $\varphi(t;t_I)$  may diverge, 
but the obtained  propagator has a finite limit 
and coincides with the propagator in the literature  
[see the comments following \eqref{Poin_Ginin}]. 

As a new result, we have found that a finite massless limit exists 
for the in-out propagator in the Poincar\'{e} patch. 
This is in contrast to the in-in propagator, 
where the no-go theorem is known 
that no massless limit exists for the in-in propagators 
without breaking the de Sitter invariance \cite{Allen:1985ux}. 
The same functional form had been obtained for the in-out propagators 
without precise numerical coefficients
in \cite{Polyakov:2007mm,Akhmedov:2009ta} from other approaches,  
and the massless case also had been considered in \cite{Polyakov:2007mm}. 
However, one cannot discuss the existence of a finite massless limit   
without knowing the precise numerical coefficients. 
Indeed, our in-out propagator in the global patch 
diverges in the massless limit 
just because the numerical coefficient diverges.

We have argued that our in-out propagator for a given foliation 
coincides with the Feynman propagator obtained by a path integral 
with the $\ii\varepsilon$ prescription, 
provided that the foliation is effectively noncompact 
in the temporal direction. 
We also have shown that 
both the Poincar\'e and the global patches meet the condition, 
and have confirmed the coincidence by numerical calculations.

We have also shown that the in-out propagators 
in both the Poincar\'{e} and the global patches satisfy Polyakov's composition law, 
demonstrating that the in-out propagators can be expressed 
as a sum over paths of a relativistic particle. 
It should be interesting to investigate 
whether the composition law holds universally 
for the in-out propagators in any spacetime. 
Furthermore, as a more fundamental issue, 
it must be important to clarify the meaning of 
(or to try to give an interpretation to) 
the relativistic {\em particle} 
in the language of quantum field theory  
in curved spacetime, 
where it is known that the concept of particles 
is not always possible to be introduced.

Our in-in propagator in the global patch has a finite value for $m\geq (d-1)/2$\,, 
but it diverges for $m<(d-1)/2$\,. 
It will be important to compare the in-in propagators 
with those obtained (numerically) by the path integral of the Schwinger-Keldysh type  
\cite{Schwinger:1960qe,Keldysh:1964ud}
(see also \cite{Weinberg:2005vy}).

As an important application of our construction, 
it should be interesting to investigate a thermodynamic property 
intrinsic to de Sitter space, 
especially its nonequilibrium property \cite{Fukuma:2013uxa}. 
It would be also interesting to investigate some physical quantities 
such as the rate of vacuum decay at finite times 
on the basis of our formalism. 
As another future direction, it should be interesting 
to apply our method to quantum field theories in spacetimes with horizon, 
such as a spacetime with black hole and de Sitter space in the static patch. 
For such a spacetime, one needs to carefully study 
the consistency of our formalism 
with boundary conditions at the horizon.

It should be important to consider interacting fields 
in generic nonstatic spacetimes 
and to establish perturbation theory on the basis of our formalism. 
It will be also interesting 
to investigate the in-out propagators for gravitons, 
since our method can be applied to field theory of higher spins
without any essential modifications.


\begin{acknowledgments}
The authors thank an anonymous referee 
for valuable comments 
on the first version of the manuscript. 
This work was supported by the Grant-in-Aid for the Global COE program 
``The Next Generation of Physics, Spun from Universality and
Emergence" from the Ministry of Education, Culture, Sports, 
Science and Technology (MEXT) of Japan. 
This work was also supported by MEXT (Grant No.\,23540304).
\end{acknowledgments}

\appendix

\section{Proof of Eq.~\eqref{in-in_a0}}
\label{appendix:in-in-matrix}

Setting $t=T_s$ in \eqref{c12-a} and using the hermiticity 
$q^\dagger(T_s)=q(T_s)$ and $p^\dagger(T_s)=p(T_s)$, 
we obtain
\begin{align}
 \begin{pmatrix} c_1^\dagger \\ c_2^\dagger
 \end{pmatrix}
 &= \frac{1}{\bigl(W_\rho[f,g]\bigr)^\ast}\,
   \begin{pmatrix}
    -V_\rho[g^\ast,f] & -V_\rho[g^\ast,g] \\
    V_\rho[f^\ast, f] & V_\rho[f^\ast, g]
   \end{pmatrix}\!(T_s)\,
 \begin{pmatrix} c_1 \\ c_2
 \end{pmatrix}\nn\\
 &\equiv M_s
 \begin{pmatrix} c_1 \\ c_2
 \end{pmatrix}\qquad 
 \bigl(V_\rho[f,g]\equiv \rho\,f\,\dot{g}-\rho^\ast\,\dot{f} \,g\bigr)\,.
\end{align}
Then, from \eqref{Ct} and \eqref{Ct_inv}, we obtain
\begin{align}
 \begin{pmatrix}
  a^\dagger(t) \\ \bar{a}^\dagger(t)
 \end{pmatrix}
 &= [C^{-1}(t)]^\ast\, M_s\,C(t')
 \begin{pmatrix}
  a(t') \\ \bar{a}(t')
 \end{pmatrix}\nn\\
 &\equiv \Lambda(t;t')
 \begin{pmatrix}
  a(t') \\ \bar{a}(t')
 \end{pmatrix}\,.
\label{in-in_matrix1}
\end{align}
A straightforward calculation shows that 
\begin{align}
 &\Lambda(t;t') \nn\\
 &= \ii\Bigl(
 \begin{smallmatrix}
  {}-  V_\rho[\bar{\varphi}^\ast(T_s,t),\varphi(T_s,t')] 
    & {}-  V_\rho[\bar{\varphi}^\ast(T_s,t),\bar{\varphi}(T_s,t')] \\
  V_\rho[\varphi^\ast(T_s,t),\varphi(T_s,t')] 
    & V_\rho[\varphi^\ast(T_s,t),\bar{\varphi}(T_s,t')] 
 \end{smallmatrix}\Bigr)\,.
\label{in-in_matrix2}
\end{align}
Note that $\det\Lambda(t;t')=-1$ 
due to the commutation relations $[a^\dagger(t),\bar{a}^\dagger(t)]=-1$ 
and $[a(t'),\bar{a}(t')]=1$ 
(this can also be checked by a direct calculation). 
Then, setting $t=t'=t_0$ in \eqref{in-in_matrix1} and \eqref{in-in_matrix2}, 
we find that 
\begin{align}
 \bar{a}^\dagger_0 
  &=  {}- \frac{V_\rho[\varphi^\ast(T_s;t_0),\bar{\varphi}(T_s;t_0)]}
       {V_\rho[\bar{\varphi}^\ast(T_s;t_0),\bar{\varphi}(T_s;t_0)]}\,a^\dagger_0\nn\\
  &\quad\ + \frac{\ii}{V_\rho[\bar{\varphi}^\ast(T_s;t_0),\bar{\varphi}(T_s;t_0)]}\,a_0\,.
\end{align}

\section{Asymptotically Minkowski space}
\label{appendix:asymptotically_Minkowski}

In this appendix, 
we reinvestigate within our framework a well-studied case 
where spacetime is asymptotically Minkowski 
in both the remote past and the remote future \cite{BD}. 

\subsection{Setup}

For brevity we consider the two-dimensional spacetime 
with the metric 
\begin{align}
 \rmd s^2 = a^2(t)\,\bigl( {}- \rmd t^2 + \rmd x^2\bigr)\,,
\end{align}
where the scale factor $a^2(t)$  now depends on time 
and takes the form
\begin{align}
 a^2(t) = a_0^2\,\frac{1-\tanh t}{2} + a_1^2\,\frac{1+\tanh t}{2}\,.
\end{align}
This spacetime is asymptotically Minkowski 
with scale $a_0$ in the remote past 
and with scale $a_1$ in the remote future. 
By expanding a scalar field $\phi(t,x)$ as 
\begin{align}
 \phi(t,x)=\sum_{k\geq 0}\sum_a\,\phi_{k,\,a}(t)\,Y_{k,\,a}(x)
\end{align}
as in section \ref{sec:Minkowski}, 
the action becomes%
\footnote{
Since $\ii\varepsilon$ plays no essential role in this appendix, 
we have eliminated it from the action. 
} 
\begin{align}
 &S[\phi(t,x)] \nn\\
 &= \int\!\rmd t\,\rmd x\,\sqrt{-g}\,\Bigl[\,
  {}- \frac{1}{2}\,g^{\mu\nu}\,\partial_\mu\phi\,\partial_\nu\phi
  -\frac{m^2}{2}\,\phi^2\,\Bigr] \nn\\
 &=\sum_{k\geq 0}\,\sum_a\,\int\!\rmd t \,\,
  \frac{1}{2}\,\bigl[\,\dot\phi_{k,\,a}^2(t) - \bigl(k^2+m^2\,a^2(t)\bigr)\,
  \phi_{k,\,a}^2(t)\,\bigr]\,.
\end{align}
Thus, the correspondence with the ingredients of section \ref{sec:general}
is given by
\begin{align}
 q(t) &= \phi_{k,\,a}(t)\,,\quad
 \rho(t) = 1\,,\\
 \omega(t) &= \sqrt{ \omega_0^2\,\frac{1-\tanh t}{2}+\omega_1^2\,\frac{1+\tanh t}{2}}\,,
\end{align}
where
\begin{align}
 \omega_0\equiv \sqrt{k^2 + m^2\,a_0^2}\,,\quad
  \omega_1\equiv \sqrt{k^2 + m^2\,a_1^2}\,.
\end{align}
We also introduce
\begin{align}
 \omega_\pm \equiv \frac{1}{2}\bigl(\omega_1\pm \omega_0\bigr)\,.
\end{align}

\subsection{Wave functions}

The equation of motion takes the form 
\begin{align}
 0&= \ddot{q} + \frac{\dot\rho}{\rho}\,q + \omega^2\,q \nn\\
  &= \ddot{q} 
  +\Bigl(\,
  \frac{\omega_1^2+\omega_0^2}{2}
  + \frac{\omega_1^2-\omega_0^2}{2}\,\tanh t\,\Bigr)\,q\,,
\end{align}
which can be solved analytically with the hypergeometric function. 
We set a pair of independent solutions $\{f(t),g(t)\}$ as
\begin{align}
 f(t) &= \Bigl(\frac{1-\zeta}{2}\Bigr)^{\ii\omega_1/2}\,
  \Bigl(\frac{1+\zeta}{2}\Bigr)^{-\ii\omega_0/2}\,\nn\\
  &\quad\times F\Bigl(\ii\omega_-,\,
  1+\ii\omega_-\,;
  1-\ii\omega_0\,;\frac{1+\zeta}{2}\Bigr)\,,
\\
 g(t) &= \Bigl(\frac{1-\zeta}{2}\Bigr)^{-\ii\omega_1/2}\,
  \Bigl(\frac{1+\zeta}{2}\Bigr)^{\ii\omega_0/2}\,\nn\\
  &\quad\times F\Bigl(-\ii\omega_-,\,
  1-\ii\omega_-\,;
  1+\ii\omega_0\,;\frac{1+\zeta}{2}\Bigr)\,,
\end{align}
where $\zeta\equiv \tanh t$ and $F(a,b;c;\,z)$ is the hypergeometric function. 
Their asymptotic forms for $t=t_0\sim -\infty$ (or $\zeta=\zeta_0\sim -1$) 
are easily found to be
\begin{align}
 f_0=f(t_0)\sim e^{-\ii\omega_0 \,t_0}\,,\quad
 g_0=g(t_0)\sim e^{\ii\omega_0 \,t_0}\,,
\end{align}
and the weighted Wronskian $ W_\rho[f,g]= \rho(t)\,W[f,g](t)$ is found to be 
\begin{align}
 W_\rho[f,g] =2 \ii \omega_0\,.
\end{align}
From this we find that the functions in \eqref{uuvv-1} and \eqref{uuvv-2} have the asymptotic forms
\begin{align}
 u_0&\sim0\,,~~~~~~~~~~~~~~~\,\bar{u}_0\sim-2\ii\omega_0 \,e^{-\ii\omega_0 \,t_0}\,,
\label{BD_u0}
\\
 v_0&\sim2\ii\omega_0 \,e^{\ii\omega_0 \,t_0}\,,~~~\,\bar{v}_0\sim 0\,.
\label{BD_v0}
\end{align}
The wave functions in the limit $t_0\to-\infty$ then take the form
\begin{align}
 \varphi(t;t_0) &\sim\frac{1}{\sqrt{2\omega_0}}e^{\ii\omega_0 t_0}f(t)
\nn\\
 &\sim\frac{1}{\sqrt{2\omega_0}}
  e^{\ii\omega_0 t_0-\ii\omega_+t-\ii\omega_- \log(2\cosh t)}\nn\\
 &\times F\Bigl(\ii\omega_-,\,
  1+\ii\omega_-\,;
  1-\ii\omega_0\,;\frac{1+\tanh t}{2}\Bigr)\,,\\[-0.2cm]
\bar{\varphi}(t;t_0) &\sim\frac{1}{\sqrt{2\omega_0}}e^{-\ii\omega_0 t_0}g(t)
\nn\\
  &\sim\frac{1}{\sqrt{2\omega_0}}
  e^{-\ii\omega_0 t_0+\ii\omega_+t+\ii\omega_- \log(2\cosh t)}\nn\\
 &\times F\Bigl(-\ii\omega_-,\,
  1-\ii\omega_-\,;
  1+\ii\omega_0\,;\frac{1+\tanh t}{2}\Bigr)\,.
\end{align}

In order to calculate the asymptotic forms of various functions for $t\sim +\infty$\,, 
it is convenient to rewrite  $f(t)$ and $g(t)$ 
by using the formula
\begin{align}
 &F(a,b;c;\,z)\nn\\
 &= \frac{\Gamma(c)\,\Gamma(c-a-b)}{\Gamma(c-a)\,\Gamma(c-b)}\,
  F(a,\,b;\,a+b-c+1;\,1-z)
\nn\\
 &~~~+\,\frac{\Gamma(c)\,\Gamma(a+b-c)}{\Gamma(a)\,\Gamma(b)}\,
  (1-z)^{c-a-b}\,\nn\\
 &\qquad \times F(c-a,\,c-b;\,c-a-b+1;\,1-z)\,.
\label{hypergeom_anal}
\end{align}
We then obtain
\begin{align}
 &f(t)\nn\\
 &= \Bigl(\frac{\omega_0}{\omega_1}\Bigr)^{1/2}\Bigl(\frac{1-\zeta}{2}\Bigr)^{\ii\omega_1/2}
  \Bigl(\frac{1+\zeta}{2}\Bigr)^{-\ii\omega_0/2}\nn\\
  &\times\Bigl[\tilde{\alpha}^{\ast}
  F\Bigl(\ii\omega_-,\,
  1+\ii\omega_-\,;
  1+\ii\omega_1\,;\frac{1-\zeta}{2}\Bigr)
\nn\\
  &-\tilde{\beta} \Bigl(\frac{1-\zeta}{2}\Bigr)^{-\ii\omega_1}F\Bigl(1-\ii\omega_+,\,
  -\ii\omega_+\,;
  1-\ii\omega_1\,;\frac{1-\zeta}{2}\Bigr) 
  \Bigr]\,,
\\
 &g(t) \nn\\
 &= \Bigl(\frac{\omega_0}{\omega_1}\Bigr)^{1/2}\Bigl(\frac{1-\zeta}{2}\Bigr)^{-\ii\omega_1/2}
  \Bigl(\frac{1+\zeta}{2}\Bigr)^{\ii\omega_0/2}\nn\\
 &\times\Bigl[\tilde{\alpha}
  F\Bigl(-\ii\omega_-,\,
  1-\ii\omega_-\,;
  1-\ii\omega_1\,;\frac{1-\zeta}{2}\Bigr)
\nn\\
  &-\tilde{\beta}^{\ast} \Bigl(\frac{1-\zeta}{2}\Bigr)^{\ii\omega_1}F\Bigl(1+\ii\omega_+,\,
  \ii\omega_+\,;
  1+\ii\omega_1\,;\frac{1-\zeta}{2}\Bigr) 
  \Bigr]\,,
\end{align}
where
\begin{align}
\tilde{\alpha}&\equiv \Bigl(\frac{\omega_1}{\omega_0}\Bigr)^{1/2}
\frac{\Gamma(1+\ii\omega_0)\Gamma(\ii\omega_1)}{\Gamma(1+\ii\omega_+)\Gamma(\ii\omega_+)}\,,\\
\tilde{\beta}&\equiv -\Bigl(\frac{\omega_1}{\omega_0}\Bigr)^{1/2}
\frac{\Gamma(1-\ii\omega_0)\Gamma(\ii\omega_1)}{\Gamma(1+\ii\omega_-)\Gamma(\ii\omega_-)}\,.
\end{align}
With these, the asymptotic forms for $t_1\sim +\infty$ can be obtained easily as
\begin{align}
 &f_1=f(t_1)\sim \Bigl(\frac{\omega_0}{\omega_1}\Bigr)^{1/2}
  \bigl(\tilde{\alpha}^{\ast}e^{-\ii\omega_1 t_1}
 -\tilde{\beta}e^{\ii\omega_1 t_1}\bigr)\,,
\\ 
 &g_1=g(t_1)\sim \Bigl(\frac{\omega_0}{\omega_1}\Bigr)^{1/2}
 \bigl(\tilde{\alpha}e^{\ii\omega_1 t_1}
 -\tilde{\beta}^{\ast}e^{-\ii\omega_1 t_1}\bigr)\,,
\end{align}
and 
\begin{align}
 u_1&\sim \Bigl(\frac{\omega_0}{\omega_1}\Bigr)^{1/2}
  (-2\ii\tilde{\beta}\omega_1)e^{\ii\omega_1 t_1}\,,\\
  \bar{u}_1&\sim \Bigl(\frac{\omega_0}{\omega_1}\Bigr)^{1/2}
  (-2\ii\tilde{\alpha}^{\ast}\omega_1)e^{-\ii\omega_1 t_1}\,,
\\
 v_1&\sim\Bigl(\frac{\omega_0}{\omega_1}\Bigr)^{1/2}
 (2\ii\tilde{\alpha}\omega_1)e^{\ii\omega_1 t_1}\,,\\
 \bar{v}_1&\sim\Bigl(\frac{\omega_0}{\omega_1}\Bigr)^{1/2}
  (2\ii\tilde{\beta}^{\ast}\omega_1)e^{-\ii\omega_1 t_1}\,.
\end{align}
We then find that the Bogoliubov coefficients take the asymptotic forms
\begin{align}
 \alpha(t_1, t_0)&\sim\tilde{\alpha}\,e^{-\ii(\omega_0 t_0-\omega_1 t_1)} \equiv\alpha_1
 \quad \bigl(\bar{\alpha}(t_1,t_0)\sim\alpha_1^{\ast}\bigr)\,,
\label{2dal}
\\
 \beta(t_1, t_0)&\sim\tilde{\beta}\,e^{\ii(\omega_0 t_0+\omega_1 t_1)} \equiv\beta_1
 \quad \bigl(\bar{\beta}(t_1,t_0)\sim\beta_1^{\ast}\bigr)\,.
\label{2dbe}
\end{align}
They coincide with the well-known values in the literature 
(see, e.g., in \cite{BD}) up to a phase. 
It is easy to see 
\begin{align}
 \abs{\alpha_1}^2&=\frac{\sinh^2(\pi\omega_+)}{\sinh(\pi\omega_0)\sinh(\pi\omega_1)}\,,\\
 \abs{\beta_1}^2&=\frac{\sinh^2(\pi\omega_-)}{\sinh(\pi\omega_0)\sinh(\pi\omega_1)}\,,
\label{be^2}
\end{align}
and thus the relation $\abs{\alpha_1}^2-\abs{\beta_1}^2=1$ actually holds.

Using the asymptotic forms of $u_1$ and $v_1$\,, 
we can calculate the wave function $\varphi(t,t_1)$ 
for $t_1\to +\infty$:
\begin{align}
 \varphi(t;t_1)=\frac{1}{\sqrt{2\omega_0}}e^{\ii\omega_1 t_1}
 \Bigl[\tilde{\alpha}f(t)+\tilde{\beta}g(t)\Bigr]\,. 
\end{align}
This can be further rewritten by using Kummer's relation
\begin{align}
 F(a,b;c;\,z)=(1-z)^{c-a-b}\,F(c-a,\,c-b;\,c;\,z)
\label{Kummers_relation}
\end{align} 
into the form
\begin{align}
 \varphi(t;t_1)&=\frac{1}{\sqrt{2\omega_1}}
  e^{\ii\omega_1 t_1-\ii\omega_+t-\ii\omega_- \log(2\cosh t)}\nn\\
 &\times F\Bigl(\ii\omega_-,\,
  1+\ii\omega_-\,;
  1-\ii\omega_0\,;\frac{1-\tanh t}{2}\Bigr)\,.
\end{align}
This certainly coincides up to a phase 
with the positive-energy wave function in the remote future given in \cite{BD}. 
One can easily see that $\varphi(t;t_1)$ actually has the form 
$\varphi(t;t_1)\sim (1/\sqrt{2\omega_1})\,e^{-\ii\omega_1 (t-t_1)}$ 
when $t$ is also very large. 

\section{Propagator in Minkowski space}
\label{appendix:propagator_Minkowski}

In order to evaluate the integral \eqref{Minkowski1}, 
we introduce the polar coordinates for the wave vector as
\begin{align}
 \rmd \bk^2 = \rmd k^2 + k^2\,\bigl(
  \rmd\theta^2+\sin^2\theta\,\rmd\Omega_{d-3}^2\bigr)\,,
\end{align}
where $\theta$ is chosen such that $\theta=0$ corresponds 
to the direction $\bx-\bx'$\,, i.e., 
$\bk\cdot(\bx-\bx')=k\,\abs{\bx-\bx'}\,\cos\theta$\,. 
The volume element is then given by
\begin{align}
 \rmd^{d-1}\bk &= \rmd k\,k^{d-2}\,\rmd\theta\,\sin^{d-3}\theta\,\rmd\Omega_{d-3}\nn\\
 &=\rmd k\,k^{d-2}\,\rmd\cos\theta\,(1-\cos^2\theta)^{(d-4)/2}\,\rmd\Omega_{d-3}\,,
\end{align}
and \eqref{Minkowski1} becomes
\begin{align}
 G(x,x')&=\frac{\abs{\Omega_{d-3}}}{(2\pi)^{d-1}}\,
  \int_0^\infty \!\!\rmd k\,
  k^{d-1}\, G_k(t,t')\nn\\
 &\times \int_{-1}^{1} \!\!\rmd s\,(1-s^2)^{(d-4)/2}\,\cos\bigl(k\abs{\bx-\bx'}\,s\bigr)\,,
\end{align}
where $\abs{\Omega_{n-1}}$ is the area of the unit sphere 
in $n$-dimensional Euclidean space, 
$\abs{\Omega_{n-1}}=\int\rmd \Omega_{n-1}=2\,\pi^{n/2}/\Gamma(n/2)$\,.
The integration with respect to $s=\cos\theta$ 
can be carried out by using the formula (8.411-8 in \cite{GR}) 
\begin{align}
 &\int_{-1}^1\!\!\rmd s\,(1-s^2)^{\nu}\,\cos (z s)\nn\\
  &=\sqrt{\pi}\,\Gamma(\nu+1)\,\Bigl(\frac{z}{2}\Bigr)^{-\nu-\frac{1}{2}}
  J_{\nu+\frac{1}{2}}(z) \quad [\,\Ree\nu > -1\,]\,,
\end{align}
and we obtain
\begin{align}
 &G(x,x')\nn\\
 &= \frac{1}{(2\pi)^{\frac{d-1}{2}}\,\abs{\bx-\bx'}^{\frac{d-3}{2}}}\,\nn\\
 &\quad  \times\int_0^\infty\!\! \rmd k\,k^{\frac{d-1}{2}}\,
    G_k(t,t')\,J_{\frac{d-3}{2}}\bigl(k\abs{\bx-\bx'}\bigr)\, 
\nn\\
 &=\frac{2^{-(d+1)/2}\,\pi^{-(d-1)/2}}{\abs{\bx-\bx'}^{(d-3)/2}}\,
   \int_0^\infty \!\!\rmd k\,\frac{k^{(d-1)/2}}{\omega_{k}}\, \nn\\
 &\qquad\qquad\qquad \times
  e^{-\ii\omega_{k,\varepsilon}\,(t_>-t_<)}\,
  J_{\frac{d-3}{2}}\bigl(k\abs{\bx-\bx'}\bigr) 
\nn\\
 &= \frac{2^{-(d+1)/2}\,\pi^{-(d-1)/2}}{\abs{\bx-\bx'}^{(d-3)/2}}\,m^{\frac{d-1}{2}}\,
    \int_1^\infty \!\!\rmd\lambda\,(\lambda^2-1)^{\frac{d-3}{4}}\, \nn\\
 &\qquad\quad \times
  e^{{}- m\,\lambda\,e^{\ii(\pi-\varepsilon)/2}\,(t_>-t_<)}\,
  J_{\frac{d-3}{2}}\bigl(k\abs{\bx-\bx'}\bigr) \,,
\end{align}
where, assuming $m>0$\,, we have set $\lambda=\omega_k/m=\sqrt{k^2+m^2}/m$ 
to obtain the last expression. 
Then applying the formula (6.645-2 in \cite{GR}):
\begin{align}
 &\int_1^\infty\rmd\lambda\,(\lambda^2-1)^{\frac{\nu}{2}}\,e^{{}- \alpha\,\lambda}\,
  J_\nu\bigl(\beta\,\sqrt{\lambda^2-1}\bigr) \nn\\
 &~~ =\sqrt{\dfrac{2}{\pi}}\,\beta^\nu\,
  \bigl(\alpha^2+\beta^2\bigr)^{-\frac{\nu}{2}-\frac{1}{4}}\,
  K_{\nu+\frac{1}{2}}\bigl(\sqrt{\alpha^2+\beta^2}\bigr)
  \nn\\
 &\qquad\qquad [\,\Ree\alpha>0\,,~\beta\in\bRR\,]
\end{align}
with $\nu=(d-3)/2$\,, $\alpha=m\,e^{\ii(\pi-0)/2}\,(t_>-t_<)$\,, 
$\beta=m\,\abs{\bx-\bx'}$\,, 
we obtain
\begin{align}
 G(x,x')&=\frac{m^{(d-2)/2}}{(2\pi)^{d/2}}\,
  \bigl[e^{\ii(\pi-0)}\,\Delta t^2 +\Delta\bx^2\bigr]^{-\frac{d-2}{4}}\,\nn\\
 &\quad \times K_{\frac{d-2}{2}}\bigl(m\sqrt{e^{\ii(\pi-0)} \Delta t^2 +\Delta\bx^2}\bigr)\,,
\end{align}
where $\Delta t\equiv t-t'$ and $\Delta\bx\equiv \bx-\bx'$\,.
This expression can be further rewritten by separately investigating 
the cases for different sign of 
$\sigma\equiv (x-x')^2 = -\Delta t^2+\Delta\bx^2$\,.

\noindent
\underline{(1) spacelike ($\sigma>0$)\,:} \\[1mm]
The modified Bessel function is readily evaluated as
$K_{(d-2)/2}\bigl(m\sqrt{e^{\ii(\pi-0)}\,\Delta t^2 +\Delta\bx^2}\bigr)
=K_{(d-2)/2}\bigl(m\,\sqrt{\sigma}\bigr)$\,, 
and we have
\begin{align}
 G(x,x')=\frac{m^{(d-2)/2}}{(2\pi)^{d/2}}\,\sigma^{-\frac{d-2}{4}}\,
  K_{\frac{d-2}{2}}\bigl(m\,\sqrt{\sigma}\bigr)\,.
\end{align}

\noindent
\underline{(2) timelike ($\sigma<0$)\,:} \\[1mm]
By using the relations $e^{\ii(\pi-0)}\,\Delta t^2+\Delta\bx^2
=e^{\ii\pi}\,(-\sigma)$ and 
$K_\nu(e^{\ii\pi/2}\,z)={}- (\ii\pi/2)\,H_\nu^{(2)}(z)$\,,
we have
\begin{align}
 G(x,x')=\frac{\pi}{2}\,\frac{m^{(d-2)/2}}{(2\pi)^{d/2}\,\ii^{d-1}}\,
  (-\sigma)^{-\frac{d-2}{4}}\,
  H_{\frac{d-2}{2}}^{(2)}\bigl(m\,\sqrt{\sigma}\bigr)\,.
\end{align}

\noindent
\underline{(3) null ($\sigma\to 0+$)\,:} \\[1mm]
By using the expansion 
\begin{align}
 K_\nu(z) = \frac{\Gamma(\nu)}{2}\,
            \Bigl(\frac{z}{2}\Bigr)^{-\nu}\,\bigl(1+O(z^2)\bigr)\,,
\end{align}
we obtain
\begin{align}
 G(x,x')\rightarrow \frac{\Gamma\bigl((d-2)/2\bigr)}{4\,\pi^{d/2}}\,
 (\sigma+\ii0)^{-\frac{d-2}{2}}\,.
\end{align}
The right-hand side actually coincides with the massless propagator. 

It is easy to see that all of the expression for three cases 
can be derived from a single expression,
\begin{align}
 G(x,x')=\frac{m^{(d-2)/2}}{(2\pi)^{d/2}\,
  \bigl(\sigma+\ii0\bigr)^{(d-2)/4}}\,
  K_{\frac{d-2}{2}}\bigl(m\sqrt{\sigma+\ii0}\bigr)\,.
\end{align}

\section{Proofs of Eqs.~\eqref{Poincare_in_out_full} and \eqref{Poincare_in_in_full}}
\label{appendix:Poincare}

In order to show Eq.~\eqref{Poincare_in_out_full}, 
we use the following equation (6.578-11 in \cite{GR}):
\begin{align}
 &\int_0^\infty\rmd x\,x^{\mu+1}\,
     K_\nu\bigl(\hat{a}\,x\bigr)\,
     I_\nu(\hat{b}\,x)\,
     J_\mu(c\,x) \nn\\
 &\qquad = \frac{c^\mu\,e^{-\ii\pi (\mu+\frac{1}{2})}}
         {\sqrt{2\pi}\,(\hat{a}\,\hat{b})^{\mu+1}}\,
    \bigl(u^2-1\bigr)^{-\frac{1}{2}\, (\mu+\frac{1}{2})}\,
    \LQ_{\nu-1/2}^{\mu+\frac{1}{2}}(u)
\nn\\
 &\Biggl[ \begin{array}{l}
u\equiv \frac{\hat{a}^2+\hat{b}^2+c^2}{2\hat{a}\,\hat{b}}\,,\quad 
 \Ree \hat{a}>\abs{\Ree \hat{b}}+\abs{\Imm c}\,, \\[1mm]
 \Ree \mu >-1\,,\quad \Ree(\mu+\nu)>-1 \end{array} \Biggr] \,.
\end{align}
By setting $\hat{a}=e^{\ii\pi/2}\,a$ and $\hat{b}=e^{\ii\pi/2}\,b$ 
for $-\pi/2<\arg a\leq \pi$ and $-\pi<\arg b\leq \pi/2$\,, 
and by using the identities
\begin{align}
 H^{(2)}_\nu(a\,x) 
  &= \frac{2\ii}{\pi}\,e^{\ii\pi\nu/2}\,K_\nu\bigl(\hat{a}\,x\bigr)\,,\\
 J_\nu(b\,x)
  &= e^{-\ii\pi\nu/2}\,I_\nu\bigl(\hat{b}\,x\bigr)\,,
\end{align}
the equation is rewritten to the form
\begin{align}
 &\int_0^\infty\rmd x\,x^{\mu+1}\,
     H^{(2)}_\nu (a\,x)\,
     J_\nu(b\,x)\,
     J_\mu(c\,x) \nn\\
 &\quad= \frac{\sqrt{2}\,c^\mu\,e^{-2\ii\pi(\mu+\frac{1}{2})}}
         {\pi^{3/2}\,(a\,b)^{\mu+1}}\,
    \bigl(u^2-1\bigr)^{-\frac{1}{2}\, (\mu+\frac{1}{2})}\,
    \LQ_{\nu-1/2}^{\mu+\frac{1}{2}}(u) 
\nn\\
 &\Biggl[ \begin{array}{l}
 u\equiv \frac{a^2+b^2-c^2}{2a\,b}\,,\quad 
 (-\Imm a) >\abs{\Imm b}+\abs{\Imm c}\,,\\[1mm]
 \Ree \mu >-1\,,\quad \Ree(\mu+\nu)>-1
 \end{array} \Biggr] \,.
\label{Poincare_in_out_formula}
\end{align}
We substitute to this $x=k$, $a=e^{-\ii\varepsilon}(-\eta_<)$, $b=e^{-\ii\varepsilon}(-\eta_>)$, 
$c=\abs{\bx-\bx'}$ and $\mu=(d-3)/2$\,.  
We then obtain Eq.~\eqref{Poincare_in_out_full} 
with $u= Z(x,x')-\ii\,0$ for infinitesimal $\varepsilon$\,.

In order to show Eq.~\eqref{Poincare_in_in_full}, 
we start from the following equation (6.578-10 in \cite{GR}):
\begin{align}
 &\int_0^\infty\rmd x\,x^{\mu+1}\,
     K_\nu\bigl(\hat{a}\,x\bigr)\,
     K_\nu(\hat{b}\,x)\,
     J_\mu(c\,x) \nn\\
 &\qquad = \frac{\sqrt{\pi}\,c^\mu\,\Gamma(\mu+\nu+1)\,\Gamma(\mu-\nu+1)}
         {2^{3/2}\,(\hat{a}\,\hat{b})^{\mu+1}}\,\nn\\
 &\qquad\qquad\qquad \times\bigl(u^2-1\bigr)^{-\frac{1}{2}\,(\mu+\frac{1}{2})}\,
    \LP_{\nu-1/2}^{-(\mu+\frac{1}{2})}(u)
\nn\\
 &\Biggl[  \begin{array}{l}
 u\equiv \frac{\hat{a}^2+\hat{b}^2+c^2}{2\hat{a}\,\hat{b}}\,,\quad 
 \Ree (\hat{a}+\hat{b})> \abs{\Imm c}\,,\\[1mm]
 \Ree (\mu\pm\nu) >-1\,,\quad \Ree \mu>-1 \end{array}
 \Biggr] \,.
\end{align}
By setting $\hat{a}=e^{-\ii\pi/2}\,a$ and $\hat{b}=e^{\ii\pi/2}\,b$ 
with $-\pi/2<\arg a\leq \pi$ and $-\pi/2<\arg b\leq \pi$\,, 
and by using the identities
\begin{align}
 H^{(1)}_\nu(a\,x) 
  &= -\frac{2\ii}{\pi}\,e^{-\ii\pi\nu/2}\,K_\nu\bigl(\hat{a}\,x\bigr)\,,\\
 H^{(2)}_\nu(b\,x) 
  &= \frac{2\ii}{\pi}\,e^{\ii\pi\nu/2}\,K_\nu\bigl(\hat{b}\,x\bigr)\,,
\end{align}
the equation is rewritten to the form
\begin{align}
 &\int_0^\infty\rmd x\,x^{\mu+1}\,
     H^{(1)}_\nu (a\,x)\,
     H^{(2)}_\nu (b\,x)\,
     J_\mu(c\,x) \nn\\
 &\qquad = \frac{\sqrt{2}\,c^\mu\,\Gamma(\mu+\nu+1)\,\Gamma(\mu-\nu+1)}
         {\pi^{3/2} \,(a\,b)^{\mu+1}}\,\nn\\
 &\qquad\qquad\qquad\times  \bigl(u^2-1\bigr)^{-\frac{1}{2}\,(\mu+\frac{1}{2})}\, 
    \LP_{\nu-1/2}^{-\mu-\frac{1}{2}}(u) 
\nn\\
 &\Biggl[ \begin{array}{l}
 u\equiv \frac{-a^2-b^2+c^2}{2a\,b}\,,\quad 
 \Imm (a-b) > \abs{\Imm c}\,,\\[1mm]
 \Ree (\mu\pm\nu) >-1\,,\quad \Ree \mu>-1
 \end{array} \Biggr] \,.
\label{Poincare_in_in_formula}
\end{align}
We substitute to this $x=k$, $a=e^{\ii\varepsilon}(-\eta_>)$, $b=e^{-\ii\varepsilon}(-\eta_<)$, 
$c=\abs{\bx-\bx'}$ and $\mu=(d-3)/2$. 
We then obtain Eq.~\eqref{Poincare_in_in_full} 
with $u= -Z(x,x')+\ii\,0$ for infinitesimal $\varepsilon$\,.

\section{Associated Legendre functions and the addition formulae}
\label{appendix:Legendre_addition}

In this appendix, we give several formulae of the associated Legendre functions 
which are used in subsection \ref{sec:global}. 
For details of the associated Legendre functions, see \cite{GR} and \cite{Erdelyi}.

The associated Legendre functions $\LP^{\mu}_{\nu}(z)$ and $\LQ^{\mu}_{\nu}(z)$ 
are defined over the complex $z$-plane other than the cut 
along the real axis to the left of the point $z=1$ 
(running from $-\infty$ to $1$), 
while the associated Legendre functions $\LPc^{\mu}_{\nu}(x)$ and $\LQc^{\mu}_{\nu}(x)$ 
are defined only on the interval $-1<x<1$:
\begin{align}
 \LP^{\mu}_{\nu}(z) 
  &\equiv \frac{1}{\Gamma(1-\mu)}\,\Bigl(\frac{z+1}{z-1}\Bigr)^{\frac{\mu}{2}}
     F\Bigl(-\nu,\,\nu+1;\,1-\mu;\,\frac{1-z}{2}\Bigr) \,,
\\
 \LQ^{\mu}_{\nu}(z) 
  &\equiv \frac{e^{\ii \pi\mu}\,\pi^{\frac{1}{2}}\,\Gamma(\nu+\mu+1)}
          {2^{\nu+1}\,\Gamma(\nu+3/2)}\,
     z^{-\nu-\mu-1}\,\bigl(z^2-1\bigr)^{\frac{\mu}{2}}\,\nn\\
  &\quad\times F\Bigl(\frac{\nu+\mu+2}{2},\,\frac{\nu+\mu+1}{2};\,\nu+\frac{3}{2};\,\frac{1}{z^2}\Bigr) \,,
\\
 \LPc^{\mu}_{\nu}(x) 
  &\equiv \frac{1}{2}\, \Bigl[e^{\frac{1}{2}\,\ii\pi\mu}\,\LP^{\mu}_{\nu}(x+\ii 0)
                       +e^{-\frac{1}{2}\,\ii\pi\mu}\,\LP^{\mu}_{\nu}(x-\ii 0)\Bigr] 
\nn\\
 &= \frac{1}{\Gamma(1-\mu)}\,\Bigl(\frac{1+x}{1-x}\Bigr)^{\frac{\mu}{2}}\,
  F\Bigl(-\nu,\,\nu+1;\,1-\mu;\,\frac{1-x}{2}\Bigr)\,,
\\
 \LQc^{\mu}_{\nu}(x) 
  &\equiv \frac{1}{2}\,e^{-\ii \pi\mu}\,
    \Bigl[e^{-\frac{1}{2}\,\ii \pi\mu}\,\LQ^{\mu}_{\nu}(x+\ii 0)
         +e^{\frac{1}{2}\,\ii \pi\mu}\,\LQ^{\mu}_{\nu}(x-\ii 0)\Bigr]
\nn\\
 &= \frac{\pi}{2\sin\pi\mu}\,\Bigl[
  \cos\pi\mu\,\LPc_\nu^\mu(x)-\frac{\Gamma(\nu+\mu+1)}{\Gamma(\nu-\mu+1)}\,\LPc_\nu^{-\mu}(x)
  \Bigr]\,.
\end{align}

\subsubsection*{Functional relations}

The four functions $\LP(z)$, $\LQ(z)$, $\LPc(x)$ and $\LQc(x)$ 
are related to each other as  (3.4 and 3.3.1 in \cite{Erdelyi})
\begin{align}
 &e^{\ii\pi\frac{\mu}{2}}\,\LP^{\mu}_{\nu}(x+\ii 0)=
 e^{-\ii\pi\frac{\mu}{2}}\,\LP^{\mu}_{\nu}(x-\ii 0) = \LPc^{\mu}_{\nu}(x) \,,
\label{fun1} 
\\
 & e^{-\ii\pi\mu}\,
    \Bigl[e^{-\ii\pi\frac{\mu}{2}}\,\LQ^{\mu}_{\nu}(x+\ii 0)
    \pm e^{\ii\pi\frac{\mu}{2}}\,\LQ^{\mu}_{\nu}(x-\ii 0)\Bigr]\nn\\
 &\qquad =\left\{\begin{array}{l}
  2\LQc^{\mu}_{\nu}(x) \\
  -\ii\pi\LPc^{\mu}_{\nu}(x)
  \end{array}\right.\,,
\label{fun2}
\\
 &\LQ^{\mu}_{\nu}(-z) 
  =\left\{\begin{array}{ll}
           {}-e^{\ii\pi\nu}\LQ^{\mu}_{\nu}(z)  &\quad [\Imm z>0] \\
           {}-e^{-\ii\pi\nu}\LQ^{\mu}_{\nu}(z) &\quad [\Imm z<0] 
           \end{array}\right. \,.
\label{fun12}
\end{align}
We also have (8.73 in \cite{GR})
\begin{align}
 \LP_\nu^{-\mu}(z)&= \frac{\Gamma(\nu-\mu+1)}{\Gamma(\nu+\mu+1)}\,\nn\\
  &\times\Bigl[\,\LP_\nu^\mu(z)-\frac{2}{\pi}\,e^{-\ii\pi\mu}\,\sin\pi\mu\,\LQ_\nu^\mu(z)\Bigr]\,,
\label{LP_mmu}
\\
 \LQ_\nu^{-\mu}(z)&= e^{-2\ii\pi\mu}\,
  \frac{\Gamma(\nu-\mu+1)}{\Gamma(\nu+\mu+1)}\,\LQ_\nu^\mu(z)\,,
\label{LQ_mmu}
\\
 \LPc_\nu^{-\mu}(x)&= \frac{\Gamma(\nu-\mu+1)}{\Gamma(\nu+\mu+1)}\,\nn\\
  &\times\Bigl[\,\cos\pi\mu\,\LPc_\nu^\mu(x)-\frac{2}{\pi}\,\sin\pi\mu\,\LQc_\nu^\mu(x)\Bigr]\,,
\label{LPc_mmu}
\\
 \LQc_\nu^{-\mu}(x)&= \frac{\Gamma(\nu-\mu+1)}{\Gamma(\nu+\mu+1)}\,\nn\\
  &\times\Bigl[\,\frac{\pi}{2}\,\sin\pi\mu\,\LPc_\nu^\mu(x)
  +\cos\pi\mu\,\LQc_\nu^\mu(x)\Bigr]\,,
\label{LQc_mmu}
\\
 \LPc_{-\nu-1}^\mu(x)&= \LPc_\nu^\mu(x)\,,
\\
 \LQc_{-\nu-1}^\mu(x)&= \frac{1}{\sin\pi(\nu-\mu)}\nn\\
 &\times\bigl[{}-\pi\cos\pi\nu\,\LPc_\nu^\mu(x)+\sin\pi(\nu+\mu)\,\LQc_\nu^\mu(x)\bigr]\,.
\end{align}
Their Wronskians have the forms (8.741 in \cite{GR})
\begin{align}
 \left| \begin{array}{cc}
  \LPc_\nu^\mu(x) & \LQc_\nu^\mu(x) \\
  \frac{\rmd}{\rmd x}\LPc_\nu^\mu(x) & \frac{\rmd}{\rmd x}\LQc_\nu^\mu(x) 
 \end{array} \right|
  = \frac{1}{1-x^2}\,\frac{\Gamma(\nu+\mu+1)}{\Gamma(\nu-\mu+1)}\,,
\\
 \left| \begin{array}{cc}
  \LPc_\nu^{-\mu}(x) & \LPc_\nu^{\mu}(x) \\
  \frac{\rmd}{\rmd x}\LPc_\nu^{-\mu}(x) & \frac{\rmd}{\rmd x}\LPc_\nu^{\mu}(x) 
 \end{array} \right|
  = \frac{1}{1-x^2}\,\frac{2\sin\pi\mu}{\pi}\,.
\end{align}
We also have (8.733-1 in \cite{GR})
\begin{align}
 (1-x^2)\,\frac{\rmd}{\rmd x}\,\LPc_\nu^\mu(x)
  &= (\nu+1)\,x\,\LPc_\nu^\mu(x) - (\nu-\mu+1)\,\LPc_{\nu+1}^\mu(x)\,, \\
 (1-x^2)\,\frac{\rmd}{\rmd x}\,\LQc_\nu^\mu(x)
  &= (\nu+1)\,x\,\LQc_\nu^\mu(x) - (\nu-\mu+1)\,\LQc_{\nu+1}^\mu(x)\,.
\end{align}

\subsubsection*{Asymptotic forms}

The associated Legendre functions have the following asymptotic forms near boundaries 
(3.9.2 in \cite{Erdelyi}):
\begin{align}
 \LP^{\mu}_{\nu}(z)
  &\overset{z\sim \infty}{\sim}
   \left\{\begin{array}{l}
      2^{\nu}\,\pi^{-\frac{1}{2}}\,
      \dfrac{\Gamma(\nu+1/2)}{\Gamma(\nu-\mu+1)}\,z^{\nu} \\[3mm]
   \qquad\qquad\qquad [\,\Ree \nu > -1/2\,]\\[3mm]
      2^{-\nu-1}\,\pi^{-\frac{1}{2}}\,
      \dfrac{\Gamma(-\nu-1/2)}{\Gamma(-\nu-\mu)}\,z^{-\nu-1}\\[3mm]
   \qquad\qquad\qquad [\,\Ree \nu < -1/2\,]
     \end{array} \right.\,, 
\label{LP_as}
\\
 \LQ^{\mu}_{\nu}(z)
  &\overset{z\sim \infty}{\sim}
   e^{\ii\pi\mu}\,2^{-\nu-1}\,\pi^{\frac{1}{2}}\,
   \frac{\Gamma(\nu+\mu+1)}{\Gamma(\nu+3/2)}\,z^{-\nu-1} \,,
\label{LQ_as}
\\
 \LPc^{\nu}_{k}(x) 
  &\overset{x\sim +1}{\sim}
   \frac{2^{\nu/2} \sin(\pi\nu)\,\Gamma(\nu)}{\pi}
   \,\bigl(1-x^2\bigr)^{-\frac{\nu}{2}} \,,
\label{p1}
\\
 \LQc^{\nu}_{k}(x) 
  &\overset{x\sim +1}{\sim}
   2^{\nu-1}\,\cos(\pi\nu)\,\Gamma(\nu)\,\bigl(1-x^2\bigr)^{-\frac{\nu}{2}} \nn\\
  &\qquad\qquad\qquad\qquad\qquad\qquad [\,\Ree \nu > 0\,] \,,
\label{q1}
\\
 \LPc^{\nu}_{k}(x) 
  &\overset{x\sim -1}{\sim}
   \left\{\begin{array}{l}
           2^{-\nu}\,\cos(\pi k)\,
           \dfrac{\Gamma(k+\nu+1)}{\Gamma(\nu+1)\,\Gamma(k-\nu+1)} \\[3mm]
  \quad\times \bigl(1-x^2\bigr)^{\frac{\nu}{2}} \qquad [\,k \in \mathbb{Z}\,] \\[3mm]
          {}-\dfrac{2^{\nu}\,\sin(\pi k)\,\Gamma(\nu)}{\pi}\,
          \bigl(1-x^2\bigr)^{-\frac{\nu}{2}}\\[3mm]
  \qquad\quad [\,k \in \mathbb{Z}+1/2\,,\ \Ree \nu >0\,] \\
   \end{array}\right.\,,
\label{pm1}
\\
 \LQc^{\nu}_{k}(x) 
  &\overset{x\sim -1}{\sim} 
   \left\{\begin{array}{l}
    {}-2^{\nu-1}\cos(\pi k) \,\Gamma(\nu)\,\bigl(1-x^2\bigr)^{-\frac{\nu}{2}} \\[3mm]
   \qquad\qquad\qquad [\,k \in \mathbb{Z}\,,\ \Ree \nu >0\,] \\[3mm]
    {}-\dfrac{2^{-\nu-1}\,\pi\,\Gamma(k+\nu+1)}{\sin(\pi k)\,\Gamma(\nu+1)\,\Gamma(k-\nu+1)} \\[3mm]
   \quad\times \bigl(1-x^2\bigr)^{\frac{\nu}{2}}\qquad [\,k \in \mathbb{Z}+1/2\,]
  \end{array} \right.\,.
\label{qm1}
\end{align}

\subsubsection*{Addition formulae}

We find the following formulae, 
which are useful in obtaining the propagators in the global patch:
\begin{align}
 &\frac{\ii\pi(\cos\varphi_1\,\cos\varphi_2)^{\frac{d-1}{2}}}{2\,(d-2)\,
  \abs{\Omega_{d-1}}\,\sin(\pi\nu)} \,
  \sum_{L=0}^{\infty}(2L+d-2) \,\nn\\
 &\qquad\qquad\times\LPc^{-\nu}_{k}(\sin\varphi_1)\,\LPc^{\nu}_{k}(\sin\varphi_2)\,
  C_L^{\frac{d-2}{2}}(\cos\theta) \nn\\
 &~~
 = \frac{{}-e^{-\ii\pi\frac{d-2}{2}}}{2\,(2\pi)^{\frac{d}{2}}\,\sin(\pi\nu)}\,
  \biggl[ (u_{+}^2-1)^{-\frac{d-2}{4}}\,\LQ^{\frac{d-2}{2}}_{\nu-\frac{1}{2}}(u_{+})\nn\\
 &\qquad\qquad\qquad\qquad\qquad
  -(u_{-}^2-1)^{-\frac{d-2}{4}}\,\LQ^{\frac{d-2}{2}}_{\nu-\frac{1}{2}}(u_{-}) \biggr] \nn\\
 &\qquad\qquad\qquad\qquad\qquad\qquad\qquad\qquad (d:\odd)\,, 
\label{formulaodd}
\\
 &\frac{\ii(\cos\varphi_1\,\cos\varphi_2)^{\frac{d-1}{2}}}{(d-2)\,
  \abs{\Omega_{d-1}}\,\cos(\pi\nu)}\, 
  \sum_{L=0}^{\infty}(2L+d-2) \,\nn\\
 &\qquad\qquad\times
  \LPc^{-\nu}_{k}(\sin\varphi_1)\,\LQc^{\nu}_{k}(\sin\varphi_2)\,
  C_L^{\frac{d-2}{2}}(\cos\theta) 
\nn\\
 &~~
 = \frac{\ii e^{-\ii\pi\frac{d-2}{2}}}{2\,(2\pi)^{\frac{d}{2}}\,\cos(\pi\nu)}
  \biggl[(u_{+}^2-1)^{-\frac{d-2}{4}}\,
         \LQ^{\frac{d-2}{2}}_{\nu-\frac{1}{2}}(u_{+})\nn\\
 &\qquad\qquad\qquad\qquad\qquad
        +(u_{-}^2-1)^{-\frac{d-2}{4}}\,
         \LQ^{\frac{d-2}{2}}_{\nu-\frac{1}{2}}(u_{-}) \biggr] \nn\\
 &\qquad\qquad\qquad\qquad\qquad\qquad\qquad\qquad (d:\even)\,, 
\label{formulaeven}\\
 &\frac{\ii (\cos\varphi_1\,\cos\varphi_2)^{\frac{d-1}{2}}}{(d-2)\,\abs{\Omega_{d-1}}}\,  
  \sum_{L=0}^{\infty}(2 L+d-2) \,\nn\\
 &\qquad\times\frac{\Gamma(k-\nu+1)}{\Gamma(k+\nu+1)}\, 
  \LPc^{\nu}_{k}(\sin\varphi_1)\,\LQc^{\nu}_{k}(\sin\varphi_2)\,
  C_L^{\frac{d-2}{2}}(\cos\theta) 
\nn\\
 &~~
 = \frac{\ii e^{-\ii\pi\frac{d-2}{2}}}{2\,(2\pi)^{\frac{d}{2}}}
  \biggl[e^{-\ii\pi\nu}\,(u_{+}^2-1)^{-\frac{d-2}{4}}\,
         \LQ^{\frac{d-2}{2}}_{-\nu-\frac{1}{2}}(u_{+})\nn\\
 &\qquad\qquad\qquad\qquad
        +e^{\ii\pi\nu}\,(u_{-}^2-1)^{-\frac{d-2}{4}}\,
         \LQ^{\frac{d-2}{2}}_{-\nu-\frac{1}{2}}(u_{-}) \biggr] \nn\\
 &\qquad\qquad\qquad\qquad\qquad\qquad\qquad\qquad (d:\even)\,, 
\label{formulaeven2}\\
 &\frac{2\,\ii (\cos\varphi_1\,\cos\varphi_2)^{\frac{d-1}{2}}}{\pi\,(d-2)\,\abs{\Omega_{d-1}}}\, 
  \sum_{L=0}^{\infty}(2L+d-2) \,\nn\\
 &\qquad\qquad\times\LQc^{-\nu}_{k}(\sin\varphi_1)\,\LQc^{\nu}_{k}(\sin\varphi_2)\,
  C_L^{\frac{d-2}{2}}(\cos\theta) 
\nn\\
 &~~
 = \frac{e^{-\ii\pi\frac{d-2}{2}}}{2\,(2\pi)^{\frac{d}{2}}}\,
    \Bigl\{\,
     (u_+^2-1)^{-\frac{d-2}{4}}\,\LQ_{\nu-\frac{1}{2}}^{\frac{d-2}{2}}(u_+)\nn\\
 &\qquad\qquad\qquad - (u_-^2-1)^{-\frac{d-2}{4}}\,\LQ_{\nu-\frac{1}{2}}^{\frac{d-2}{2}}(u_-) \nn\\
 &\qquad\qquad  + \frac{\ii\,\pi}{\cos(\pi\nu)}\,
     \bigl[e^{-\ii\pi\nu}\,(u_+^2-1)^{-\frac{d-2}{4}}\,
                             \LP_{\nu-\frac{1}{2}}^{\frac{d-2}{2}}(u_+)\nn\\
 &\qquad\qquad\qquad\qquad  + e^{\ii\pi\nu}\,(u_-^2-1)^{-\frac{d-2}{4}}\, 
                             \LP_{\nu-\frac{1}{2}}^{\frac{d-2}{2}}(u_-)
     \bigr]\Bigr\} \nn\\
 &~~
 ={}- \frac{e^{-\ii\pi\frac{d-2}{2}}}{2\,(2\pi)^{\frac{d}{2}}}\,
    \Bigl\{\,
     (u_+^2-1)^{-\frac{d-2}{4}}\,\LQ_{\nu-\frac{1}{2}}^{\frac{d-2}{2}}(u_+)
\nn\\
 &\qquad\qquad\qquad\qquad - (u_-^2-1)^{-\frac{d-2}{4}}\,\LQ_{\nu-\frac{1}{2}}^{\frac{d-2}{2}}(u_-)\nn\\
 &\qquad\qquad  {}- \frac{\pi}{\cos(\pi\nu)}\,
     \bigl[(u_+^2-1)^{-\frac{d-2}{4}}\,
                             \LP_{\nu-\frac{1}{2}}^{\frac{d-2}{2}}(-u_+)\nn\\
 &\qquad\qquad\qquad\qquad - (u_-^2-1)^{-\frac{d-2}{4}}\, 
                             \LP_{\nu-\frac{1}{2}}^{\frac{d-2}{2}}(-u_-)
     \bigr]\Bigr\} \nn\\
 &\qquad\qquad\qquad\qquad\qquad\qquad\qquad\qquad (d:\even)\,, 
\label{formulaeven3}
\end{align}
where $-\pi/2<\varphi_{2}<\varphi_{1}<\pi/2$\,,\,\ $0\leq\theta\leq\pi$\,,\,\ 
$k\equiv L+(d-3)/2$\,, and
\begin{align}
 &u_{\pm}(\varphi_1,\varphi_2,\theta) 
  \equiv -Z(\varphi_1,\varphi_2,\theta) \pm \ii 0 \nn\\
 &\mbox{with}\qquad 
 Z(\varphi_1,\varphi_2,\theta) 
 \equiv \frac{-\sin\varphi_1\,\sin\varphi_2+\cos\theta}{\cos\varphi_1\,\cos\varphi_2} \,.
\label{formulacond}
\end{align}
We prove Eqs.~\eqref{formulaodd} and \eqref{formulaeven} for the rest of this appendix. 
Equation \eqref{formulaeven2} can be proved in a similar way, 
and \eqref{formulaeven3} is readily obtained from \eqref{formulaeven} and \eqref{formulaeven2}.

\begin{figure}[htbp]
\vspace{3ex}
\begin{center}
\includegraphics[scale=0.8]{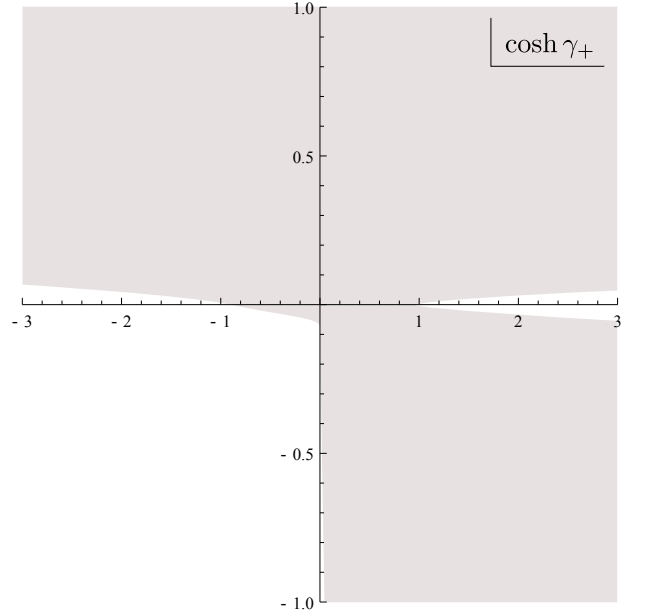}
\begin{quote}
\vspace{-5mm}
\caption{
Schematic view illustrating the range of $\cosh\gamma_+$ 
for $\varepsilon_1=0.001$ and $\varepsilon_2=0.02$\,.
Here, $\varphi_1$ and $\varphi_2$ run over the range 
$-\pi/2<\varphi_2<\varphi_1<\pi/2$\,, and 
$\theta$ runs over its full range $0\leq\theta\leq\pi$\,.  
The range of $\cosh\gamma_-$ can be obtained 
by turning the figure by $180^\circ$ over the horizontal axis. 
\label{fig:chgplus}}
\end{quote}
\end{center}
\vspace{-6ex}
\end{figure}

We start from Eq.~(12) of \cite{Beloozerov:1976}:%
\footnote{
Equation \eqref{formulaeven2} can be proved by replacing 
$\LQ_k^\nu(\cosh\beta_2)$ in \eqref{formula1} 
by $\LQ_k^{-\nu}(\cosh\beta_2)$ 
with the help of \eqref{LQ_mmu}. 
} 
\begin{align}
 &(\sinh\gamma)^{-\frac{d-2}{2}}\,\LQ^{\frac{d-2}{2}}_{\nu-\frac{1}{2}} (\cosh\gamma)\nn\\
 &=2^{\frac{d}{2}-2}\,\Gamma\Bigl(\frac{d-2}{2}\Bigr)\,
 e^{\ii\pi\bigl(-\nu+\frac{d-2}{2}\bigr)}\,
 (\sinh\beta_1\,\sinh\beta_2)^{\frac{d-1}{2}} \nn\\
 &\times\sum_{L=0}^{\infty} (2L+d-2)\, \LP^{-\nu}_{k}(\cosh \beta_1)\,
 \LQ^{\nu}_{k}(\cosh \beta_2)\,
 C_L^{\frac{d-2}{2}}(\cos\theta) \,,
\label{formula1}
\end{align}
where
\begin{align}
 \cosh\gamma(\beta_1,\beta_2,\theta)
  &\equiv\frac{\cosh\beta_1\cosh\beta_2-\cos\alpha}{\sinh\beta_1\sinh\beta_2}  \,, 
\nn\\
  \Ree\beta_2 &> \abs{\Ree\beta_1}+\abs{\Imm \theta} \,.
\label{chg}
\end{align}
Both sides of Eq.~\eqref{formula1} should be understood 
as the quantities that are continued analytically from the region where 
$\beta_1$,\ $\beta_2$, and $\theta$ take all real values (for which $\cosh\gamma >1$). 
We reparametrize the variables in Eq.~\eqref{formula1} as 
\begin{align}
 \beta^{\pm}_{\alpha} &\equiv \pm\ii
  \Bigl(\frac{\pi}{2}-\varphi_{\alpha}\Bigr)+\varepsilon_{\alpha} \quad 
 (\alpha =1,2)  \nn\\
 \Bigl[\,-\frac{\pi}{2}&<\varphi_{2}<\varphi_{1}<\frac{\pi}{2} \,, \quad 
       0<\varepsilon_1<\varepsilon_2 \ll 1\,\Bigr]\,,
\label{bepm}
\end{align}
and only keep the contributions from $\varepsilon_{\alpha}$ to the linear order. 
We then have
\begin{align}
 \cosh\beta^{\pm}_{\alpha}
  &=\sin\varphi_{\alpha}\pm\ii \varepsilon_\alpha \cos\phi_\alpha \,,\\
 \sinh\beta^{\pm}_{\alpha}
  &=\pm\ii\cos\varphi_{\alpha}+ \varepsilon_\alpha\sin\phi_\alpha \,,
\end{align}
and
\begin{align}
 \cosh\gamma_{\pm} &\equiv \cosh\gamma(\beta^{\mp}_1,\beta^{\pm}_2,\theta) \nn\\
 &=-Z(\varphi_1,\varphi_2,\theta)+\ii\mathcal{O}(\varepsilon_1,\varepsilon_2) \,.
\end{align}
If we fix the parameters $\varepsilon_1$ and $\varepsilon_2$, 
and vary $\varphi_1$, $\varphi_2$, and $\theta$ within the regions 
$-\pi/2<\varphi_{2}<\varphi_{1}<\pi/2$ and $0\leq\theta\leq\pi$\,, 
then $\cosh\gamma_{\pm}$ ranges in the region 
depicted in Fig.~\ref{fig:chgplus}. 

In the following, we divide the parameter region of 
$\varphi_1$,\ $\varphi_2$ and $\theta$ into three parts, 
where $Z$ takes values in (1) $Z>1$, (2) $1>Z>-1$, and (3) $Z<-1$, respectively. 
We then derive a simpler expression of Eq.~\eqref{formula1} for each case, 
and show that the obtained expressions for the three cases 
can be summarized in the form \eqref{formulaodd} and \eqref{formulaeven}.

\noindent{\bf\underline{(1) $Z>1$}}

In this case, 
$\cosh\gamma_{\pm}=-Z\pm\ii0$ as can be seen from Fig.~\ref{fig:chgplus}.%
\footnote{
$\varepsilon_1$ and $\varepsilon_2$ are to be taken to zero 
with keeping $\varepsilon_2>\varepsilon_1$. 
} 
Thus, we have
\begin{align}
 \sinh\gamma_{\pm}
 &=e^{\pm\ii\pi}\,\bigl(Z^2-1\bigr)^{\frac{1}{2}}\,, \\
 \LQ^{\frac{d-2}{2}}_{\nu-\frac{1}{2}}(\cosh\gamma_{\pm}) 
 &= {}-e^{\mp\ii\pi(\nu-\frac{1}{2})}\,
     \LQ^{\frac{d-2}{2}}_{\nu-\frac{1}{2}}(Z\mp\ii 0)\nn\\
 &= {}-e^{\mp\ii\pi(\nu-\frac{1}{2})}\,
     \LQ^{\frac{d-2}{2}}_{\nu-\frac{1}{2}} (Z) \,,
\end{align}
where we have used Eq.~\eqref{fun12} and the fact that 
$\LQ^{\mu}_{\nu} (z)$ does not have a cut in the region $\Ree z>1$.
Then, Eq.~\eqref{formula1} becomes
\begin{align}
 &{}-e^{\mp\ii\pi(\frac{d-3}{2}+\nu)}\,
  \bigl(Z^2-1\bigr)^{-\frac{d-2}{4}}\,
  \LQ^{\frac{d-2}{2}}_{\nu-\frac{1}{2}} (Z) \nn\\
 &=2^{\frac{d}{2}-2}\,\Gamma\Bigl(\frac{d-2}{2}\Bigr)\,
   e^{\ii\pi\bigl(-\nu\mp\frac{\nu}{2}+\frac{d-2}{2}\bigr)}\,
   (\cos\varphi_1\,\cos\varphi_2)^{\frac{d-1}{2}} \nn\\
 &\quad\times\sum_{L=0}^{\infty} (2L+d-2)\,\nn\\
 &\quad\quad\times\LPc^{-\nu}_{k}(\sin\varphi_1)\,
  \LQ^{\nu}_{k}(\sin\varphi_2\pm\ii 0)\,C_L^{\frac{d-2}{2}}(\cos\theta) \,.
\end{align}
By taking the difference between the above equations 
with the upper and the lower signs, 
we obtain
\begin{align}
 &2\ii\sin\Bigl[\pi\Bigl(\frac{d-3}{2}+\nu\Bigr)\Bigr]\,
  \bigl(Z^2-1\bigr)^{-\frac{d-2}{4}}\,
  \LQ^{\frac{d-2}{2}}_{\nu-\frac{1}{2}} (Z) \nn\\
 &={}-\ii\pi 2^{\frac{d}{2}-2}\,\Gamma\Bigl(\frac{d-2}{2}\Bigr)\,
   e^{\ii\pi\frac{d-2}{2}}
   (\cos\varphi_1\,\cos\varphi_2)^{\frac{d-1}{2}} \nn\\
 &\times\sum_{L=0}^{\infty} (2L+d-2)\,\LPc^{-\nu}_{k}(\sin\varphi_1)\,
  \LPc^{\nu}_{k}(\sin\varphi_2)\,C_L^{\frac{d-2}{2}}(\cos\theta)\,,
\end{align}
where Eq.~\eqref{fun2} has been used. 
Similarly, by taking their sum, 
we obtain
\begin{align}
 &{}-2\cos\Bigl[\pi\Bigl(\frac{d-3}{2}+\nu\Bigr)\Bigr]\,
   \bigl(Z^2-1\bigr)^{-\frac{d-2}{4}}\,\LQ^{\frac{d-2}{2}}_{\nu-\frac{1}{2}} (Z) \nn\\
 &=2^{\frac{d}{2}-1}\,\Gamma\Bigl(\frac{d-2}{2}\Bigr)\,e^{\ii\pi\frac{d-2}{2}}\,
   (\cos\varphi_1\,\cos\varphi_2)^{\frac{d-1}{2}} \nn\\
 &\times\sum_{L=0}^{\infty}(2L+d-2)\,\LPc^{-\nu}_{k}(\sin\varphi_1)\,
  \LQc^{\nu}_{k}(\sin\varphi_2)\,C_L^{\frac{d-2}{2}}(\cos\theta) \,.
\end{align}
The right hand side of Eqs.~\eqref{formulaodd} and \eqref{formulaeven}
can then be written as
\begin{align}
 \left\{\begin{array}{l}
  {}-\displaystyle\frac{\ii\sin\pi(\frac{d-2}{2})}
       {(2\pi)^{\frac{d}{2}}}\,e^{-\ii\pi\frac{d-2}{2}}\,
  \bigl(Z^2-1\bigr)^{-\frac{d-2}{4}}\,
  \LQ^{\frac{d-2}{2}}_{\nu-\frac{1}{2}}(Z) \cr
  \qquad\qquad\qquad\qquad\qquad\qquad\qquad\qquad (d:\odd) \,, \cr
  {}-\displaystyle\frac{\ii\cos\pi(\frac{d-2}{2})\tan\pi\nu}
       {(2\pi)^{\frac{d}{2}}}\,e^{-\ii\pi\frac{d-2}{2}}\,
  \bigl(Z^2-1\bigr)^{-\frac{d-2}{4}}\,
  \LQ^{\frac{d-2}{2}}_{\nu-\frac{1}{2}}(Z) \cr
  \qquad\qquad\qquad\qquad\qquad\qquad\qquad\qquad (d:\even)\,.
 \end{array}\right. 
\label{case1}
\end{align}

\noindent{\bf\underline{(2) $1>Z>-1$}}

In this case, as can be seen from Fig.~\ref{fig:chgplus}, 
$\cosh\gamma_+$ crosses the branch cut between $-1<z<1$ from above 
and move to another Riemann sheet.
On the other hand, $\cosh\gamma_-$ crosses the branch cut 
between $-1<z<1$ from below. 
Thus, in this region, we have
\begin{align}
 &\sinh\gamma_{\pm}
   =e^{\pm\ii\frac{\pi}{2}}\,
    \bigl(1-Z^2\bigr)^{\frac{1}{2}}\,, \\
 &\LQ^{\frac{d-2}{2}}_{\nu-\frac{1}{2}}(\cosh\gamma_{\pm}) 
   = \LQ^{\frac{d-2}{2}}_{\nu-\frac{1}{2}} (-Z\pm\ii 0)\nn\\
  &= e^{\ii\pi\frac{d-2}{2}(1\pm\frac{1}{2})}\, 
    \Bigl[\LQc^{\frac{d-2}{2}}_{\nu-\frac{1}{2}}(-Z)
       \mp\frac{\ii\pi}{2}\,\LPc^{\frac{d-2}{2}}_{\nu-\frac{1}{2}}(-Z)\Bigr]\,,
\end{align}
where we have used Eq.~\eqref{fun2}. 
Then, Eq.~\eqref{formula1} becomes
\begin{align}
 &e^{\ii\pi\frac{d-2}{2}}\,\bigl(1-Z^2\bigr)^{-\frac{d-2}{4}}\,
  \Bigl[\LQc^{\frac{d-2}{2}}_{\nu-\frac{1}{2}}(-Z)
    \mp\frac{\ii\pi}{2}\,\LPc^{\frac{d-2}{2}}_{\nu-\frac{1}{2}}(-Z)\Bigr]\nn\\
 &=2^{\frac{d}{2}-2}\,\Gamma\Bigl(\frac{d-2}{2}\Bigr)\,
   e^{\ii\pi\bigl(-\nu\mp\frac{\nu}{2}+\frac{d-2}{2}\bigr)}\,
   (\cos\varphi_1\,\cos\varphi_2)^{\frac{d-1}{2}} \nn\\
 &\quad\times\sum_{L=0}^{\infty} (2L+d-2)\,\nn\\
 &\qquad\times \LPc^{-\nu}_{k}(\sin\varphi_1)\,
  \LQ^{\nu}_{k}(\sin\varphi_2\pm\ii 0)\, C_L^{\frac{d-2}{2}}(\cos\theta) \,.
\end{align}
By taking the difference of the equations 
with the upper and the lower signs, we obtain
\begin{align}
 &\bigl(1-Z^2\bigr)^{-\frac{d-2}{4}}\,
  \LPc^{\frac{d-2}{2}}_{\nu-\frac{1}{2}}(-Z) \nn\\
 &=2^{\frac{d}{2}-2}\,\Gamma\Bigl(\frac{d-2}{2}\Bigr)\,
   (\cos\varphi_1\,\cos\varphi_2)^{\frac{d-1}{2}}\, \nn\\
 &\times \sum_{L=0}^{\infty} (2L+d-2)\,\LPc^{-\nu}_{k}(\sin\varphi_1)\,
   \LPc^{\nu}_{k}(\sin\varphi_2)\,C_L^{\frac{d-2}{2}}(\cos\theta) \,.
\end{align}
Similarly, by taking their sum, we obtain
\begin{align}
 &\bigl(1-Z^2\bigr)^{-\frac{d-2}{4}}\,
  \LQc^{\frac{d-2}{2}}_{\nu-\frac{1}{2}}(-Z) \nn\\
 &=2^{\frac{d}{2}-2}\,\Gamma\Bigl(\frac{d-2}{2}\Bigr)\,
   (\cos\varphi_1\,\cos\varphi_2)^{\frac{d-1}{2}} \,\nn\\
 &\times \sum_{L=0}^{\infty} (2L+d-2)\,\LPc^{-\nu}_{k}(\sin\varphi_1)\,
   \LPc^{\nu}_{k}(\sin\varphi_2)\,C_L^{\frac{d-2}{2}}(\cos\theta)\,.
\end{align}
The right hand side of Eqs.~\eqref{formulaodd} and \eqref{formulaeven} 
can then be written as follows:
\begin{align}
 \left\{\begin{array}{cl}
  \displaystyle\frac{\ii\pi}{2\,(2\pi)^{\frac{d}{2}}\sin(\pi\nu)}\,
  \bigl(1-Z^2\bigr)^{-\frac{d-2}{4}}\,
  \LPc^{\frac{d-2}{2}}_{\nu-\frac{1}{2}} (-Z)\quad
  & (d:\odd)\,, \cr
  \displaystyle\frac{\ii}{(2\pi)^{\frac{d}{2}}\cos(\pi\nu)}\,
  \bigl(1-Z^2\bigr)^{-\frac{d-2}{4}}\,\LQc^{\frac{d-2}{2}}_{\nu-\frac{1}{2}} (-Z)
  & (d:\even)\,.
 \end{array}\right.
\label{case2}
\end{align}

\noindent{\bf\underline{(3) $Z<-1$}}

In this case, from Fig.~\ref{fig:chgplus}, 
we know that, in the region $\Ree z>1$, 
$\cosh\gamma_+$ runs above the real axis, 
or below the real axis in the next Riemann sheet 
after passing through the cut on $-1<z<1$ from above. 
On the other hand, in the region $\Ree z>1$, 
$\cosh\gamma_-$ runs below the real axis, 
or above the real axis in another sheet 
after passing through the cut on $-1<z<1$ from below.

When $\cosh\gamma_\pm=-Z\pm\ii 0$, we have
\begin{align}
 \sinh\gamma_{\pm}
  = \bigl(Z^2-1\bigr)^{\frac{1}{2}} \,,\quad 
 \LQ^{\frac{d-2}{2}}_{\nu-\frac{1}{2}} (\cosh\gamma_{\pm})
  = \LQ^{\frac{d-2}{2}}_{\nu-\frac{1}{2}} (-Z) \,,
\end{align}
and then the following equation is obtained:
\begin{align}
 &\bigl(Z^2-1\bigr)^{-\frac{d-2}{4}}\,\LQ^{\frac{d-2}{2}}_{\nu-\frac{1}{2}} (-Z) \nn\\
 &=2^{\frac{d}{2}-2}\,\Gamma\Bigl(\frac{d-2}{2}\Bigr)\,
   e^{\ii\pi\bigl(-\nu\mp\frac{\nu}{2}+\frac{d-2}{2}\bigr)}\,
   (\cos\varphi_1\,\cos\varphi_2)^{\frac{d-1}{2}} \nn\\
 &\times\sum_{L=0}^{\infty} (2L+d-2)\,\nn\\
 &\qquad \times\LPc^{-\nu}_{k}(\sin\varphi_1)\,
   \LQ^{\nu}_{k}(\sin\varphi_2\pm\ii 0)\,C_L^{\frac{d-2}{2}}(\cos\theta)\,.
\end{align}
By taking the difference and the sum of the above equations 
with the upper and the lower signs, we obtain
\begin{align}
 0&= {}-\ii\pi\, 2^{\frac{d}{2}-2}\,\Gamma\Bigl(\frac{d-2}{2}\Bigr)\,
     e^{\ii\pi\frac{d-2}{2}}\,(\cos\varphi_1\,\cos\varphi_2)^{\frac{d-1}{2}} \nn\\
 & \times\sum_{L=0}^{\infty} (2L+d-2)\,\LPc^{-\nu}_{k}(\sin\varphi_1)\,
     \LPc^{\nu}_{k}(\sin\varphi_2)\,C_L^{\frac{d-2}{2}}(\cos\theta)\,,  \label{odd1}\\
 &\!\!\! 2\bigl(Z^2-1\bigr)^{-\frac{d-2}{4}}\,\LQ^{\frac{d-2}{2}}_{\nu-\frac{1}{2}}(-Z)\nn\\
 &= 2^{\frac{d}{2}-1}\,\Gamma\Bigl(\frac{d-2}{2}\Bigr)\,
    e^{\ii\pi\frac{d-2}{2}}\, (\cos\varphi_1\,\cos\varphi_2)^{\frac{d-1}{2}} \nn\\
 & \times\sum_{L=0}^{\infty} (2L+d-2)\,\LPc^{-\nu}_{k}(\sin\varphi_1)\,
   \LQc^{\nu}_{k}(\sin\varphi_2)\,C_L^{\frac{d-2}{2}}(\cos\theta)\,. \label{even1}
\end{align}
On the other hand, when $\cosh\gamma_\pm=-Z\mp\ii 0$, 
we need to evaluate the functions $\sinh\gamma_{\pm}$ 
and $\LQ^{\frac{d-2}{2}}_{\nu-\frac{1}{2}} (\cosh\gamma_{\pm})$ 
on a new Riemann sheet, 
since $\cosh\gamma_\pm$ has already crossed the branch cut. 
We then have
\begin{align}
 \sinh\gamma_{\pm}&=e^{\pm\ii\pi}\,\bigl(Z^2-1\bigr)^{\frac{1}{2}}\,, \\
 \LQ^{\frac{d-2}{2}}_{\nu-\frac{1}{2}} (\cosh\gamma_{\pm}) 
 &= e^{\pm\ii\pi\frac{d-2}{2}}\,
    \LQ^{\frac{d-2}{2}}_{\nu-\frac{1}{2}}(-Z)\nn\\
 &\quad \mp\ii\pi\, e^{\ii\pi\frac{d-2}{2}}\,
    \LP^{\frac{d-2}{2}}_{\nu-\frac{1}{2}}(-Z)\,,
\end{align}
and Eq.~\eqref{formula1} takes the form
\begin{align}
 &\bigl(Z^2-1\bigr)^{-\frac{d-2}{4}}\,
  \Bigl[\LQ^{\frac{d-2}{2}}_{\nu-\frac{1}{2}} (-Z)
        \mp\ii\pi\, e^{\pm\ii\pi\frac{d-2}{2}(1\mp1)}\,
        \LP^{\frac{d-2}{2}}_{\nu-\frac{1}{2}} (-Z)\Bigr] \nn\\
 &=2^{\frac{d}{2}-2}\,\Gamma\Bigl(\frac{d-2}{2}\Bigr)\,
   e^{\ii\pi\bigl(-\nu\mp\frac{\nu}{2}+\frac{d-2}{2}\bigr)}\,
   (\cos\varphi_1\,\cos\varphi_2)^{\frac{d-1}{2}} \nn \\
 &\quad\times\sum_{L=0}^{\infty}(2L+d-2)\,\nn\\
 &\qquad\times\LPc^{-\nu}_{k}(\sin\varphi_1)\,
  \LQ^{\nu}_{k}(\sin\varphi_2\pm\ii 0)\,C_L^{\frac{d-2}{2}}(\cos\theta)\,. 
\label{pm}
\end{align}
By taking the difference of the above equations 
with the upper and the lower signs, we obtain
\begin{align}
&\cos\Bigl(\frac{d-2}{2}\pi\Bigr)\bigl(Z^2-1\bigr)^{-\frac{d-2}{4}} 
\LP^{\frac{d-2}{2}}_{\nu-\frac{1}{2}} (-Z) \nn \\
&=2^{\frac{d}{2}-2}\Gamma\Bigl(\frac{d-2}{2}\Bigr)e^{\ii\pi\frac{d-2}{2}}
(\cos\varphi_1\,\cos\varphi_2)^{\frac{d-1}{2}}\nn\\
&\times\sum_{L=0}^{\infty} (2 L+d-2) \LPc^{-\nu}_{k}(\sin\varphi_1)\LPc^{\nu}_{k}(\sin\varphi_2)
 C_L^{\frac{d-2}{2}}(\cos\theta) \,.
\end{align}
In particular, in odd dimensions, we have
\begin{align}
 &0 
  =2^{\frac{d}{2}-2}\,\Gamma\Bigl(\frac{d-2}{2}\Bigr)\,
   e^{\ii\pi\frac{d-2}{2}}\, (\cos\varphi_1\,\cos\varphi_2)^{\frac{d-1}{2}} \nn\\
 &\quad \times\sum_{L=0}^{\infty} (2L+d-2)\,\LPc^{-\nu}_{k}(\sin\varphi_1)\,
  \LPc^{\nu}_{k}(\sin\varphi_2)\,C_L^{\frac{d-2}{2}}(\cos\theta)\,, \label{odd2}
\end{align}
which is equivalent to Eq.~\eqref{odd1}. 
Similarly, by taking the sum of Eq.~\eqref{pm} 
with the upper and the lower signs, we have
\begin{align}
 &\bigl(Z^2-1\bigr)^{-\frac{d-2}{4}} \Bigl[\LQ^{\frac{d-2}{2}}_{\nu-\frac{1}{2}} (-Z)\nn\\
 &\qquad\qquad\qquad\quad -\pi e^{\ii\pi\frac{d-2}{2}}\sin\Bigl(\frac{d-2}{2}\pi\Bigr)
 \LP^{\frac{d-2}{2}}_{\nu-\frac{1}{2}} (-Z)\Bigr] \nn \\
 &=2^{\frac{d}{2}-2}\Gamma\Bigl(\frac{d-2}{2}\Bigr)e^{\ii\pi\frac{d-2}{2}}
 (\cos\varphi_1\,\cos\varphi_2)^{\frac{d-1}{2}}\nn\\
 &\times\sum_{L=0}^{\infty} (2 L+d-2) \LPc^{-\nu}_{k}(\sin\varphi_1)\LQc^{\nu}_{k}(\sin\varphi_2)
 C_L^{\frac{d-2}{2}}(\cos\theta) \,.
\end{align}
In particular, in even dimensions, we have
\begin{align}
 &\bigl(Z^2-1\bigr)^{-\frac{d-2}{4}}\,
  \LQ^{\frac{d-2}{2}}_{\nu-\frac{1}{2}}(-Z) \nn\\
 &=2^{\frac{d}{2}-2}\,\Gamma\Bigl(\frac{d-2}{2}\Bigr)\,
   e^{\ii\pi\frac{d-2}{2}}\, (\cos\varphi_1\,\cos\varphi_2)^{\frac{d-1}{2}}\,\nn\\
 &\times \sum_{L=0}^{\infty} (2 L+d-2) \,\LPc^{-\nu}_{k}(\sin\varphi_1)\,
  \LQc^{\nu}_{k}(\sin\varphi_2)\,C_L^{\frac{d-2}{2}}(\cos\theta)\,, \label{even2}
\end{align}
which is equivalent to Eq.~\eqref{even1}. 
Thus, when $Z<-1$, the right hand sides of Eqs.~\eqref{formulaodd} and 
\eqref{formulaeven} always take the forms
\begin{align}
 \left\{\begin{array}{r}
  \displaystyle 0 \qquad\qquad\qquad\qquad (d:\odd) \,, \\[3mm]
  \displaystyle 
  \frac{\ii}{(2\pi)^{\frac{d}{2}}\,\cos(\pi\nu)}\,e^{-\ii\pi\frac{d-2}{2}}\,
  \bigl(Z^2-1\bigr)^{-\frac{d-2}{4}}\,
  \LQ^{\frac{d-2}{2}}_{\nu-\frac{1}{2}}(-Z)\cr
  (d:\even)\,. 
 \end{array}\right.
\label{case3}
\end{align}

We thus have obtained simplified expressions for Eq.~\eqref{formula1} 
for three different regions of $Z$ 
in the form Eqs.~\eqref{case1}, \eqref{case2}, and \eqref{case3}. 
One can readily see that three equations 
can be obtained from Eqs.~\eqref{formulaodd} and \eqref{formulaeven}.
This completes the proof of our assertion. 

\section{Integral representation of the associated Legendre functions}
\label{appendix:Legendre_integral}

In this appendix, we give a proof of Eq.~\eqref{Legendre_integral_rep} 
[we write it again here for convenience], 
\begin{align}
 &\LQ_{\nu-1/2}^{\frac{d-2}{2}}(u)\nn\\
 &= e^{\ii\pi\frac{d-2}{2}} \,
 \int_0^\infty\rmd\lambda\, 
 \frac{\Gamma\bigl(\frac{d-1}{2}+\ii\lambda\bigr)\,
       \Gamma\bigl(\frac{d-1}{2}-\ii\lambda\bigr)}
      {\Gamma(\ii\lambda)\,\Gamma(-\ii\lambda)}\,
 \frac{\LP_{\ii\lambda-1/2}^{-\frac{d-2}{2}}(u)}{\nu^2+\lambda^2} \nn\\
 &\qquad\qquad\qquad\qquad\qquad \bigl[d\in\mathbb{Z}\,,\quad \Ree\nu>0\bigr]\,, 
\label{Legendre_integral_rep2}
\end{align}
treating the odd and even dimensional cases separately. 
Our discussion is heavily based on the derivation of the heat kernel 
in Euclidean AdS space performed in \cite{Grosche:1987de} 
(see also \cite{Akhmedov:2009ta}).

\subsubsection*{Odd dimensions}

When $d$ is odd, using the direct relation between the associated Legendre functions 
and the Gegenbauer functions, 
\begin{align}
 \LP_{\ii\lambda-\frac{1}{2}}^{-\frac{d-2}{2}}(u) 
 &= \frac{2^{\frac{d-2}{2}}\,\pi^{1/2}\, \Gamma\bigl(\frac{d-1}{2}\bigr)}
         {\sin\bigl[\pi\,\bigl(\frac{d-1}{2}-\ii\lambda \bigr)\bigr]\,
  \Gamma\bigl(\frac{d-1}{2}+\ii\lambda\bigr)\,\Gamma\bigl(\frac{d-1}{2}-\ii\lambda\bigr)}\,\nn\\
 &\qquad\qquad \times (u^2-1)^{\frac{d-2}{4}}\, C_{\ii\lambda-\frac{d-1}{2}}^{\frac{d-1}{2}}(u)\,,
\end{align}
we can rewrite the integral on the right-hand side of \eqref{Legendre_integral_rep2} as
\begin{align}
 &\int_0^\infty\rmd\lambda\, 
 \frac{\Gamma\bigl(\frac{d-1}{2}+\ii\lambda\bigr)\,\Gamma\bigl(\frac{d-1}{2}-\ii\lambda\bigr)}
      {\Gamma(\ii\lambda)\,\Gamma(-\ii\lambda)}\,
 \frac{\LP_{\ii\lambda-1/2}^{-\frac{d-2}{2}}(u)}{\nu^2+\lambda^2} \nn\\
 &=\frac{2^{\frac{d-2}{2}}\,e^{\ii\pi d/2}\,\Gamma\bigl(\frac{d-1}{2}\bigr)}{\pi^{1/2}}\,
    \bigl(u^2-1\bigr)^{\frac{d-2}{4}} \,\nn\\
 &\quad\times\int_0^\infty\rmd\lambda\, \frac{\lambda}{\nu^2+\lambda^2} \, 
 C_{\ii\lambda-\frac{d-1}{2}}^{\frac{d-1}{2}}(u) \,.
\label{Legendre-Gegenbauer}
\end{align}
By further using the identities for the Gegenbauer function  
\begin{align}
 \ii\lambda\,C_{\ii\lambda-n}^n(\cosh\gamma)
 =\frac{2^{1-n}}{\Gamma(n)}\,
  \Bigl[\frac{\rmd}{\rmd (\cosh\gamma)}\Bigr]^n\cos(\lambda\gamma)
\end{align}
with nonnegative integers $n$\,, 
Eq.~\eqref{Legendre-Gegenbauer} can be rewritten as follows:
\begin{align}
 &=\frac{2^{1/2}\,e^{\ii\pi \frac{d-1}{2}}}{\pi^{1/2}}\,
   \bigl(\cosh^2\gamma-1\bigr)^{\frac{d-2}{4}}\,\nn\\
 &\qquad\times\Bigl[\frac{\rmd}{\rmd (\cosh\gamma)}\Bigr]^{\frac{d-1}{2}}\,
  \int_0^\infty\rmd\lambda\, \frac{\cos(\lambda\gamma)}{\nu^2+\lambda^2} \nn\\
 &= \frac{\pi^{1/2}\,e^{\ii\pi \frac{d-1}{2}}}{2^{1/2}\,\nu}\,
  \bigl(\cosh^2\gamma-1\bigr)^{\frac{d-2}{4}}\,\nn\\
 &\qquad\times\Bigl[\frac{\rmd}{\rmd (\cosh\gamma)}\Bigr]^{\frac{d-1}{2}}\,
  e^{\mp\gamma\,\nu} \quad [\,\Ree\gamma\gtrless0\,,\quad \Ree\nu>0\,]\nn\\
 &= e^{-\ii\pi \frac{d-1}{2}}\,\LQ_{\nu -1/2}^{\frac{d-2}{2}}(\cosh\gamma)
 \quad \quad [\,\Ree\nu>0\,]\,.
\end{align}
Here, in order to show the second equality, we have used the formula [3.723-2] of \cite{GR}:
\begin{align}
 &\int_0^\infty\rmd\lambda\, \frac{\cos(\lambda\gamma)}{\nu^2+\lambda^2}\nn\\
 &= \left\{\begin{array}{ll}
   \displaystyle 
   \frac{\pi}{2\nu}\,e^{-\gamma\,\nu}  & [\,\Ree\gamma>0\,,\quad \Ree\nu>0\,] \\[3mm]
   \displaystyle 
   \frac{\pi}{2\nu}\,e^{\gamma\,\nu}  & [\,\Ree\gamma<0\,,\quad \Ree\nu>0\,]
\end{array}\right.
 \,,
\end{align}
and to show the third equality 
we have used the following formula [see (A.20) of \cite{Grosche:1987de}]: 
\begin{align}
 &e^{-\ii\pi \frac{d-2}{2}}\,\LQ_{\nu -1/2}^{\frac{d-2}{2}}(\cosh\gamma)\nn\\
 &= \left\{\begin{array}{ll}
   \displaystyle \frac{\pi^{1/2}\,e^{\ii\pi \frac{d-1}{2}}}{2^{1/2}\,\nu}\,
    \bigl(\cosh^2\gamma-1\bigr)^{\frac{d-2}{4}}\, & \cr
   \displaystyle
    \qquad\times \Bigl[\frac{\rmd}{\rmd (\cosh\gamma)}\Bigr]^{\frac{d-1}{2}}\,
    e^{-\gamma\,\nu} & [\Ree \gamma>0]\cr
  \displaystyle \frac{\pi^{1/2}\,e^{\ii\pi \frac{d-1}{2}}}{2^{1/2}\,\nu}\,
                \bigl(\cosh^2\gamma-1\bigr)^{\frac{d-2}{4}}\, & \cr
  \displaystyle 
    \qquad \times \Bigl[\frac{\rmd}{\rmd (\cosh\gamma)}\Bigr]^{\frac{d-1}{2}}\,
    e^{\gamma\,\nu} & [\Ree \gamma<0]
\end{array}\right.
 \,.
\end{align}

\subsubsection*{Even dimensions}

When $d$ is even, the equality
\begin{align}
 \LP_{\ii\lambda-\frac{1}{2}}^{-\frac{d-2}{2}}(u) 
 = \frac{\Gamma\bigl(\ii\lambda-\frac{d-1}{2}+1\bigr)}
        {\Gamma\bigl(\ii\lambda+\frac{d-1}{2}\bigr)}\,
   \LP_{\ii\lambda-\frac{1}{2}}^{\frac{d-2}{2}}(u) 
\end{align}
holds, and we can show \eqref{Legendre_integral_rep2} as follows:
\begin{align}
 &\int_0^\infty\rmd\lambda\, 
 \frac{\Gamma\bigl(\frac{d-1}{2}+\ii\lambda\bigr)\,\Gamma\bigl(\frac{d-1}{2}-\ii\lambda\bigr)}
      {\Gamma(\ii\lambda)\,\Gamma(-\ii\lambda)}\,
 \frac{\LP_{\ii\lambda-1/2}^{-\frac{d-2}{2}}(u)}{\nu^2+\lambda^2}\nn\\
 &=(-1)^{\frac{d-2}{2}} \int_0^\infty \rmd \lambda \, 
 \frac{\lambda\,\tanh(\pi\lambda)}{\nu^2+\lambda^2}\,
 \LP_{\ii\lambda-1/2}^{\frac{d-2}{2}}(u) \nn\\
 &= (-1)^{\frac{d-2}{2}} \,\bigl(u^2-1\bigr)^{\frac{d-2}{4}} \, \nn\\
 &\quad\times\Bigl(\frac{\rmd}{\rmd u}\Bigr)^{\frac{d-2}{2}}
 \int_0^\infty \rmd \lambda \, 
 \frac{\lambda\,\tanh(\pi\lambda)}{\nu^2+\lambda^2}\,
 \LP_{\ii\lambda-1/2}(u)\nn\\
 &= (-1)^{\frac{d-2}{2}} \,\bigl(u^2 -1\bigr)^{\frac{d-2}{4}}\, 
 \Bigl(\frac{\rmd}{\rmd u}\Bigr)^{\frac{d-2}{2}}
 Q_{\nu-1/2}(u)\nn\\
 &= e^{-\ii\pi\frac{d-2}{2}} \, Q_{\nu-1/2}^{\frac{d-2}{2}}(u)\,,
\end{align}
where we have used the identities for the associated Legendre functions 
with integer order, 
\begin{align}
 \bigl(u^2 -1\bigr)^{n/2}\,
 \frac{\rmd^n}{\rmd u^n}\,\LP_{\ii\lambda-\frac{1}{2}}(u)
  &= \LP_{\ii\lambda-\frac{1}{2}}^n(u) \,,\\
 \bigl(u^2 -1\bigr)^{n/2}\,
 \frac{\rmd^n}{\rmd u^n}\,\LQ_{\nu-\frac{1}{2}}(u)
  &= \LQ_{\nu-\frac{1}{2}}^n(u) \,,
\end{align}
and the formula [7.213] of \cite{GR}:
\begin{align}
 \int_0^\infty \rmd \lambda \, 
 \frac{\lambda\,\tanh(\pi\lambda)}{\nu^2+\lambda^2}\,
 \LP_{\ii\lambda-1/2}(u) = Q_{\nu-1/2}(u)
\nn\\ 
 [\Ree \nu>0]\,.
\end{align}

\section{$\alpha$-vacua}
\label{appendix:alpha}

Since the in-in and in-out propagators in de Sitter space always 
have de Sitter invariant forms, 
it is natural to expect that the in- and out-vacua 
both in the Poincar\'e and the global patch 
belong to a family of de Sitter invariant vacua, 
i.e., the $\alpha$-vacua (or the Mottola-Allen vacua) 
\cite{Mottola:1984ar,Allen:1985ux}. 
In this appendix, we calculate the values of $\alpha\in\mathbb{C}$ 
associated to the in- and out-vacua 
(both in the Poincar\'e and the global patch) explicitly. 
We will set $\varepsilon=0$ in the following discussions.

Given a mode expansion, 
an $\alpha$-vacuum is defined for a complex number $\alpha$ 
with $\Ree \alpha<0$ 
such that the corresponding wave function 
for each mode $n$ is given by 
\begin{align}
 \varphi^{(\alpha)}_n(t) 
  = \frac{1}{\sqrt{1-\abs{e^{\alpha}}^2}}\,
 \big[\varphi^{\rm(E)}_n(t)+e^{\alpha}\,\varphi^{\rm(E)\,*}_n(t)\bigr]\,,
\end{align}
where $\varphi^{\rm(E)}_n(t)$ is the wave function of the Euclidean vacuum. 
The Feynman propagator associated with the $\alpha$-vacuum 
is then given (for each mode) as 
\begin{align}
 &G_n^{(\alpha)}(t,t') 
\nn\\
 &= \frac{\ii}{W_\rho[\varphi^{(\alpha)}_n,\varphi^{(\alpha)\,*}_n]}\,
   \varphi^{(\alpha)}_n(t_>) \,\varphi^{(\alpha)\,*}_n(t_<) 
\nn\\
 &= \frac{1}{1-\abs{e^{\alpha}}^2}\,\frac{\ii}{W_\rho[\varphi^{\rm(E)}_n,\varphi^{\rm(E)\,*}_n]}
\nn\\
 &\times\bigl[
   \varphi^{\rm(E)}_n(t_>) \,\varphi^{\rm(E)\,*}_n(t_<) 
  +\abs{e^{\alpha}}^2\,\varphi^{\rm(E)\,*}_n(t_>) \,\varphi^{\rm(E)}_n(t_<) 
\nn\\
 &\qquad
  +e^{\alpha^*}\,\varphi^{\rm(E)}_n(t_>) \,\varphi^{\rm(E)}_n(t_<)
  +e^{\alpha}\,\varphi^{\rm(E)\,*}_n(t_>) \,\varphi^{\rm(E)\,*}_n(t_<) 
 \bigr] \,.
\end{align}
We also have multiplied the factor 
$\ii/W_\rho[\varphi^{(\alpha)}_n,\varphi^{(\alpha)\,*}_n]$ 
with which we need not care about the normalization of wave functions. 
Thus, if we expand the in-in or out-out propagator (for each mode) 
as a quadratic form of $\varphi^{\rm(E)}_n$ and $\varphi^{\rm(E)\,*}_n$, 
we can find the values of $\alpha$ associated to the in- or out-vacua. 
Here, the in-in propagator is defined as in \eqref{G00} and \eqref{G_in_in}, 
and the out-out propagator is defined by
\begin{align}
 G^{\out/\out}(t,t') 
  &\equiv \lim_{t_1 \to t_f}G_{11}(t,t'; t_1,t_1) \,,
\nn\\
 G_{11}(t,t'; t_1,t_1)
  &\equiv \frac{\overline{\bra{0_{t_1}}}\, {\rm T}\, q(t)\, q^\dagger(t') \,\overline{\ket{0_{t_1}}}}
    {\overline{\langle 0_{t_1}\rvert 0_{t_{1}} \rangle }} \,,
\label{G_out_out}
\end{align}
which can be shown to take the form
\begin{align}
 &G^{\out/\out}(t,t')
\nn\\
 &= \lim_{t_1\to t_f}\frac{\ii}{V_\rho[\varphi(t;t_1),\,\varphi^*(t;t_1)](T_s)}\,
     \varphi(t_{>};t_1)\,\varphi^*(t_{<};t_1) \,.
\end{align}

Since the in-in propagator in the Poincar\'e patch \eqref{Poincare_in_in_full} 
coincide with the Feynman propagator in the Euclidean vacuum, 
the in-vacuum in the Poincar\'e patch is identified 
with the Euclidean vacuum (i.e., the $\alpha$-vacuum with $\alpha=-\infty$). 
On the other hand, by using \eqref{out-bar-phi}, 
we can show that the out-out propagator for each mode 
in the Poincar\'e patch [which is finite only if $m>(d-1)/2$] 
takes the following form:
\begin{align}
 &G_\bk^{\out/\out}(\eta,\eta') 
\nn\\
 &= \frac{\pi\,[(-\eta)(-\eta')]^{(d-1)/2}}{2\sinh(\pi\mu)}\,
   J_{\ii\mu}(-k_{\varepsilon}\eta_>)\,J_{-\ii\mu}(-k_{-\varepsilon}\eta_<)\,.
\end{align}
If we use the wave function associated with the Euclidean vacuum
\begin{align}
 \varphi^{\rm(E)}_k(t) 
 = \beta\, \frac{\sqrt{\pi}}{2}\,e^{-\frac{\pi\mu}{2}}\,
   (-\eta)^{\frac{d-1}{2}}\,H^{(1)}_{\ii\mu}(-k \eta)
\end{align}
with $\beta$ an arbitrary complex constant, 
we can expand the out-out propagator for each mode as follows:
\begin{align}
 &G_\bk^{\out/\out}(t,t') 
\nn\\
 &= \frac{\pi}{1-e^{-2\pi\mu}}\,\frac{\ii}{W_\rho[\varphi^{\rm(E)}_k(t),\varphi^{\rm(E)\,*}_k(t)]}
\nn\\
 &\times\bigl[
   \varphi^{\rm(E)}_n(t_>) \,\varphi^{\rm(E)\,*}_n(t_<) 
  +e^{-2\pi\mu}\,\varphi^{\rm(E)\,*}_n(t_>) \,\varphi^{\rm(E)}_n(t_<) 
\nn\\
 &\qquad
  +\frac{\beta^*}{\beta}\,e^{-\pi\mu}\,\varphi^{\rm(E)}_n(t_>) \,\varphi^{\rm(E)}_n(t_<)
\nn\\
 &\qquad
  +\frac{\beta}{\beta^*}\,e^{-\pi\mu}\,\varphi^{\rm(E)\,*}_n(t_>) \,\varphi^{\rm(E)\,*}_n(t_<) 
 \bigr] \,.
\end{align}
If we choose the constant $\beta$ real, 
the out-vacuum is shown to correspond to the $\alpha$-vacuum with $\alpha=-\pi\mu$. 

In the global patch, the in-in and out-out propagators for each mode 
in the heavy mass case ($m>(d-1)/2$) take the following forms:
\begin{align}
 &G^{\inn/\inn}_{L}(t,t') \nn\\
  &=\left\{\begin{array}{l}
    \displaystyle
     \frac{\pi}{2\sinh(\pi\mu)}\,
     \bigl[(1-t_{>}^2)\,(1-t_{<}^2)\bigr]^{\frac{d-1}{4}} \\
     \qquad\qquad\qquad\times \LPc^{-\ii\mu}_{k}(t_>)\,
     \LPc^{\ii\mu}_{k}(t_<) \qquad (d:\odd)\,,\\
    \displaystyle
     \frac{2}{\pi\sinh(\pi\mu)}\,
     \bigl[(1-t_{>}^2)\,(1-t_{<}^2)\bigr]^{\frac{d-1}{4}}\,\\
     \qquad\qquad\qquad\times \LQc^{-\ii\mu}_{k}(t_>)\,
     \LQc^{\ii\mu}_{k}(t_<) \qquad (d:\even)\,,
 \end{array}\right. \nn\\
 &G^{\out/\out}_{L}(t,t') \nn\\
 &=\frac{\pi}{2\sinh(\pi\mu)}\,
     \bigl[(1-t_{>}^2)\,(1-t_{<}^2)\bigr]^{\frac{d-1}{4}}\,
  \LPc^{-\ii\mu}_{k}(t_>)\, \LPc^{\ii\mu}_{k}(t_<)\,.
\end{align}
On the other hand, the wave function associated with the Euclidean vacuum 
is known to have the following form (see e.g., \cite{Bousso:2001mw}):
\begin{align}
 &\varphi^{\rm(E)}_L(t) 
 = \beta\, \frac{\sqrt{\pi}\,\Gamma(L+\frac{d-1}{2}+\ii\mu)\,\cosh^L\tau\,e^{(L+\frac{d-1}{2}+\ii\mu)\,\tau}}
                {2^{L+\frac{d-1}{2}}\,e^{-\ii\pi(L+\frac{d-1}{2})}\,e^{\pi\mu}\,\Gamma(L+\frac{d}{2})}
\nn\\
 &\times F\Bigl(L+\frac{d-1}{2},L+\frac{d-1}{2}+\ii\mu,2L+d-1;1+e^{2\tau}-\ii0\Bigr)
\nn\\
 &=\beta\, (1-t^2)^{\frac{d-1}{4}}\,\Bigl[\LQc_k^{\ii\mu}(t)+\frac{\ii\pi}{2}\,\LPc_k^{\ii\mu}(t)\Bigr] 
 \quad \bigl(t\equiv \tanh\tau\bigr)\,.
\end{align}
Then, by using the relations
\begin{align}
 \LPc^{\ii\mu}_{k}(t) 
  &= \frac{-\ii}{\pi\beta} \,\Bigl(\varphi^{\rm(E)}_L(t) 
   - e^{\pi\mu}\,\frac{\beta\,\Gamma(k+1+\ii\mu)}
                       {\beta^*\,\Gamma(k+1-\ii\mu)}\,\varphi^{\rm(E)\,*}_L(t)\Bigr)\,,
\nn\\
 \LQc^{\ii\mu}_{k}(t)
  &= \frac{1}{2\beta} \,\Bigl(\varphi^{\rm(E)}_L(t) 
   + e^{\pi\mu}\,\frac{\beta\,\Gamma(k+1+\ii\mu)}
                       {\beta^*\,\Gamma(k+1-\ii\mu)}\,\varphi^{\rm(E)\,*}_L(t)\Bigr)\,,
\end{align}
if we choose the constant $\beta$ such that $\arg\beta = \arg\Gamma(k+1-\ii\mu)$, 
we find that, in odd dimensions, the in- and out-vacua are the $\alpha$-vacua with $\alpha=-\pi\mu + \ii\pi$, 
while in even dimensions, 
the in-vacuum is the $\alpha$-vacuum with $\alpha=-\pi\mu$ and 
the out-vacuum is that with $\alpha=-\pi\mu+ \ii\pi$.

\section{Another $\ii\varepsilon$ prescription}
\label{appendix:iepsilon-different}

In this paper, the $\ii\varepsilon$ prescription is defined by the replacement 
\begin{align}
 \rho(t)\to e^{+\ii\varepsilon}\,\rho(t) \,,\qquad 
 \omega_n(t) \to e^{-\ii\varepsilon}\,\omega_n(t)\,,
\end{align}
which corresponds to the replacement 
$H_{n,\,s}(t)=e^{-\ii\varepsilon}\,\bigl[H_{n,\,s}(t)\rvert_{\varepsilon=0}\bigr]$\,.
Another standard definition of the $\ii\varepsilon$ prescription 
(which does not break the symmetry existing in the background spacetime) is given by
\begin{align}
 m^2\to m^2-\ii\varepsilon \,.
\end{align}
In this appendix, we comment on the difference between the two $\ii\varepsilon$ prescription. 

In fact, for global patch, there is no difference in the analytical results 
between the two $\ii\varepsilon$ prescription. 
On the other hand, for Poincar\'e patch, 
if we use the $\ii\varepsilon$ prescription given by $m^2\to m^2-\ii\varepsilon$, 
the wave functions have the form,
\begin{align}
 \varphi(\eta;\eta_0) 
  &\sim \frac{\sqrt{\pi}}{2}\, 
   e^{\ii \bigl(k\,\eta_0 +\frac{\pi(2 \nu_\varepsilon+1)}{4}\bigr)}\, 
   (-\eta)^{\frac{d-1}{2}}\,
   H^{(1)}_{\nu_\varepsilon}(-k\,\eta) \,,\\ 
 \bar{\varphi}(\eta;\eta_0)
 &\sim \frac{\sqrt{\pi}}{2}\, 
   e^{-\ii \bigl(k\,\eta_0 +\frac{\pi(2 \nu_\varepsilon+1)}{4}\bigr)}\, 
   (-\eta)^{\frac{d-1}{2}}\,
   H^{(2)}_{\nu_\varepsilon}(-k\,\eta) \,,\\
 \varphi(\eta;\eta_1)
 &\sim -\frac{\Gamma(\nu_\varepsilon)\,(k/2)^{-\nu_\varepsilon}}
           {2\sqrt{2\bar{m}_1}}\, 
    (-\eta_1)^{-\nu_\varepsilon}\,\nn\\
 &\qquad\times\Bigl(\frac{d-1}{2}- \nu_\varepsilon - \ii \bar{m}_1 \Bigr)\,
    (-\eta)^{\frac{d-1}{2}}\,J_{\nu_\varepsilon}(-k\,\eta) \,,\\
 \bar{\varphi}(\eta;\eta_1)
 &\sim -\frac{\Gamma(\nu_\varepsilon)\,(k/2)^{-\nu_\varepsilon}}
           {2\sqrt{2\bar{m}_1}}\, 
    (-\eta_1)^{-\nu_\varepsilon}\nn\\
 &\qquad\times \Bigl(\frac{d-1}{2}- \nu_\varepsilon + \ii \bar{m}_1 \Bigr)\,
    (-\eta)^{\frac{d-1}{2}}\,J_{\nu_\varepsilon}(-k\,\eta) \,.
\end{align}
These wave functions agree with \eqref{in-phi}--\eqref{out-bar-phi} 
after we take the limit $\varepsilon\to 0$, 
except for the wave function $\varphi(\eta;\eta_0)$\,. 
Since the in-in or in-out propagator does not use $\varphi(\eta;\eta_0)$, 
the propagators in the Poincar\'e patch do not depend on the manner 
of the $\ii\varepsilon$ prescription. 
Thus, in the both patches, there is no difference in the analytical results 
between the two $\ii\varepsilon$ prescription. 
However, for the numerical calculations given in section \ref{sec:path_integral}, 
these two prescription give slightly different results, 
and the $\ii\varepsilon$ prescription used in this paper seems to be better 
in comparison with the analytical results.




\end{document}